# AN EVENT DETECTION TECHNIQUE USING SOCIAL MEDIA DATA

A THESIS

Submitted to the
## FACULTY OF ENGINEERING AND TECHNOLOGY
## PANJAB UNIVERSITY, CHANDIGARH
for the degree of

## DOCTOR OF PHILOSOPHY
2019

MUSKAN

UNIVERSITY INSTITUTE OF ENGINEERING AND TECHNOLOGY
PANJAB UNIVERSITY, CHANDIGARH - 160014

# Abstract

People post information about different topics which are in their active vocabulary over social media platforms (like Twitter, Facebook, PInterest and Google+). They follow each other and it is more likely that the person who posts information about current happenings will receive better response. Manual analysis of huge amount of data on social media platforms is difficult. This has opened new research directions for automatic analysis of user-contributed social media documents. Automatic social media data analysis is difficult due to abundant information shared by users. Many researchers use Twitter data for Social Media Analysis (SMA) as the Twitter data is freely available in the public domain. One of the most studied research areas in SMA is 'Topic Detection and Tracking' (TDT). Among five different research directions in TDT, 'Event Detection' is being studied in this research work. Event Detection from social media data is used for different applications like traffic congestion detection, disaster and emergency management, and live news detection.

Nature of the information which is shared on twitter platform is short-text, noisy, and ambiguous. Thus, event detection and extraction of event phrases from user-generated and ill-formed data becomes challenging. To address these challenges, events are extracted from streaming social media data in the form of keyphrases using different cognitive properties. The motivation behind this research work is to provide substantial improvements in the lexical variation of event phrases while detecting events and sub-events from twitter data. In this research work, the approach towards event detection from social media data is divided into three phases namely: Identifying sub-graphs in Microblog Word Co-occurrence Network (WCN) which provides important information about keyphrases; Identifying multiple events from social media data; and Ranking contextual information of event phrases.

**Identifying sub-graphs in Microblog Word Co-occurrence Network (WCN) which provides essential information about keyphrases**- Microblog WCN creates the network of text in Microblogs by representing words as nodes, co-occurrences as edges, and the frequency of co-occurrences as edge weight. Traditional random walk based keyphrase extraction and ranking techniques are implemented for Microblogs WCN evolved from First

Story Detection (FSD) dataset. It is observed that there exist certain patterns among words of a topic. A comprehensive study is conducted to examine the structure of Microblog WCN using network models and network science properties. The Microblog WCN is found to be stable and scalable over increasing corpus size. It follows scale-free property for degree distribution and edge distribution. With respect to these observations, different heuristic values are used to retain high weighted edges in weighted and directed Microblog WCN. Finally, keyphrases are extracted using topological sorting of resulting sub-graphs with high weighted edges. This approach results into improved recall rate and reduced redundancy (for 'Root Two' heuristics among all) as compared to existing techniques namely Frequent Pattern Mining using Dynamic Support Values (FPM-DSV) (Alkhamees *et al.,* 2016), Soft Frequent Pattern Mining (SFPM), High Utility Pattern Mining (HUPC), and Transaction-based Rule Change Mining (TRCM). Limitation of the heuristic approach is that it is based on manual assumptions and manual judgments.

To overcome this limitation, Scale-Free (SF) edge distribution based network model 'k-bridge decomposition' is proposed to extract the most important sub-graphs. The resulting sub-graphs are obtained by iterative decomposition of Microblog WCN and give information about important phrases. In this approach, the termination of iterative decomposition depends upon controlling parameter $t$ which indicates the maximum number of nodes in resulting sub-graph. It is observed that using this parameter, resulting sub-graph may contain information about multiple events. There should be high cohesion among nodes of resulting sub-graphs to automatically extract a single event phrase. Although Microblog WCN is found to be disassortative, there exists some sub-graph which shows non-disassortative nature. These sub-graphs precisely defines event phrase. The lexical order is maintained using directed characteristic of Microblog WCN. Thus, a non-parametric k-Bridge decomposition and Assortativity based keyphrase extraction and Ranking measure (BArank) is proposed. BArank outperforms all the existing keyphrase extraction techniques namely TextRank, NErank, and TopicRank in terms of ROUGE scores, Precision, Recall, and F-measure. Experiments have been performed for FSD dataset. BArank is independent of controlling parameters and has improved Keyword Precision (K-PREC) and Keyword Recall (K-REC). and is used for event detection from social media data.

**Identifying multiple events from social media data**- Using k-bridge decomposition and topological sorting based keyphrase extraction technique, an event detection technique is proposed. In this approach, events are identified using keywords and key-phrases as obtained from resulting sub-graphs. The keywords are all the words of resulting sub-graph, but key-phrases refer to the phrase obtained using topological sorting. Thus, comparative analysis is performed for both KeyWord Based Results (KWBR) and Key-Phrase Based Results (KPBR) using First Story Detection (FSD) dataset. The performance evaluation measures used for this study are Topic Recall (T-REC), Keyword Recall (K-REC), and Keyword Precision (K-PREC). It is observed that KPBR performs better in terms of Topic Recall (T-REC) and K-REC. However, KWBR performs better in terms of K-PREC. This is because keywords give more precise information than Keyphrases. As observed earlier that k-bridge decomposition technique depends upon controlling parameter, BArank is proposed and thus, is used for event detection technique. A discrete set of streaming social media is extracted and summaries are obtained using BArank keyphrase extraction. As there are multiple event phrases which are obtained from a discrete set, there is a need to rank the context of the event phrases so obtained.

**Ranking contextual information of event phrases**- For ranking the context of event phrases which were obtained from resulting sub-graphs of Microblog WCN, an optimization technique, Analytical Hierarchical Process (AHP) is used. Four different attributes for different alternatives of event phrases are proposed based on the structure of Microblog WCN. These attributes are Edge Strength Density (ESD), Strength Difference (SD), Phrase Degree (PD), and Degree Computation (DC). The top-ranked event phrases are different topics as observed in a discrete set of streaming Microblogs. The contextual information which is obtained using AHP optimization techniques segregates and rank keyphrases in three forms namely topic (top-ranked keyphrases), detailed information (mid-ranked keyphrases), and relevant phrases (bottom ranked keyphrases). Using BArank and AHP optimization technique, a novel event detection technique is proposed. The proposed event detection technique outperforms existing techniques namely Latent Dirichlet Allocation (LDA) approach, SFPM, BNgram, and Nguyen method in terms of K-PREC and K-REC, and shows comparative performance in terms of T-REC.

In a nutshell, it is observed that the study of the structure of word co-occurrence networks evolved from Microblogs of social media data shows cognitive patterns. These cognitive patterns can be used for identifying significant relation among words which are being used by users. Users often use words from the active language which is used within his social community. The proposed event detection technique is highly effective in detecting events and extracting event phrases using latent cognitive patterns in words of human-generated textual data.

# Acknowledgement

I extend my thanks to my guide Dr. Mukesh Kumar, Assistant Professor, Department of Computer Science & Engineering, UIET, Panjab University, Chandigarh, for providing me this golden opportunity to pursue my doctorate under his supervision. His presence gave me support, advice gave me direction and his belief gave me motivation to pursue research with utmost dedication. I thank Dr. Sudarshan Iyengar, Assistant Professor, Department of Computer Science, IIT Ropar, for giving me valuable guidance and for conducting a workshop on "Networks and Games - 2015" which gave me direction during initial stages of my research work. I would like to thank Prof. Savita Gupta, Director, UIET, Panjab University, Chandigarh for providing infrastructure and access to research facilities. I am also very grateful to my country and CSIR for providing me 'Direct SRF' grant during my research work. I would also like to thank Dr. Maunendra Desarkar from IIT Hyderabad, Dr. Swades De from IIT Delhi, Dr. R. Venkata Rao from NIT Surat, and Dr. Hirok Chaudhuri from NIT Durgapur for conducting great workshops which acted as catalyst during my research work.

I wish to thank all my friends for supporting me in this journey. I would like to thank Ms. Annu Mor, Ms. Tanu Goyal, Ms. Rashmi Vishraj, Ms. Prakriti Sodhi, Mr. Saurabh Sharma, Ms. Seema Rani, Mr. Subodh Bansal, Mr. Dinesh Vij, Mr. Sahil Kansal, Mr. Jaskirat Singh and all well-wishers with whom I had fruitful discussions and great learning experiences during my research work. I would like to thank my senior research fellows Dr. Raman Kumar Goyal, Dr. Amrith Krishna and Dr. Sumit Kalra for their useful inputs. I thank all my friends and acquaintances for constant support and valuable feedback during most critical times of my research work.

Finally, I thank almighty God for giving me immense strength and belief to pursue my research. I extend sincere thanks to my parents having faith in me. I would also like to thank my sister and my husband for their consistent love and care.

Muskan

# Table of Contents

# Publications

**Journals**

- Muskan Garg and Mukesh Kumar, "The Structure of Word Co-occurrence Network for Microblogs", *Physica A: Statistical Mechanics and its Applications*, *512*, pp.698-720, 2018. (**SCI indexed; Impact Factor – 2.132**)
- Muskan Garg and Mukesh Kumar, "Identifying Influential Segments from Word Co-occurrence Networks using AHP", *Cognitive Systems Research 47*, pp.28-41, 2018. (**SCI Indexed; Impact Factor – 1.182**)
- Muskan Garg and Mukesh Kumar, "Review on Event Detection Techniques in Social Multimedia", *Online Information Review*, Vol. 40, issue 3, pp. 347-361, 2016. (**SCIE Indexed; Impact Factor – 1.534**)

**Conferences**

- Muskan Garg and Mukesh Kumar, "Performance Analysis of Graph based Keyphrase Extraction metrics for uncertain User-generated data", ICACC 2018, Kochi India.
- Muskan Garg and Mukesh Kumar, "TWCM: Twitter Word Co-occurrence Model for Microblogs", ICACC 2018, Kochi, India.
- Muskan Garg and Mukesh Kumar, "Identifying Events using k-Bridge Decomposition from Twitter Feeds", ICACIE 2018, Bhubaneswar.
- Muskan Garg and Mukesh Kumar, "Comprehensive Study of Keyphrase Extraction metrics for uncertain User-generated data", MARC 2018, New Delhi.

**Communicated**

- Muskan Garg and Mukesh Kumar, "Cognitive Pattern based Keyphrase Extraction from Social Media Data using Markov Decision Process", IEEE Transaction on Knowledge and Data Engineering. (under Major Revision)

# List of Acronyms

| | |
|---|---|
| AHP | Analytical Hierarchical Process |
| API | Application Program Interface |
| ASPL | Average Shortest Path Length |
| BArank | k-Bridge decomposition and Assortativity based Ranking |
| CC | Clustering Coefficient |
| CI | Consistency Index |
| CR | Consistency Ratio |
| DAG | Directed Acyclic Graph |
| DARPA | Defense Advanced Research Projects Agency |
| DC | Degree Computation |
| ER random network | Erdos – Renyi random network |
| FPM – DSV | Frequent Pattern Mining – Dynamic Support Vector |
| FSD | First Story Detection |
| GM | Geometric Mean |
| HIN | Heterogeneous Information Network |
| HITS | Hypertext Induced Topic Search |
| HUPC | High Utility Pattern Clustering |
| IR | Information Retrieval |
| kBD_AHP | k-Bridge Decomposition using Analytical Hierarchical Process |
| K – PREC | Keyword Precision |
| K – REC | Keyword Recall |
| KECNW | Keyword Extraction using Collective Node Weight |
| KPBR | Key-Phrase Based Results |
| KWBR | Key-Word Based Results |
| LDA | Latent Dirichlet Allocation |
| MLS | Multiple Lexical Sequence |
| NED | New Event Detection |
| NLP | Natural Language Processing |
| PCM | Pairwise Comparison Matrix |
| PD | Phrase Degree |
| RI | Random Index |
| ROUGE | Recall-Oriented Understudy for Gisting Evaluation |
| RT | Re-Tweet |
| SBKE | Selectivity Based Keyword Extraction |
| SF | Scale – Free |
| SFPM | Soft Frequent Pattern Mining |
| SMA | Social Media Analysis |
| T – REC | Topic Recall |
| TDT | Topic Detection and Tracking |
| TF-IDF | Term Frequency – Inverse Document Frequency |
| TKG | Twitter Keyword Graph |
| TRCM | Transaction-based Rule Change Mining |
| TWCM | Twitter Word Co-occurrence Model |
| WCN | Word Co-occurrence Network |
| WMD | Word Mover's Distance |
| WNTM | Word Network Topic Model |
| WWW | World Wide Web |
| URL | Unified Resource Locator |

# List of Figures

# List of Tables

# List of Algorithms

# List of Algorithms

# Chapter 1

# Introduction

People use social media platform to share views and information out of their interest. It is observed that people tend to share updated information about current happenings, outbreak, emergencies, and real-time events over social media data. When updated information is posted in interest of online users, they tend to follow each other. The problem of sharing a huge amount of information on the online platform is called "Information Overloading". It is difficult to study and analyze the huge amount of social media data, manually. Thus, automatic Social Media Analysis (SMA) is gaining much attention from academic researchers in recent years. Some of the most widely studied problem domains of SMA are emotion and sentiment analysis, Topic Detection and Tracking (TDT) (James Allan, 2002), information dissemination, and information credibility. Among these, TDT has gained much potential due to its use in various industrial applications. Event detection from social media data is one of the major research directions of TDT.

## 1.1. Motivation

Social media platforms like Twitter have become the most important means of communication in recent years. As per the reports of October 2017, China, India, and United States have highest numbers of internet users: 772 million, 462.12 million, and 312.32 million users, respectively (Statista, 2017). It is observed that 9 out of 10 Twitter users use Twitter to post current happenings, news, and real-time events as *Tweet*. Tweet is a short textual message of 140 characters which contains different types of modalities on data. Among these, 77% of them keep up with the news at least once a day which is similar to non-Twitter users (76%) (Rosenstiel *et al.*, 2015). The live and instant information is shared over social media which is potentially first-hand witnessed by a user. Automatic identification of events from social media might give more credible results than traditional news media, as users actively participate in contributing the information and generating content. The automatic event detection from social media data is significant because there are many controversies over traditional and popular news media about the credibility and unbiased news. Also, information related to real-time events is usually broadcasted by traditional news media only if it interests the audience and leads to high TRP (Television Rating Point). There are many such events which are not discussed in public by traditional media and remain unnoticed.

Researchers observed that 95% of the events can be identified as compared to the traditional media, even within the small amount of data which is nearly 1% of the Tweets (Petrovic *et al.,* 2010). Other than this, many researchers proved that the disaster management information is obtained using the live Tweets, much earlier than that of traditional news channels. The automatic event detection from social media data is applicable to various government and industrial applications like emergency management during disasters, spread of contagious and infectious disease, campaigning events, instant outbreaks like floods, earthquake, and bomb blast; quickly spreading communicable diseases like swine flu and bird flu; public gatherings like conferences, ceremonies, fest; en-route traffic by detecting traffic congestion due to public events, and election campaigns and protests. Many business applications used news data for trading and block-chain analysis. All these applications act as motivation to carry this research work.

## 1.2. Event detection

The research direction of identifying events automatically is the sub-domain of Topic Detection and Tracking (TDT). TDT is a broader research domain which was initially studied (James Allan, 2003) for well-formed data. In TDT, the five different research areas are as follows:

- **Segmentation of the source stream:** The process of dividing news into individual stories.
- **Tracking previously known topics:** New streams are extracted and synchronized with previously known stories.
- **Detecting new and previously unseen topics:** The stories are monitored to detect the events which have not been seen before and are also called New Event Detection (NED).
- **First Story Detection (FSD):** Identifying the first story of the previously unknown event.
- **Story Link Detection:** Linking a given pair of stories with each other.

This research work is carried out to detect news and previously unseen topics from user-generated information over social media. Earlier, these unseen topics and news were detected from well-formed data using traditional algorithms of Information Retrieval. Identifying topics, trends, and events from social media data have opened new research directions for building interactive applications.

**Definition 1.1: Topic:** A point of discussion may refer to any theoretical or practical subject which can be discussed verbally. Any phrase (set of words) which represents meaningful information about the point of discussion is known as a *topic*.

**Definition 1.2: Trend:** A trend is a general direction in which something is changing or developing. The change which becomes popular because of its frequency of discussion and which is consistent over a specified time period is called *trend*.

The definitions for Topic and Trend are widely accepted by research community. But there is no widely used common definition of *Events* for event detection from social media data. Some academic researchers consider current happenings as an event; some others mentioned anything happened in specific time and location; many worked with the identification of sub-events in an event, and others have found more frequently discussed topic/ news as an event. In this research work, Definition 1.3 and Definition 1.4 have been used for Event and Sub-Events, respectively.

**Definition 1.3: Event:** An *event* is significant story and has impact on certain group of people due to which people tend to Tweet. In this research work, different topics or sub-topics which are obtained as information about anything happened around are considered as an event and are described by using event-phrases. The sub-events can be extracted for domain-specific set of Tweets.

**Definition 1.4: Sub-Events:** Small stories which give information about instances during a big event are called *sub-events*. These sub-events are of public interest and represent different stories, for instance, the goal, and foul during a football match.

**Definition 1.5: Event Phrases:** An Event is described in a phrase or sentence as an *Event Phrase*. The resulting event-phrases are the set of words or a summarized text as obtained from the discrete set of similar Tweets.

The resulting *event-phrases* should be human interpretable by keeping words in ordered lexical sequence. In this research work, event detection technique is proposed for social media data by finding summaries from discrete set of Tweets. The SMA is performed by using the syntax and semantics of Microblog WCN evolved from user-generated data on social media platforms.

## 1.3. Social Media Analysis (SMA)

Social network is defined as "web-based services that allow individuals to (1) construct a public or semi-public profile within a bounded system, (2) articulate the list of other users with whom they share a connection, and (3) view and traverse their list of connections and those made by others within the system" (Boyd *et al.,* 2007). There are many definitions for social networks. The idea of SMA is to automatically analyze the huge amount of social media data which have been evolved from social networks and is used for various industrial applications. The SMA gives new research directions on the basis of multiple modalities of data and different social media platforms.

### 1.3.1. Social media data

Social media data is present in multiple modalities on cross-media platforms. The different modalities for SMA are textual information, visual (image and video) information, audio information, and spatio-temporal information. It is difficult to map all the modalities and analyze social media data because huge storage space is required to store information. Also, high software and hardware configurations are required for processing the multimodal data. This data analysis gives high complexity and thus, low speed. The limited information is available for different modalities (like location, time zone) for event detection. There are many Tweets for which visual and audio information is absent. Also, people may share information from different places due to wide range of spatial connections of users. The information might not be shared from the place of event happening because of high mobility of users. In this research work, only textual information is used for SMA of real-time industrial applications.

### 1.3.2. Different social media platforms

The multimodal content on social media websites is uncertain, user-generated and is available over cross-media. Different social networking platforms are business-based (LinkedIn), location-based (Foursquare), content sharing (PInterest, Blogs), photo sharing (Flickr, Instagram), Microblogging (Twitter, Sina Weibo), and video platforms (Youtube, Vimeo) (Mainka, Hartmann, Stock, & Peters, 2014). In past few years, the use of different social media platforms has increased. The number of users for different social media platforms is shown in Figure 1.1. It is observed (Statista, 2018) that Facebook is the leading social media platform.

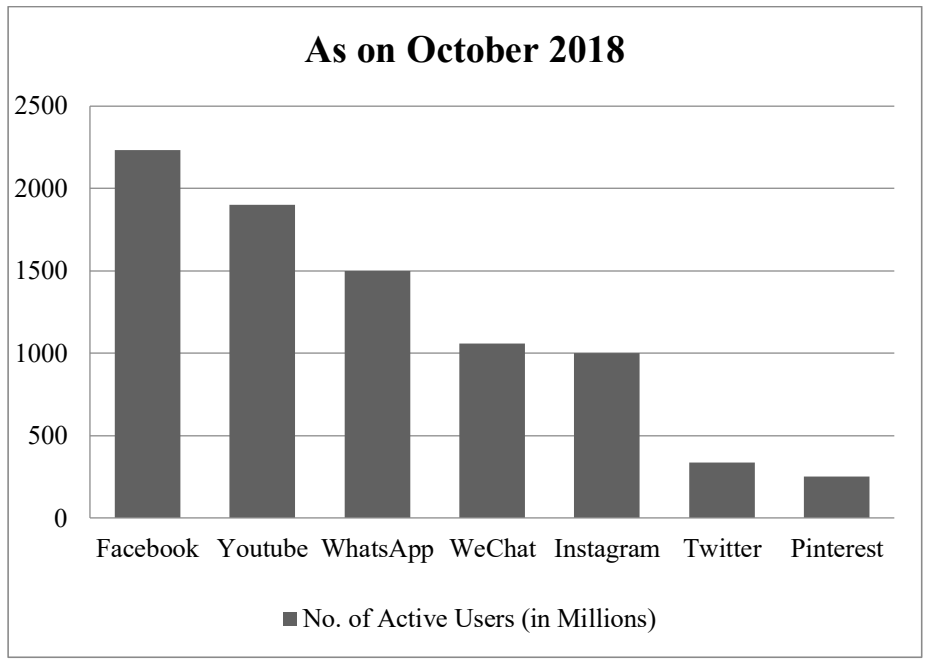

**Figure 1.1**: Number of active users for different social media platforms

Facebook contains sharing of personal information and viewpoints. This is because Facebook is secure platform and have tendency to create personal groups and other communities via two-way networking. Unlike Facebook, the Twitter social media platform is one-way connectivity of Follower – Followee relation. The Twitter is used for commercial, marketing, and awareness purpose. Many commercial offers, news, government awareness programs, convocations, academic conferences, and symposium, sale offers, cultural and technical fests, recruitment or openings, and other events of public interest is posted for broadcasting information on social media platform. The informative profiles which act as source of information for its followers become popular. The Twitter has the policy of making its Tweets available for public domain. This ease of availability of Twitter data has increased the use of Twitter for SMA.

## 1.4. Twitter: A social media platform

Twitter is one of the most widely used media among all other social media platforms. Twitter social media have one-way connectivity and create a directed graph among profiles for the Follower-Followee relationship. Many people follow each other, preferably those with similar interest and those who are more active. To pitch their Tweets to a wider audience, users Tweet information about current happenings or live events. The user is a first-hand witness of the current happenings. Only useful information is embedded in a Tweet because for each Tweet, length is limited to 140 characters. The words which are used for Tweet are informative and contain named entities and activity, as verbs. Words of a Tweet are ill-formed, user-generated and created from the active vocabulary of the user. The words in active vocabulary are learnt from the offline social community or from the information obtained over social media. The Twitter have two entities namely *users* and *Tweet*. Many existing techniques used user-user interaction, and user-Tweet interaction to identify events due to availability of meta-data of Tweet (Hromic *et al.,* 2015). The sample of metadata of any Tweet is given as follows:

{"created_at", "id", "id_str", "text", "source", "truncated", "name", "screen_name", "location", "URL", "description", "protected", "verified", "followers_count", "friends_count", "listed_count", "favourites_count", "statuses_count", "created_at", "utc_offset", "time_zone", "geo_enabled", … "lang"}

This metadata gives information about each Tweet. For every Tweet, *the user id, number of followers of the user who posted it, number of times it was re-tweeted, the time and location of posting, the time zone and other information* is recorded. The selected information from metadata is used as feature set for different event detection techniques. In this research work, only textual Twitter data is used which have different characteristics.

### 1.4.1. Characteristics of Twitter data

The characteristics of the text of Tweet describe the type of information which is shared in Twitter. In this research work, the behaviour of text as a unit (word) and relation among words in Tweet is analysed using the characteristics of Twitter data which are given below:

- Every Tweet is of maximum 140 characters.

- It contains abbreviations, a special *Hashtag* with concatenations of two or more words, SMS language, slangs, emoticons, spelling mistakes, typing mistakes, and location-specific jargons.
- The information shared is human understandable due to active encoding and decoding of ill-formed text. But it is difficult to train the system so that it can interpret user-generated information.
- Words act as a unit of Tweet. The words which are used to Tweet about the same event are usually repetitive and contain important words.
- Other than words, Tweets contain *Hashtags, At-Mentions, Uniform Resource Locator (URL), ReTweet (RT)* and other emoticon symbols which are not useful for research areas of TDT. However, this information may be useful for sentiment and emotion analysis.

The hypothesis and assumptions which are used in this research work are:

- It is assumed that the information which is available on the internet is credible and there is no intentionally induced misinformation by users.
- The Tweets which are extracted using API are original and user-generated. All Tweets are related to some event and there is no such Tweet which is not related to any event.

Tweets contain some important information in the form of special words which are used as significant word and are marked as *keywords*. The keywords or combination of these words are used to emphasise the topic as *Hashtags,* for instance '*#SomethingNew*' and '*#Trending2019*', or to refer someone as At_Mention, for instance, '*@CnnNews*'. Various Twitter specific features which are considered for this study are:

- **Textual Information**: Text of Twitter posts in which all the words are tokenized and used for further analysis of social media data.
- **Hashtags**: Hashtags are used for different types of SMA but in this research work, the Hashtags are used as common term of Tweet.
- **At_mentions**: Also, in this research work, *At_mentions* are used as the common terms by removing '@' prefix from *at_mention* in Tweet.

In this research work, only textual information is used for identifying events from social media data using the characteristics of Twitter data and features of text in Tweets. Tweets are tokenized into words after being extracted from Twitter.

### 1.4.2. Tweet extraction from Twitter

The Tweets are extracted from the Twitter social media platform using Application Program Interface (API). Python is used as a scripting language for extracting Tweets using Tweepy API with combination of *keys* and *tokens* namely *access token, access token secret, consumer-key, and consumer-secret*. These keys and tokens are obtained from an application which can be created at the developer platform of Twitter by user. The developer platform of Twitter allows user to use the information which is available on Twitter to build personalised and non-commercial applications. Thus, the textual data is extracted from Twitter and instances of events are analysed from Tweets.

### 1.4.3. Instances of event specific Tweets

Despite of ill-formed and uncertain text generated by users, many Tweets provide useful information. This information may or may not contribute to extract *event-phrases*. For instance, a Tweet is considered from 72 Twitter TDT – FA Cup dataset which is described in Section 1.5. The text of Tweet is given below:

*"Ohhhhh yeaaaaaahhhhhhh!!!!! #goal #CFC"*

This Tweet gives precise information to the user during the time of FA Football Cup. It indicates that Chelsea Football Club has scored a goal and the user is happy about it. In this Tweet, *'the goal'* is the story of an event 'Football Match' which is precisely known as the *sub-events* in social media data. The degree of happiness is associated with sentiment analysis which is out of the scope of this research work. Such instances give information about events and sub-events in human understandable format. The system should undergo pre-processing of Tweet for tokenizing the text of Tweet in different words which act as a unit for automatic SMA.

### 1.4.4. Pre-processing Tweets

Pre-processing is an initial step to clean the text and tokenize it into different words. It is observed that *emoticons* and *URLs* are not required for research areas of TDT. Thus, the special characters are truncated and stopwords are removed from the Tweet. Following are the two types of tokenization process:

- **Automatic Tokenization:** Automatic tokenization is the process of automatically removing irrelevant symbols and tokenizing the words of Tweets into separate units.

These units form the path as they appear in Tweet in the ordered lexical sequence. The path of words is used for creating Microblog WCN.

This process of *automatic tokenization* is carried out using regular expressions to remove irrelevant characters and symbols automatically. However, some ill-formed text containing spelling mistakes, jargons, location specific information cannot be removed using regular expression.

- **Manual Tokenization:** The *manual tokenization* approach is used to convert the ill-formed text into well-formed text manually, and separating each word of Tweet as a unit. Users can understand the sense of in-formal text but it is difficult to train a system for SMA on in-formal text which is usually ill-formed. Manual tokenization gives well-formed Tweets as output which can be used for SMA using traditional *Information Retrieval (IR)* algorithms. The time consumed for *manual tokenization* is much more than that of *automatic tokenization*. Thus, it is difficult to handle huge number of social media Tweets, manually.

Though ill-formed text is present in Tweets, researchers (Rudra *et al.,* 2015) found that for event detection in Twitter, most of the information is centred around content words in network of words. These content words are usually named entities and activity/ verbs which are also referred as well-formed words in this research work. Thus, we used *automatic tokenization* process for pre-processing of Tweets to identify events from social media data. The social media data which is used for this research work are publicly available annotated datasets.

## 1.5. Datasets and annotations

As the information on Twitter is in public domain by default, users can make their Tweets private and secure. Many users and Tweets which are obtained in real-time may get deleted. Thus, information may or may not get fixed and the behaviour of Tweet Corpus may change with time. In this research work, the Tweets which are available in public domain are used rather than those which have been deleted. Thus, the datasets which are used in this research work are as follows:

- **'First Story Detection (FSD)':** The FSD dataset (Petrovic *et al.,* 2010) is used for identifying key-phrases and event detection. This dataset has Tweets with a list of Tweet IDs. Another list of 27 topics is marked as list of events which is used for annotated data.

The number of Tweet ID in this dataset is 51,879,318. Among all Tweet ID, 3034 Tweets are marked with corresponding topic id in a file named *'relevance_judgements'*. This file is considered as *ground truth* for experiments and evaluation.

- **'FA Cup' in '72 Twitter TDT':** From *72-Twitter-TDT*, *FA Final Cup dataset* is used for identifying events and sub-events in Football Association Challenge Cup. In this dataset, Tweets are extracted during *'Football Final'* match. There are *'13 one-minute windows'* which are highly informative and are thus, marked as *ground truth* (Aiello *et al.*, 2013). This information is used for experimental results and evaluation.

To handle the given datasets, different Software and Hardware configurations are required for automatic SMA.

## 1.6. Hardware and software requirements

The programming interface is required to extract Tweets using Tweepy API from Twitter. Different hardware and software requirements are discussed in this section. Although huge amount of data needs high processing software, in this research work, summaries are extracted from discrete set of Tweets to reduce complexity. The software which is used for this research work are:

- **Python 2.7:** This is the scripting language and programming interface which is used for implementing the proposed and existing algorithms. The implementation of existing techniques is used to compare and validate the proposed technique.
- **Tweepy API:** This API is used for extracting Tweets from a Twitter social media platform. This API takes some information about the user and provide access to Tweets and application development platform.
- **Networkx Module:** This module is a Python library which contains in-built functions to handle networks. The Microblog Word Co-occurrence Network (WCN) is studied using Networkx module. Also, different network metrics, network science properties and network models are implemented on Microblog WCN using this Networkx library.
- **NLTK Module:** This module is used for pre-processing and tokenization of user-generated and ill-formed text which is extracted from Twitter.

Good Hardware specifications are required to handle and process huge amount of data. The networks of words are analyzed which needs RAM of minimum 4GB for processing summaries from discrete set of Twitter dataset. The Hardware specifications which are used for this research work are:

- CPU @ 2.90 GHz
- Intel Core i7-7500 CPU
- 64-bit Operating System
- 8.00 GB RAM
- x64-based Processor

Given Software and Hardware configurations are used and experiments are carried out on Twitter datasets. The performance is evaluated for keyphrase extraction and event detection techniques for social media data using different performance measures.

## 1.7. Performance measures for event detection

For event detection from Twitter dataset, *Event-phrases* are extracted and validated over annotated dataset using different performance measures. The different performance measures which are used in this research work are:

- **Recall:** *Recall* is the ratio of the number of relevant records retrieved to the total number of correct relevant records in the database as shown in Equation 1.1.

$$recall = \frac{Number\ of\ words\ (Resulting\ phrases\ \cap\ Ground\ truth\ topic)}{Number\ of\ correct\ words\ in\ ground\ truth} \quad 1.1$$

- **Precision:** *Precision* shows the ratio of relevant records to total number of words obtained. Precision is defined as shown in Equation 1.2.

$$precision = \frac{Number\ of\ words\ (Resulting\ phrases\ \cap\ Ground\ truth\ topic)}{Total\ Number\ of\ words\ obtained} \quad 1.2$$

- **F-measure:** *F-measure* is defined as the harmonic mean of precision and recall. It is represented as shown in Equation 1.3

$$F - measure = \frac{2 * precision * recall}{precision + recall} \quad 1.3$$

- **ROUGE Scores:** *Recall-Oriented Understudy for Gisting Evaluation (ROUGE)* score is used for validating the quality of automatic text summarization and keyphrase extraction from Microblogs. Different types of *ROUGE* scores which are used for this research work are ROUGE – 1, ROUGE – 2, and ROUGE – L.

- **Topic Recall (T-REC):** *Topic Recall (T-REC)* gives the percentage of the number of topics detected from ground truth topics.
- **Keyword Recall (K-REC):** *Keyword Recall (K-REC)* is the percentage of correctly detected keywords relative to the total number of keywords for the ground truth topics that matched any candidate topic.
- **Keyword Precision (K-PREC):** *Keyword Precision (K-PREC)* is the percentage of keywords detected correctly, relative to the total number of keywords for the topics that matched any ground truth topic.
- **Redundancy:** *Redundancy* is used to identify if resulting keyphrases give redundant information. It defines the percentage of repetitive information obtained in the resulting list of event phrases. Lower the redundancy better will be performance.

Many researchers carried out research on event detection for social media data and have given different research directions. After extensive literature survey, as discussed in upcoming Chapter 2, some research gaps were identified.

## 1.8. Research gaps and challenges

Event detection from social media data is the subjective and abstract research area. The research gaps in existing event detection techniques are identified and enlisted in this Section. The four significant research gaps are:

- *The cognitive behaviour of short-text and human-generated information:* The information shared by users on Twitter is related to current happenings. Cognitive patterns in Microblog WCN need to be analyzed for various applications of SMA because words and relations among words may have latent patterns.
- *Ordered lexical key-phrase extraction technique for social media data:* The key-phrase extraction from the batch of short-text may play the pivotal role in identifying influential segments from the network of words. This area should be studied based on the structure of social media data. The lexical variation of words in event-phrases is not ordered in existing techniques.
- *Normalization of ill-formed text:* The data shared over social media data is ill-formed and user-generated which may contain abbreviations, special Hashtags with concatenations of two or more words, SMS language, slangs, emoticons, spelling mistakes, typing mistakes, and location-specific jargons. Thus, *text normalization* is required to test and validate the traditional IR algorithms for social media data.
- *Classifying Events and Non-Events:* The classification of Tweets is missing in traditional approaches which tells whether the Tweet is event related or not. This classification is performed manually which is difficult for huge amount of data. Thus, it should be automated to speed up SMA.

First two research gaps are explored in this research work. *Automatic tokenization* and event related Tweets are used for this research work. The solution of a problem domain should handle research challenges to demonstrate its ability in solving a research problem of event detection. The primary research challenges for event detection techniques are

- How many general events are detected.
- Is the proposed method scalable for the arrival of batch-based streaming data.
- Is the proposed technique supervised (given set of parameter values/ data/ keywords) or unsupervised (independent of the query/ data given).
- The event detection technique is meant for retrospective data or new data.

## 1.9. Objectives of the proposed work

Objectives describe the path of the research work. After an extensive literature survey, and research findings, the following objectives are formulated for this research work

- Analysis of existing event detection techniques for social media data and identification of different keyphrase extraction metrics for uncertain user-generated data.
- Propose the keyphrase extraction algorithm for uncertain user-generated data.
- Design and development of event detection technique using proposed keyphrase extraction algorithm for social media data.
- Validation of the proposed event detection technique.

Research objectives define the significance of the research work. These research objectives are achieved in this research work and different contributions are made to give new directions to the research community.

## 1.10. Contributions of research work

As per the research objectives, an event detection technique for social media data is proposed. Based on the research gaps, challenges, and unexplored areas, different contributions made in this research work are described as follows:

- Study of *the structure of word co-occurrence networks for Microblogs* has contributed towards identifying cognitive patterns among Microblog WCN.
- The comprehensive study of existing random-walk based key-phrase extraction techniques is carried out. The existing techniques and network science metrics are implemented over *Microblog WCN* for performance evaluation.
- Using the insights from the study of *'the structure of Microblog WCN'*, an edge-decomposition based key-phrase extraction technique is proposed for social media data.
- The events are identified by decomposition of Microblog WCN and topological sorting of words of resulting sub-graph using heuristic values. This heuristic approach has outperformed the existing techniques in terms of recall and redundancy and act as preamble for the proposed event detection technique for social media data.
- The network model based k-bridge decomposition is proposed as a keyphrase extraction technique to extract influential segments from the Microblog WCN. The k-bridge decomposition used controlling parameters to terminate the decomposition.
- To remove parameter dependency for termination of decomposition, a keyphrase extraction technique is proposed for decomposition of Microblog WCN using network science property of assortativity,
- An optimization technique *AHP* is used for contextual ranking of the information which is obtained in the form of key-phrases. Top-ranked keyphrases are termed as event-phrases.
- An event detection technique is proposed to find summaries from the discrete set of streaming social media data using *BArank* keyphrase extraction technique. This approach eliminates the use of dependency parameters as observed in heuristic based event detection. The proposed approach outperformed all the existing techniques in terms of K-REC and K-PREC.

## 1.11. Thesis organization

The remaining part of the thesis consists of seven different chapters. These chapters are structured as follows:

**Chapter 2** describes the extensive literature survey of existing techniques for keyphrase extraction and event detection from social media data. This research work gives the evolution of historical perspective for different modules like network science, graphical keyword extraction, graphical keyphrase extraction, and event detection from social media data. Also, the baseline techniques are discussed for both keyphrase extraction and event detection technique.

**Chapter 3** describes the study of the structure of word co-occurrence networks for Microblogs. This study is carried out for the network science properties which include the scale-free property, the small-world feature, the hierarchical organization, the assortativity, and the spectral analysis. The experimental results and evaluation is performed over 100K Tweets of First Story Detection (FSD) dataset.

**Chapter 4** elaborates the study of existing keyword and keyphrase extraction techniques for social media data. The comparisons are performed and analyzed for all the existing techniques and network metrics using different performance measures.

**Chapter 5** identifies the influential segments from social media data. Heuristic values are used for decomposition of Microblog WCN as a preamble approach. The *k-bridge decomposition* is proposed to decompose the massive amount of information in Microblog WCN into small but useful sub-graphs. To reduce the number of computations, a *threshold based approach for k-bridge decomposition* is used for decomposition of Microblog WCN. The graphical keyphrase extraction technique, *BArank*, is proposed using *assortativity* network science property which follows automatic termination and has no static controlling parameter dependency for termination of decomposition.

**Chapter 6** is used for contextual ranking of keyphrases. The keyphrases are obtained by preserving the lexical sequence of Microblogs. In this research work, contextual ranking of keyphrases is performed using AHP optimization technique which results into *topic headlines, brief description and relevant keywords*. This contextual ranking is used to identify top ranked keyphrases as *event-phrases*.

**Chapter 7** discuss about the proposed event detection technique for social media data. The preamble is set for identifying event-phrases using heuristic values based approach for decomposition of Microblog WCN into small sub-graphs. The keyphrases are extracted from sub-graphs using topological sorting. An event detection technique is proposed as BArank, to obtain summaries from discrete set of Tweets.

**Chapter 8** concludes the research work and highlights the important findings. The scope of future work and possible extensions to this work are discussed in this Chapter to give new research directions in this area.

# Chapter 2.
# Literature Survey

The event detection from social media data has received much attention from academic researchers which is evident from the sheer number of research papers published in this field. An event detection framework can be beneficial for social lifestyle, business perspective and many more applications. An extensive literature survey is carried out for graph-based event detection techniques and keyword/ keyphrase extraction from short- text documents.

The research for event detection and tracking was initialized (Allan, 1998) with an introduction to Topic Detection and Tracking (TDT). The TDT is Defense Advanced Research Projects Agency (DARPA) sponsored initiative which was started to investigate new events in the news stories which were broadcasted. The Pilot study of TDT was started in September 1996 and it was conducted till October 1997. Different researchers used statistical models like language modelling, and machine learning models like decision trees (Yang *et al.*, 1999) for TDT. This study was demonstrated using experiments which were conducted for text retrieval and clustering. This information was extracted from streaming data of news stories which were broadcasted (Yang *et al.*, 1998). The evolution of historical perspective of event detection techniques is shown in Figure 2.1. The bursty segment concept was introduced for identifying events from well-formed textual data (Fung et al. 2005). This was further improved by introducing the word co-occurrence measure for terms in the document with respect to time.

After this, many different approaches were proposed to identify events from social media data using Frequent Pattern Mining (FPM) (Huang *et al.*, 2015; Petkos *et al.*, 2014), peak keywords in Twitter data (Fung *et al.*, 2005; Mathoudakis *et al.*, 2010), using KeyGraph algorithm (Ohsawa *et al.*, 1998; Sayyadi *et al.*, 2009), keyword evolving graph (Kwan *et al.*, 2011), topic modelling (Ferrari *et al.*, 2011; Zhou *et al.*, 2014), using named entities and temporal clustering (Li C *et al.*, 2012; Li R *et al.*, 2012), Locality Sensitive Hashing (LSH) (Kaleel *et al.*, 2015), hypergraph based (Gao *et al.*, 2015), wavelet transforms and Discrete

Fourier Transformations (DFT) (He *et al.,* 2007; Weng *et al.,* 2007), Heterogeneous Information Network (HIN) (Prangnawarat *et al.,* 2015; Hromic *et al.,* 2015), and many other miscellaneous approaches.

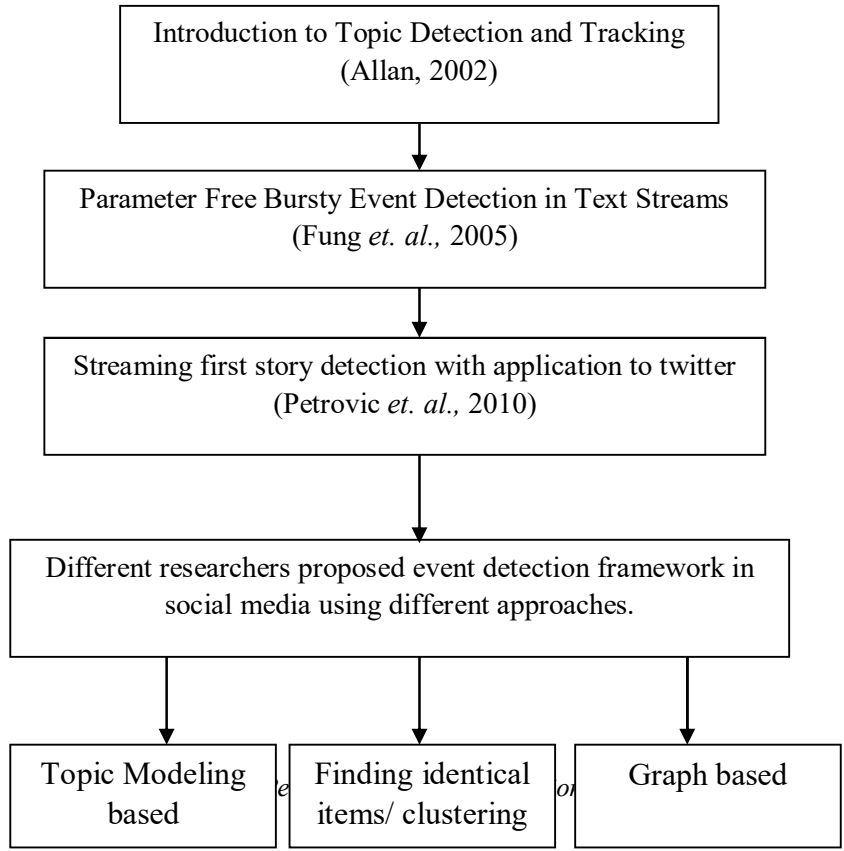

**Figure 2.1:** Historical Perspective of Event detection in Social Media Data

Although there are many different approaches to extract events from social media data, but none of the existing techniques extract event-phrases in ordered lexical sequence. The major challenge and research gap in the literature is the nature of social media data. The nature of textual information in Microblogs is different from the text of well-formed documents. The Word Co-occurrence Network (WCN) is the network of words which have been used in textual documents and has received much attention from academic researchers for keyword extraction from Twitter (Abilhoa *et al.,* 2014; Biswas *et al.,* 2018); still, it is unexplored for keyphrase extraction and event detection from short-text and user-generated data.

## 2.1. Network science and Word Co-occurrence Network

The network is evolved out of data of many real-time applications in different phases of life. Researchers have studied the structure and dynamics of many complex networks like power network, social networks, airlines networks, network of the mobility of people, network of roads and railways, proteins networks, information networks, and many other real-time complex networks. The structure of the static Word Co-occurrence Networks (WCN) for Microblogs is studied in this research work. Although complex networks (Boccaleti *et al.,* 2006) have received much attention from researchers in recent years, yet Microblog WCNs are minimally explored for keyphrase extraction from social media data.

The study of WCNs has been carried out in recent years for English and Chinese (Liang *et al.,* 2009) languages. The authors have discussed commonalities and differences among different types of articles. The research was improved over English and Chinese Poems (Liang *et al.,* 2015), spectral analysis of Chinese language (Liang *et al.,* 2016) from literary genres, and English language (Liang *et al.,* 2017). The author (Gao *et al.,* 2014) compared directed and weighted WCN of six different languages which are Arabic, Chinese, English, French, Russian, and Spanish. It is observed that the French and Spanish languages share many commonalities like those of Chinese and English languages. The word connections for English, Arabic and Russian languages are sparse, and for Chinese language follow uniform distribution.

In contrast to this, the WCN dynamics are uncovered for well-formed English book data, Tweets, and Facebook data (Turker *et al.,* 2016). The author calculated network science metrics for undirected and weighted edges namely average degree, modularity, average clustering coefficient, diameter, path length, edge weights, and maximum degree differences of pairing nodes. Further, 12 Slavic and 2 non-Slavic languages were used to study WCN using syntactic dependency of complex networks (Liu *et al.,* 2013). The co-occurrence network was studied for biological analysis of the structure of the protein domain network (Witchy *et al.,* 2005). The major applications where WCNs are used are keyword identification, keyphrase extraction, and authorship attribution.

## 2.1.1. Graphical keyword extraction

There are various applications for keyword or keyphrase extraction from uncertain user-generated data (Carretero-Campos *et al.*, 2013) namely sentiment analysis, opinion mining, TDT (Allan el at., 2003), stock market prediction (Bollen *et al.*, 2011), topic detection and events detection (Petrovic *et al.*, 2010). SMA is the major source for latent pattern analysis of human behaviour by using Tweets posted by users. As shown in Figure 2.2, among different keyword extraction approaches, context-independent unsupervised graph-based keyword extraction techniques have been studied for this research work.

Traditional graph based automatic keyword extraction methods are used for short text to identify keywords from social media data (Abilhoa *et al.*, 2014; Biswas *et al.*, 2018) and Croatian News (Beliga *et al.*, 2016). Graphical approaches can be of different types namely network science metric based, knowledge representation based, and rule induction based. The WCN is generated from documents using the co-occurrence of frequent words in Microblogs. This Microblog WCN is used for identifying keywords using network metrics.

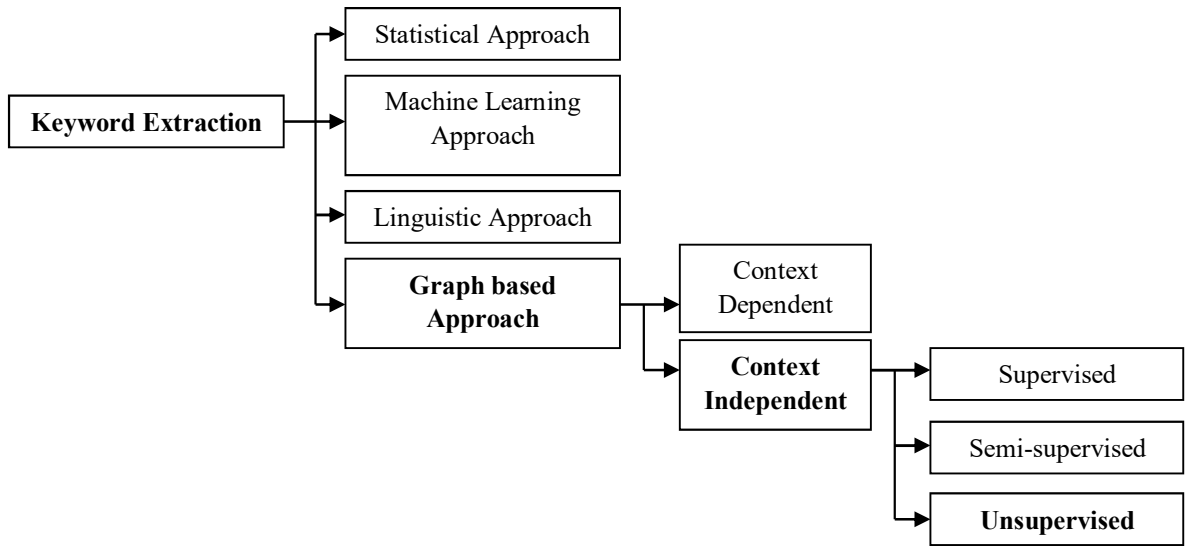

**Figure 2.2**: Classification of Keyword Extraction Methods

The Twitter Keyword Graph (TKG) was proposed to extract keywords from Twitter data using different network metrics (Abilhoa *et al.*, 2014). The author introduced all-pair neighbor edging and nearest-neighbor edging for constructing a graph; frequency based and inverse-frequency based weights in the graph; and different centrality measures for different types of networks. Additionally, they also used the third type of edge weight given as inverse co-occurrence frequency which has proved to be superior parameter than same edge weight, and co-occurrence frequency edge weight for keyword extraction.

A WCN analysis is performed (Beliga *et al.,* 2016) for structural analysis of structured well-formed languages on Croatian News. The Selectivity-Based Keyword Extraction (SBKE) was proposed (Beliga *et al.,* 2016) to extract keywords from Croatian news using F1 and F2 measure as performance measures. In F1-measure, K-REC and K-PREC are equally important for calculating F-measure. F2-measure is the F-measure in which K-PREC is two times more important than that of K-REC.

Recently, Graph of words (GoW) (Batziou *et al.,* 2017) and centrality measures for keyword extraction have been used to identify the largest community containing key nodes (keywords) in the graph of words. The proposed measures are two parameters namely Mapping Entropy Betweenness (MEB), and Mapping Entropy Centrality (MEC). These are tested over Jaccard, average precision, and P@N measures for linking the number of words (N) for N=2 and N=3. It is observed that MEC and MEB outperform other baseline measures after closeness and performs better for N=3. The CoreRank (Meladianos *et al.,* 2017) is proposed to identify keywords from WCN using k-core decomposition model (Alvarez-Hamelin *et al.,* 2017). The ROUGE-1 and Word Mover's Distance (WMD) are used as the performance measure. The ROUGE-1 computes similarity based on unigram overlap, while the WMD takes into account semantic similarity between the terms, and is, therefore, more robust to the fact that the abstractive summaries contain words that were never actually spoken.

Recently, the graph-based keyword extraction model is proposed using Collective Node-Weight (KECNW) (Biswas *et al.,* 2018). The authors used five parameters namely distance from central node, position of the node, selectivity of a node, importance of neighbouring nodes, and term frequency for identifying importance of the node in Microblog WCN evolved from social media data. The KECNW outperformed TKG and SBKE.

### 2.1.2. Graphical keyphrase extraction

Multiple key-phrases are extracted from a batch of Twitter posts which gives trending information about events from social media data. Since, Tweets are the short text which contains not more than 140 characters; Tweets related to single event are summarized to obtain key-phrases. Thus, the key-phrases and text/ tweet summarization are used interchangeably in this research work.

**Definition 2.2: Key-Phrase:** A set of keywords which occurs in ordered lexical sequence to represent the topic/ summary of the document/ short- text is called keyphrase.

The key-phrase extraction is the process of identifying important phrases from the given textual data. It is used to summarize the short text to identify the topic of discussion from a set of Twitter posts which are related to the same topic or domain. The topic or domain is very subjective decision: whether the extracted keyphrase is about main topic or sub-topics under that topic. For instance, *'floods in Kerala 2018'* is the main topic but *'having nothing to eat in XYZ village'*, *'save us! Water all around'*, and *'our house drowned in flood'* are sub-events. These sub-events help us to determine the emergency conditions and take necessary actions. The key-phrase extraction is an intermediary step to identify events and sub-events from social media data.

Keyphrase is extracted either from the set of Tweets which are classified into different topics or by representing the Tweet Corpus in such a way that words which belong to same domain and have some relation among each other. In this research work, Microblog WCN is used for extracting keyphrases by identifying cognitive patterns in relations among words. Recent progress on word co-occurrence network has added statistical and computational significance over information processing (Akimushkin *et al.*, 2017; Liang *et al.*, 2015; Liang *et al.*, 2016). Graph-based keyphrase extraction techniques are the random-walk based measures which are studied for well-formed data.

In existing literature, the key-phrase extraction was studied for well-formed text which was grammatically correct. The traditional key-phrase extraction techniques have proved to be useful in text summarization. There are many random-walk based approaches, as shown in Table 2.1, which have been evolved with introduction to PageRank (Page *et al.*, 1999) algorithm.

The word score is used to propose a random-walk based keyphrase extraction technique for WCN and is introduced as TextRank (Mihalcea *et al.*, 2004). It gives 36.2 F-measure for un-directed graph with co-occurrence window of 2. In earlier studies, the TextRank was used for well-formed data of Hulth 2003. The affinity score and transition matrix were used (Wan *et al.*, 2008) to propose the ExpandRank in WCN of publicly available datasets. The well-formed datasets namely DUC 2009 and TREC 9 gives 31.6 F-measure for top 10 keyphrases. The NErank (Bellaachia *et al.,* 2012) was introduced for keyphrase extraction from social media data of Tweet Corpus and performance is evaluated using Bpref measure.

**Table 2.1:** Historical Perspective of different keyphrase extraction techniques for Microblog WCN: F is F measure, CW is co-occurrence window, UD is undirected, k is number of top key-phrases

| S. No. | Keyphrase Extraction Algorithm | Metrics Proposed/ Used | Datasets Used | Results |
|---|---|---|---|---|
| Mihalcea *et al.,* 2004 | TextRank | Word score and PageRank | Hulth 2003 | F=36.2 for UD, CW=2 |
| Wan *et al.,* 2008 | ExpandRank | Affinity score and transition matrix | DUC2001 and TREC 9 | F=31.6 for k=10 |
| Bellaachia *et al.,* 2012 | NE Rank | Introduced word weight in PageRank | Tweet collection | Bpref=85.7 for k=10 |
| Bougouin *et al.,* 2013 | TopicRank | Distance between topic related words | Inspec, SemEval (S), WikiNews (W), DEFT (D) | F=12.1, 35.6, 15.1 for k=10 for S,W,D |
| Wang *et al.,* 2014 | WordAttractionRank | Word Attraction Score | Hulth2003 (H), DUC2001 (D), SemEval2010 (S) | F=42.68 (H) F=26.92 (D) F=13.61 (S) |
| Florescu *et al.,* 2017 | PositionRank | Position of word occurrences based measure for PageRank | KDD, WWW, Nguyen | F=10 to 12 for k=4 to 8 |
| Bennani-Smires *et al.,* 2018 | EmbedRank | Maximal Marginal Relevance | Inspec, DUC, Nguyen | F=38.39 and 31.52 on Inspec and DUC |

The TopicRank (Bougouin *et al.,* 2013) was proposed by introducing the metric *'distance between topic-related words'* over datasets namely Inspec, SemEval, Wikinews, DEFT which gives F-measure 12.1, 35.6, and 15.1, respectively. After that, a word attraction score was introduced to propose WordAttractionRank (Wang *et al.,* 2014). It identifies keyphrases from Hulth 2003, DUC 2001, and SemEval 2010 datasets which give 42.68, 26.92, and 13.61 F-measure, respectively.

The position of a word is considered to be important by many researchers (Biswas *et al.,* 2018) for keyword extraction technique. This measure was also used to propose another keyphrase extraction approach for well-formed data named as PositionRank (Florescu *et al.,* 2017). It is implemented on KDD, WWW, and Nguyen datasets which gives 10-12 F-measure for top 4 to 8 keywords extracted from given datasets. The EmbedRank is proposed using maximal marginal relevance for Inspec, DUC, and Nguyen datasets which gives 38.39 F-measure for Inspec, and 31.52 F-measure for DUC. The evaluation for keyphrase extraction techniques for well-formed WCN is summarized in Table 2.1.

Further, trend and event detection are important applications which use Tweet summarization from WCN. In 2004, the LexRank was (Erkan *et al.*, 2004) proposed which is insensitive to noise in text and calculates the importance of sentence (or word) using eigenvector centrality. The spectral analysis is used for community detection of keywords which belongs to the same class (Ferrer *et al.*, 2007) for keyphrase extraction. In 2008, the HITS-based algorithm was proposed (Litvak *et al.*, 2008) for keyphrase extraction using hub or authority scores in WCN. The eccentricity, centrality, and proximity measures were used to propose a keyword extraction technique (Palishkar *et al.*, 2007).

A Graph based keyphrase extraction technique was proposed for Twitter data named as NErank. In this approach, the node score and edge score were proposed and random-walk was used for Microblog WCN (Bellaachia *et al.*, 2012). Next, the GRAPHSUM was proposed (Baralis *et al.*, 2013) as a novel and general-purpose summarizer which represents correlations among multiple terms by discovering association rules.

Although there are many different WCN based graphical keyphrase extraction techniques which have been studied in existing literature, but there is need to study the nature of language and relations among words in human generated language. The semantic of the graph of words has yet not been explored for social media data of Twitter. To study the structure of Microblog WCN, different network science properties and network metrics should be implemented on Microblog WCN. The key idea in this research work is to implement existing keyphrase extraction techniques for uncertain user-generated data and find useful insights about the structure of WCN.

### 2.1.3. Author attribution using WCN

**Definition 2.1: Text Authorship:** The text authorship is the process of identifying the authorship in disputed documents by studying textual or stylistic features. The text authorship was calculated (Akimushkin *et al.*, 2017) using the dynamics of the WCNs.

**Definition 2.2: Network Motifs:** The network motifs are the repeating sub-network of a set of words which can distinguish the writing style of authors.

During 2013, the word adjacency network was used for authorship attribution analysis from textual documents (Segarra *et al.,* 2013). Later in 2015, researchers (Amancio *et al.,* 2015) used a fluctuation analysis of network topology and word intermittency for authorship recognition. In 2016, the (Marinho *et al.,* 2016) network motifs for authorship attribution were used.

## 2.2. Event Detection from Social Media Data

Automatic event detection from social media data is one of the most challenging research areas which have been set as an important research direction due to massive amount of data in the form of Tweets. After an extensive literature survey, it is observed that there are many different approaches which were proposed for event detection from Twitter. Event detection techniques were evolved from the development of parameter-free bursty event detection in text streams (Fung *et al.,* 2005).

The bursty event detection is a document pivot clustering approach which works in three major steps namely *bursty feature identification, bursty feature grouping, and hot period of bursty event determination*. The feature of *the frequency of unigram words* was used in this approach along with *Discrete Fourier Transformation (DFT)* (He *et al.,* 2007).

The bursty feature extraction technique was proposed by constructing signals as the feature. The events were detected using wavelet analysis, computing cross-correlation and modularity-based graph partitioning (Weng *et al.,* 2010). A graph-based keyword extraction technique (Sayyadi *et al.,* 2009) was proposed using the community detection *KeyGraph algorithm* which was proposed earlier by (Ohsawa *et al.,* 1998). This approach used betweenness centrality measure in word co-occurrence network of textual data. Another graph based trend detection technique, TwitterMonitor (Mathioudakis *et al.,* 2010) was proposed using the frequency of keywords at a higher rate than usual in the data stream. Frequent keywords are grouped together using word co-occurrences.

The existing event detection techniques were found to be computationally expensive and thus, Twevent was proposed as an event detection technique (Li *et al.,* 2012a) using Symmetric Conditional Probability (SCP) to define the stickiness function of n-grams for Tweet segmentation. Event segments were detected using bursty segments. The event segments are clustered using k-Nearest neighbour, and candidate event filtering is carried out to identify event newsworthiness or segment newsworthiness. The proposed technique outperformed the EDCoW (Weng *et al.,* 2010) event detection technique. For disaster based event detection techniques, the bi-gram WCN was constructed in which each node contains bi-gram for summarization of disaster based events (Rudra *et al.,* 2015). Word Network Topic Model (WNTM) (Zuo *et al.,* 2014) is the hybrid approach which uses WCN and topic model for topic detection.

To sense the trending topics (Aiello *et al.,* 2013) in Twitter, authors used Latent Dirichlet Allocation (LDA) based technique, document pivot topic detection, graph-based feature pivot topic detection, frequent pattern mining, Soft Frequent Pattern Mining (SFPM), and BNgram for event detection. The performance evaluations which are used for event detection are Topic Recall (T-REC), Keyword Precision (K-PREC) and Keyword Recall (K-REC) for 72 Twitter TDT dataset.

A graph based trend detection approach was proposed by identifying co-occurring bi-grams (Noda *et al.,* 2015). The experimental results and evaluation are performed over 5000 Tweets, and performance is evaluated using precision, recall, and speed. With this motivation of identifying centroid words as burst words, the burst information network based event detection technique was proposed for *node-based detection model* and *area-based detection model* (Ge *et al.,* 2016). A real-time event detection technique was proposed using a sliding window for discrete social data stream (Nguyen *et al.,* 2017). The proposed technique outperforms Latent Dirichlet Allocation (LDA) based event detection technique and shows comparative performance with BNGram (Aiello *et al.,* 2013).

A frequent pattern mining based approach is introduced as Soft Frequent Pattern Mining (SFPM) (Petkos *et al.,* 2014) using selection, vector formation with co-occurrence, and expanding the set using the greedy approach. The approach is improved by introducing association rule mining and proposed Transaction based Rule Change Mining (TRCM) technique (Adedoyin-Olowe *et al.,* 2016) for event detection in Twitter. Both SFPM and TRCM are applied on 72 Twitter TDT dataset. In the existing bursty information-based event detection techniques, peak keywords are clustered and marked as an event. However, in some temporal batches of Tweets, less discussed topics may evolve which may or may not get considered as peak words due to less relative frequency. To overcome this, an event detection technique should be proposed to identify multiple event-phrases in ordered lexical sequence of words which are discussed by the sufficient number of Tweets. In this research work, graphical keyphrase extraction technique is used to extract events from Twitter social media data using cognitive patterns of Microblog WCN

## 2.3. Baselines for keyphrase extraction

The baseline techniques for comparison and validation of proposed keyphrase extraction techniques are TextRank, NErank, and TopicRank. All three keyphrase extraction techniques are graph based and used random-walk algorithm to identify influential nodes in WCN. In this research work, the TextRank is applied over Microblog WCN. However, the duplication of words with different spellings and abbreviations which are generated by users gives redundant nodes. The system is not intelligent enough to understand unique words in different form in Tweet Corpus. But this random-walk can be a good measure due to latent patterns in Microblog WCN. The baseline approaches which are studied over Microblog WCN in this research work are as follows:

### 2.3.1. TextRank

The TextRank is a keyphrase extraction technique (Mihalcea *et al.,* 2004) which is proposed using random-walk measure (Page *et al.,* 1999) for WCN. The PageRank score of any vertex $V_i$ is defined as given in Equation 2.1

$$S(V_i) = (1 - d) + d * \sum_{j \in In(V_i)} \frac{1}{Out(V_j)} S(V_j) \qquad 2.1$$

where G=(V, E) be a directed graph with the set of vertices V and set of edges E and E is the subset of $V * V$, $In(V_i)$ is the number of edges pointing to $V_i$, and $Out(V_j)$ is the number of links pointing out of node $V_j$. Here d is the damping factor which is usually set as 0.85 (Brin *et al.,* 1998). The vertex score is calculated after convergence of random-walk over WCN. The convergence is achieved when the error rate for any vertex in the graph falls below a given threshold which is given as 0.0001.

Initially, the text units are defined as vertex and relation between them is defined as an edge between two vertices. The graph-based ranking algorithm is iterated over WCN until convergence. The vertices are sorted based on their final score. The author performed two tasks namely keyword extraction task, and sentence extraction task which identifies essential sentences in the text. The selection of keywords in TextRank depends upon the presence of word and random walk, in the network. There is the need to evaluate the relationships among words in the network. Top k vertices are considered as keywords. The parameter N is chosen for the declaration of the window of the document, and is set as 2 to keep the size of WCN minimal.

### 2.3.2. TopicRank

The TopicRank (Bougouin *et al.,* 2013) is defined as keyphrase extraction techniques. The $G = (V, E)$ complete and connected graph is constructed with the set of vertices $V$, and set of edges $E$ where $E$ is the subset of $V * V$. The TopicRank is defined as shown in Equation 2.2

$$S(t_i) = (1-d) + d * \sum_{t_j \epsilon V_i} \frac{w_{j,i} * S(t_j)}{\Sigma_{t_k \epsilon V_j} w_{j,k}} \qquad 2.2$$

where $V_i$ are the topics voting for node $t_i$ and $d$ is the damping factor generally defined as 0.85 (Brin *et al.,* 1998). The $w_{i,j}$ weight of the edge is defined as shown in Equation 2.3 and calculated using the reciprocal distances $dist(c_i, c_j)$ between the offset positions of the candidate keyphrases $c_i$ and $c_j$ in the document where $pos(c_i)$ represents all the offset positions of the candidates keyphrase $c_i$. The weights and reciprocal distances are shown in Equation 2.3 and Equation 2.4, respectively.

$$w_{i,j} = \sum_{c_i \epsilon t_i} \sum_{c_j \epsilon t_j} dist(c_i, c_j) \qquad 2.3$$

$$dist(c_i, c_j) = \sum_{p_i \epsilon\ pos(c_i)} \sum_{p_j \epsilon\ pos(c_j)} \frac{1}{|p_i - p_j|} \qquad 2.4$$

### 2.3.3. NErank

It was observed that the Twitter have uncertain and user-generated data. Thus, a graph based keyphrase extraction technique was proposed for Twitter data using edge score and node score (Bellaachia *et al.,* 2012). The formula for random walk to calculate NErank is given as shown in Equation 2.5.

$$S(V_i) = (1-d) * W(V_i) + d * W(V_i) * \sum_{j:V_j \rightarrow V_i} \frac{w_{ji}}{\Sigma_{k:V_j} w_{jk}} S(V_j) \qquad 2.5$$

where $W(V_i)$ is the weight of the current node and $d$ is damping factor. The weight of the node is calculated as TF-IDF measure as shown in Equation 2.6

$$W(V_i)_{TFIDF} = tf(V_i) * \log_2 \frac{N}{df(V_i)} \qquad 2.6$$

The NErank is proposed for Twitter data and outperformed the existing techniques namely PageRank and TextRank in terms of *Precision* and *Binary preference measure (Bpref)*. The precision is defined as the number of correct keywords and out of the total number of extracted keywords. The $Bpref$ measure is defined as shown in Equation 2.7.

$$Bpref = \frac{1}{R} \sum_{r \epsilon R} 1 - \frac{|n \text{ ranked higher than } r|}{M} \qquad 2.7$$

where $R$ is the set of correct keywords within $M$ extracted keywords in a method, $r$ is a correct keyword and $n$ is incorrect keyword.

## 2.4. Baselines for event detection

BNgram is an event detection technique in which some words occur together. These words used as singleton cluster and are named as chunks. The distance between clusters is measured using similarity measures, and iteratively closest chunks are clustered together until every cluster is dissimilar from every other cluster. This is measured using threshold $\theta = 0.5$. The clusters are ranked as per the proposed highest Document Frequency – Inverse Document Frequency (DF-IDF) score. The BNgram method is evaluated for $N = 2$ where $N$ signifies top $N$ number of keyphrases obtained in the results.

Another real-time based event detection method (Nguyen Method) has been proposed by clustering the peak signals using features. The collected real-time data is extracted and terms are used to build signals. This approach used Hadoop platform to improve the processing performance and efficiency of an event detection method. For 'Nguyen method', results are obtained for both $N = 2 \text{ and } N = 10$.

The FP growth algorithm is proposed for Frequent Pattern Mining (FPM) of streaming Tweets from Twitter social media platform (Guo *et al.*, 2012). The hot topics are mined using this FP growth algorithm and experiments were performed over Twitter data to demonstrate the utility of the proposed approach for topic detection from Twitter. Although different sizes of each batch were taken, the support threshold was set as 0.03 and error bound value as 0.001.

This has been improved using Dynamic Support Values (DSV) FPM from social network streams (Alkhamees *et al.,* 2016). A new support value algorithm was designed to find support using average and median of each incoming window. The experiments for the proposed approach were performed over *UK General elections 2015* and *Greece crisis 2015* datasets.

The datasets were collected using Twitter API. The authors received more than one million Tweets for the *General Elections* event collected in weekdays from 9am- 5pm in the period of 15-4-2015 to 26-5-2015. They got more than 150k Tweets collected in the period of 29-6-2015 to 16-7-2015 from Twitter for 2 hours daily for the *Greece crisis* event. The dataset was stored in NOSQL database called *MongoDB*. The *FPM-DSV* was further demonstrated using big window batch and small window batch.

The SFPM is proposed for topic detection using term selection, co-occurrence vector formation and greedy approach which expands set S (Petkos *et al.,* 2014). The *High Utility Pattern Clustering (HUPC)* was proposed for topic detection from large scale of microblog streams (Huang *et al.,* 2015). The HUPC is proposed to remove the redundancy of the phrases obtained using FPM approaches. The *High Utility Pattern (HUP)* algorithm was proposed based on the utility of overlapping pattern of words. A rule dynamics approach TRCM is proposed (Adedoyin-Olowe *et al.,* 2016) using association rule mining. TRCM used the data from *'72 Twitter TDT dataset'* on sports *(FA Cup dataset)* and politics *(US Elections dataset)* for experiments and evaluation.

## 2.5. Concluding Remarks

The word co-occurrence network is the graph which is created using words as nodes and adjacent words having the connection between them as an edge. The pre-processing of data will clean the text by removing stop-words and other irrelevant text from Tweets. The Microblog WCN is generated using tokenized words after pre-processing of Tweet. In this research work, automatic tokenization is used for pre-processing.

The WCN for traditional well-formed data and social media data may vary due to the presence of ill-formed text in social media. Thus, the structure of WCN for Microblogs is studied using network science properties in Chapter 3. WCN shows the cognitive patterns among words in Microblogs. These cognitive patterns are used for keyphrase extraction and event detection from social media data.

# Chapter 3.
# The Structure of Word Co-occurrence Network for Microblogs

The Microblog Word Co-occurrence Network (WCN) is the graph of words which occurs together in a Tweet. The text is in-formal and ill-formed social media data. The words which are used in Tweets are tokenized and Tweet is converted into path with word as single unit. This path is represented in directed Microblog WCN. The directed Microblog WCN is used to preserve the lexical sequence of words in WCN.

**Definition 3.1: Micorblog Word Co-occurrence Network (Microblog WCN):** The Word Co-occurrence Networks (WCN) for Microblogs are defined as graphs which are having words as nodes and adjacent words have link between them called edges. The frequency of two words appearing adjacently in a *Tweet Corpus* defines edge-weight. The former words in adjacent appearance have the tail node and the later one have the head node of an edge in directed network.

**Definition 3.2: Tweet Corpus:** The set of Tweets which are extracted for SMA is defined as Tweet Corpus. This Tweet Corpus is the textual data which is used for event detection.

User-generated words which are used in Tweets are generated from the active vocabulary of users. This vocabulary is built with interaction of user in social community both online and offline. It is observed that the network science concepts might play significant role in defining the cognitive behaviour of user-generated words. Thus, analysis of network science properties is important for the study of the structure of word co-occurrence networks for Microblogs. Various network science properties which are studied for Microblog WCN are the scale-free property, the small world property, the hierarchical organization, the assortativity, and the spectral analysis.

## 3.1. Network science and Microblog Word Co-occurrence Network

The network science field is used to study the semantics and the structure of graphical representation of real-world systems (Barabasi *et al.,* 2016). There are many different real-time networks, for instance, information networks and social networking (Leskovec *et al.,* 2008) whose structure and dynamics have been studied to obtain useful insights.

Tweets are extracted using Tweet ID as mentioned in the FSD dataset which is used for studying the network science properties of Microblog WCN. First three Tweets of FSD dataset are extracted from Topic 1. These Tweets are given below as Tweet Corpus A. The Tweet Corpus A is used for constructing Microblog *WCN*. The stop-words and all the irrelevant data is removed from Microblog WCN. The tokenization is performed by automatic pre-processing and each Tweet is tokenized into different words. After pre-processing, the tokenized words are indexed as shown in Table 3.1.

***Tweets Corpus A:*** *{1: 'Hmm Sunday Mirror news ed @jasav is tweeting "confirmation" that Amy Winehouse has died.', 2: 'Maybe a rumour but Dean Piper appears to have tweeted that Amy Winehouse has died', 3: '@jlscrazymad rumours amy winehouse has died- remains rumours'}*

**Table 3.1:** Index of tokenized words as tokens generated from Tweets after pre-processing

| *Index* | *Word* | *Index* | *Word* | *Index* | *Word* |
|---|---|---|---|---|---|
| 1 | Hmm | 2 | Sunday | 3 | Mirror |
| 4 | News | 5 | Jasav | 6 | Tweeting |
| 7 | Confirmation | 8 | Amy | 9 | Winehouse |
| 10 | Died | 11 | Maybe | 12 | Rumour |
| 13 | Dean | 14 | Piper | 15 | Appears |
| 16 | Tweeted | 17 | jlscrazymad | 18 | Rumours |
| 19 | Remains | | | | |

The Microblog WCN is evolved from the indexed set of words with corresponding co-occurrences. In literature, the construction of Microblog WCN has been classified into different types (Abilhoa *et al.,* 2014) as follows:

### 3.1.1. Nearest neighbour edging and all-pair neighbour edging

In the *nearest neighbour edging* adjacent neighbouring words are linked by an edge as shown in Figure 3.1(a), Figure 3.1(b), Figure 3.1(c), and Figure 3.1(d). For Microblog WCN, the advantage of using *nearest neighbour edging* is to reduce the complexity and thus, speed up the computations. To connect all words of a Tweet with each other, *all-pair neighbour edging* is used in which every word is connected to every other node as shown in Figure 3.1(e), Figure 3.1(f), Figure 3.1(g), and Figure 3.1(h). This approach is computationally expensive. Thus, in order to perform efficient computations, *nearest neighbour edging* is used as a default approach unless or otherwise stated explicitly.

### 3.1.2. Directed and undirected Graph

There are two types of networks on the basis of edge type namely *directed* and *un-directed*. The *directed graph* have pointed edges from former word to the later word in the adjacent appearance of two words as shown in Figure 3.1(a), Figure 3.1(b), Figure 3.1(e), and Figure 3.1(f). The *undirected graph* do not give sequence of nodes in WCN as shown in Figure 3.1(c), Figure 3.1(d), Figure 3.1(g), and Figure 3.1(h). Although some of the network science properties are usually evaluated using *un-directed graph*, the *directed graph* is used to maintain the lexical sequence of words as they appear in Tweets. Thus, by default *directed graph* is used for this study.

**Algorithm 3.1:** Create Weighted DiGraph for Microblog WCN

1. Given: $D$
2. $G = nx.DiGraph()$
3. $Extract\ edges\ E_i\ from\ d_i\ joining\ adjacent\ words$
4. $For\ (u, v)\ in\ E:$
5.     $G.add\ edge(u, v, weight = nw)$ //u and v are nodes which are connected by an edge

### 3.1.3. Weighted edges and same-weight edges

*Weighted edges* are those edges which show the count of co-occurrence frequency as shown in Figure 3.1(b), Figure 3.1(d), Figure 3.1(f), and Figure 3.1(h). The same-weight edges are those edges to which weight is not assigned and thus, weight is considered as same or equivalent as shown in Figure 3.1(a), Figure 3.1(c), Figure 3.1(e), and Figure 3.1(g). The same-weight edges are referred as *un-weighted edges*. In this research work, the *weighted edges* are used for identifying influential segments using the key-phrase extraction technique and *un-weighted edges* are used for calculating assortativity.

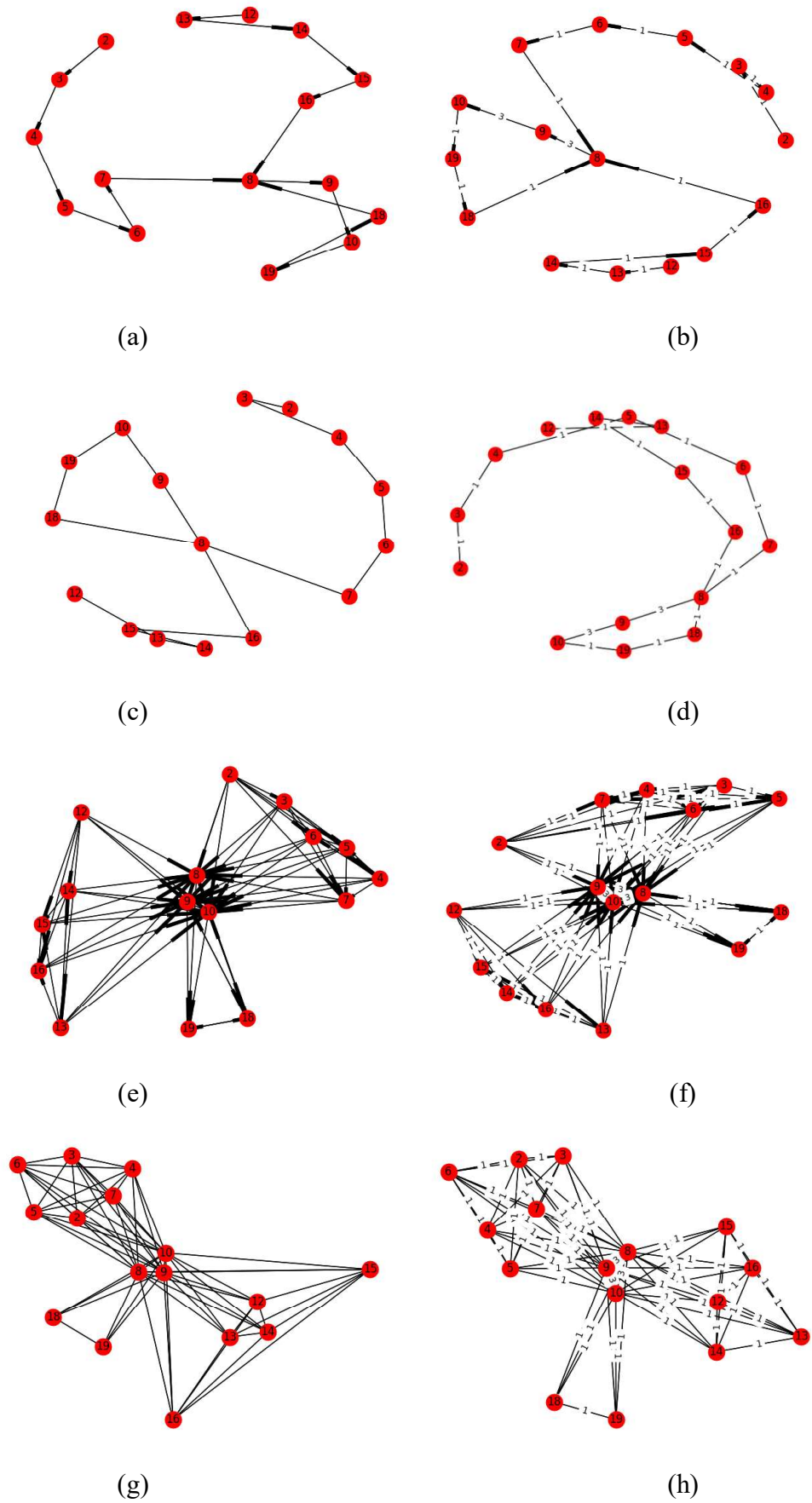

**Figure 3.1:** Classification of Microblog Word Co-occurrence Networks

An algorithm is used to create *nearest neighbour edging*, *directed*, and *the weighted* Microblog WCN as described in Algorithm 3.1. This network is shown in Figure 3.1 (b). which is evolved from index of words and co-occurrences of Tweet Corpus A, respectively. The words are tokenized and indexed in Table 3.1. It clearly indicates that edges from index 8 to index 9, and index 9 to index 10 have high edge weights. This shows that the index [8, 9, and 10] shows some pattern. As observed from Table 3.1, this indexing sequence gives the key-phrase as 'Amy Winehouse Died'. This key-phrase is the topic of discussion within the Tweet Corpus A. Different experiments is performed to apply network science properties over Microblog WCN.

## 3.2. Experimental set-up

The experiments are performed to analyze the structure of WCN using FSD dataset. For this, 100k Twitter posts are extracted from the FSD dataset and these 100k Tweets are considered as W. The Microblog WCN is constructed for the number of Tweets in Tweet Corpus as W/1000, W/100, W/20, W/10, W/2 and W to analyse the behaviour and stability of Microblog WCN. These Tweet Corpus are named as W/1000, W/100, W/20, W/10, W/2 and W, respectively.

## 3.3. Stability of Microblog Word Co-occurrence Networks

The observations are made for different sizes of Tweet corpus as obtained in the form of: number of tweets (#Tweets) which shows Tweet Corpus size, length of the network (L) which gives the number of words in the corpus, the number of nodes (N), number of edges (E), average number of nodes in the Tweet ($A_n$), average shortest path length (ASPL), and clustering coefficient (CC) as shown in Table 3.2. The Microblog WCN is found to be stable. For each corpus of $W/x$, the $x$ number of experiments are performed and average of all the results are recorded as resulting value. This is because the small amount of data may or may not give reliable results for single experiment. As the x decreases, the number of experiments are increased. As per observations, the resulting values are scalable over corpus size and thus, the Microblog WCN is stable.

It is observed that the average number of nodes in all the Tweet Corpus remains consistent. With increase in Tweet Corpus size, the number of nodes in Microblog WCN increases at lower rate than that of length of the Tweet Corpus. Also, the ratio of number of edges to number of nodes in the network is increased with increase in the size of Tweet Corpus. This observation shows that words are repetitive in active vocabulary of users.

Table 3.2: Structure of Microblog WCN for different size of corpus

| Dataset | #Tweets | L | N | E | $A_n$ | ASPL | CC |
|---|---|---|---|---|---|---|---|
| $FSD:W/1000$ | 100 | 748 | 527 | 550 | 7.480 | 2.574 | 0.0142 |
| $FSD:W/100$ | 1000 | 7359 | 3560 | 5292 | 7.359 | 4.154 | 0.0142 |
| $FSD:W/20$ | 5000 | 36369 | 11307 | 25022 | 7.273 | 3.956 | 0.0230 |
| $FSD:W/10$ | 10000 | 73243 | 18433 | 49011 | 7.324 | 3.696 | 0.0348 |
| $FSD:W/4$ | 25000 | 183466 | 34925 | 116477 | 7.338 | 3.420 | 0.0493 |
| $FSD:W/2$ | 50000 | 366243 | 56259 | 219950 | 7.324 | 3.214 | 0.0682 |
| $FSD:W$ | 100000 | 481982 | 67979 | 281802 | 7.320 | 3.144 | 0.0765 |

The number of edges increases at higher rate than that of number of nodes. This shows that words are used with different words and thus, edges make different connections among words. Although the network is stable but some deviations are observed with increase in size of the Tweet Corpus. The clustering coefficient depends upon how well the Microblog WCN is connected. As the Tweet Corpus size gets increased, the number of edges increases at higher rate than the number of nodes which shows highly connected network and thus, *clustering coefficient* increases. Different network science metrics which are used to study for Microblog WCN are explained as follows:

### 3.3.1. Degree

The sum of number of edges which connects the node to any other node is called *degree* of a node. In directed network, the number of incoming links denotes *in-degree* and number of outgoing links defines *out-degree*. *Degree* defines the importance of a node in a network. Higher the *degree*, more important is a node in a complex network.

Thus, *degree* may also represent the words which are most commonly used but may or may-not be significant. For instance*, good, morning, best, change, hello* etc. Thus, more statistical information is extracted to analyze the significance of the words in Microblog WCN. The keyword extraction technique is proposed using degree centrality as *Twitter Keyword Graph (TKG)* (Abilhoa *et al.,* 2014) which is used for identifying keywords for Twitter data.

### 3.3.2. Strength and Selectivity

The sum of the weights of weighted edges which are pointing to or from a node is said to be *Strength* of a node. The sum of edge weights of incoming link is called *in-strength* and sum of edge weights of outgoing links is called *out-strength*. The *strength* of the node indicates the total number of times the word has appeared with other words which is neither stopword nor irrelevant word. However, the concept of calculating *strength* is somewhat similar to the Term-Frequency (TF) in case of Microblog WCN as shown in Equation 3.1.

$$Degree = TF - (w_i) \qquad 3.1$$

where, $TF$ is the term frequency and $w_i$ is the number of times word $i$ was posted alone or with the stop-words/ irrelevant text only.

The *selectivity* (Beliga *et al.*, 2016) of the node is obtained with normalized edge strength which means weight that is divided among all the corresponding edges equally as defined in Equation 3.2.

$$Selectivity = \frac{Strength(w_i)}{Degree(w_i)} \qquad 3.2$$

The *selectivity* of a word $w_i$ is calculated to check the importance of word with respect to its appearance in the text. The author proposed the *Selectivity Based Keyword Extraction (SBKE)* technique (Beliga *et al.*, 2016) using *selectivity* measure.

The *edge-weight* is the number of times two words have co-occurred in Tweet Corpus. The *edge-strength* gives the information about how closely two words are knit together than their co-occurrence with other words in the network as defined in Equation 3.3.

$$Edge\ Strength = \frac{2 * freq(w_i, w_j)}{freq(w_i) + freq(w_j) - freq(w_i, w_j)} \qquad 3.3$$

The sum of edge-strength of in-coming links and out-going links for all nodes in a network is called *in-strength* and *out-strength*, respectively.

### 3.3.3. Clustering Co-efficient (CC)

The *clustering coefficient* calculates the fraction of transitive triplets in the network. It shows the degree to which the network is connected. If all the neighbours of a node are connected to each other, the *clustering coefficient* of the network is high. As per the study of the *nearest neighbour edging* based Microblog WCN, it is observed that the semantics and nature of the *word co-occurrence network* differs from the network of complex social networks. The *clustering coefficient* is defined as shown in Equation 3.4.

$$CC(k)_i = \frac{2 * E_i}{k_i(k_i - 1)} \qquad 3.4$$

where $E_i$ means the number of edges among vertices in the neighbourhood of vertex $i$ and $k_i$ denotes degree of the vertex $i$. The Microblog WCN is constructed as a path network by considering words as a unit and sequence of words as a path. The words might not connect with each other unless there is huge amount of data. If huge amount of data is obtained in Tweet Corpus, due to repetition of words in multiple Tweets, the network will start getting connected and may form word communities. Thus, *clustering coefficient* is usually low for Microblog WCN having small Tweet Corpus size and more for large size of Tweet Corpus.

### 3.3.4. Average shortest path length (ASPL)

The *Average Shortest Path Length (ASPL)*, as the name suggests, is the average of all the shortest paths from any one node to any other node in the network. In Microblog WCN, if the *clustering coefficient* is low as observed for small Tweet corpus size, the *ASPL* will be high. This is because the network of words may or may not connect many nodes (words) within the network. *ASPL* is defined as shown in Equation 3.5

$$ASPL = \frac{2 \sum_{i>j} d_{ij}}{N(N-1)} \qquad 3.5$$

where $d_{ij}$ is the *shortest path length* between two vertices $i$ and $j$. *ASPL* gives useful information for *the small-world property*. The *ASPL* decreases with increase in the amount of data due to increase in number of edges and relatively low rate of increase in number of nodes as shown in Table 3.2.

### 3.3.5. Centrality

The centrality of the network is the measure of importance of node. There are three different types of centralities. The centrality measure is used by researchers to detect keywords from the traditional well-formed text. The four types of centralities which are widely studied by researchers are *degree centrality, betweenness centrality, closeness centrality and eigenvector centrality* as explained below:

- **Degree Centrality:** *Degree centrality* is measured by calculating the degree for each node. The idea behind calculating this centrality is that the node having more degree is more important. However, in case of textual documents and WCN, this might not be the suitable measure because the network is path based network.

- **Betweenness Centrality:** *Betweenness centrality* is the total number of paths from any node (except the node $v$) to any other node that passes through a node $v$. The idea behind calculating this centrality is that the node which connects more number of nodes with each other is important. In path based network such as WCN, *betweenness centrality* is suitable measure for identifying keywords (Abilhoa *et al.,* 2014).
- **Closeness Centrality:** *Closeness centrality* is the total number of direct paths from every node to every other node passing through a node $v$. It states that the closer the node is to other nodes in the network, more important is the node (Biswas *et al.,* 2018).
- **Eigenvector Centrality:** *Eigenvector centrality* measures the influence of a node in the network with the help of neighbouring nodes. Nodes are assigned scores using different network metrics. The nodes which are connected to those nodes which have high scores, have high eigenvector centrality.

### 3.3.6. Eccentricity

*Eccentricity* is calculated as the node centrality index. For any node $v$, the reciprocal of the distance of $v$ to the most distant node is defined as *eccentricity* of node $v$. If eccentricity value is high, it means that other nodes are in high proximity. Eccentricity is used for keyword extraction from WCN of well-formed documents in existing keyword extraction techniques.

## 3.4. The Structure of Word Co-occurrence Network

As keyphrase extraction techniques have been proposed using network science metrics for Twitter data (Abilhoa *et al.,* 2014; Biswas *et al.,* 2018) and Croatian News (Beliga *et al.,* 2016), it is observed that the network science properties may define useful patterns within Microblog WCN which can help to determine keyphrases in the Corpus. The structure of Microblog WCN is analyzed using different network science properties. Different properties which are studied for this network are the scale-free property, the small-world property, the hierarchical organization, the assortativity, and the spectral analysis.

### 3.4.1. Scale-free property

The *degree distribution* is defined as the probability by which a randomly chosen node has degree $k$ (Liang *et al.,* 2015). The power law is defined as $p(k) \propto k^{-\gamma}$, where $2 < \gamma < 3$ for scale-free networks. The scale free property in Microblog WCN defines the behaviour of the network which does not scale over the chosen parameter. The value of $\gamma$ may vary as

- $\gamma \leq 2$: Shows anamolous behaviour. This implies that largest hub should have degree $> N$ which is not possible.
- $2 < \gamma < 3$: Gives scale-free regime. This indicates that the first moment is finite and second moment diverges as $N \to \infty$ where $N$ is number of nodes in the network.
- $\gamma > 3$: shows random network regime. This value shows that the second and the third moments are finite.

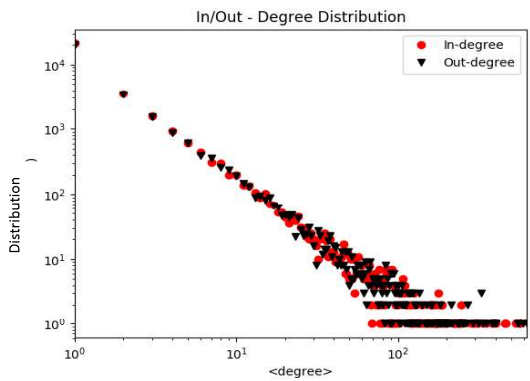

3.2.1 (a)

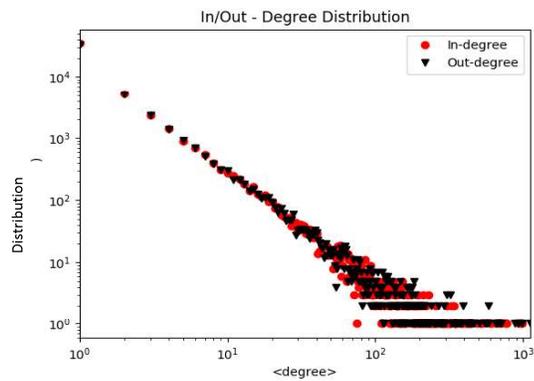

3.2.1 (b)

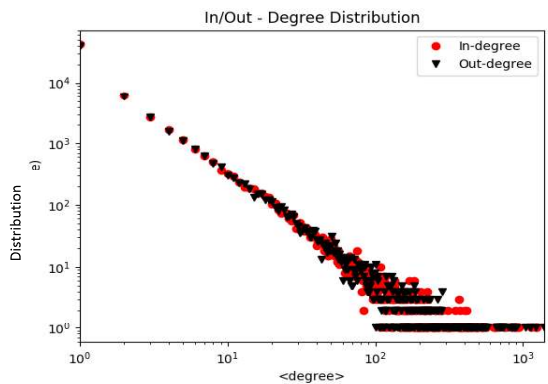

3.2.1 (c)

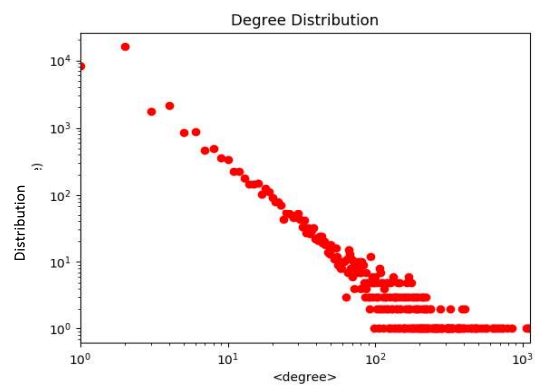

3.2.2 (a)

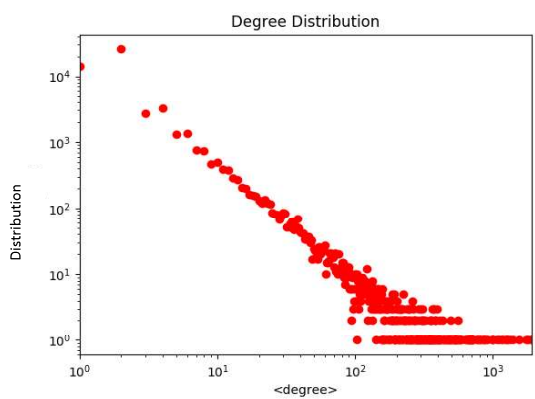

3.2.2 (b)

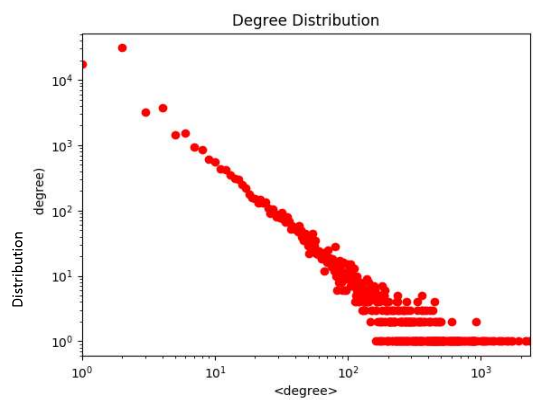

3.2.2 (c)

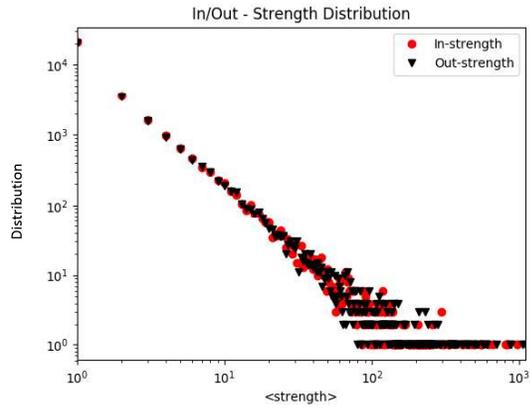
3.2.3 (a)

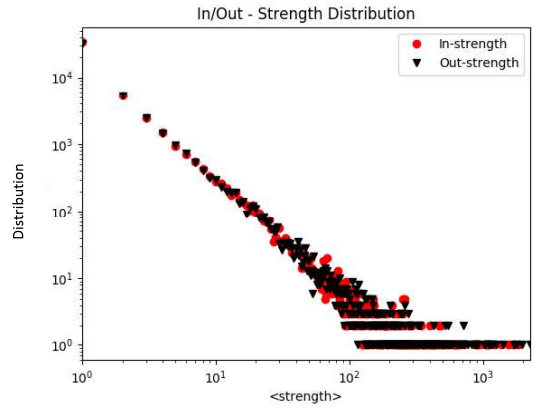
3.2.3 (b)

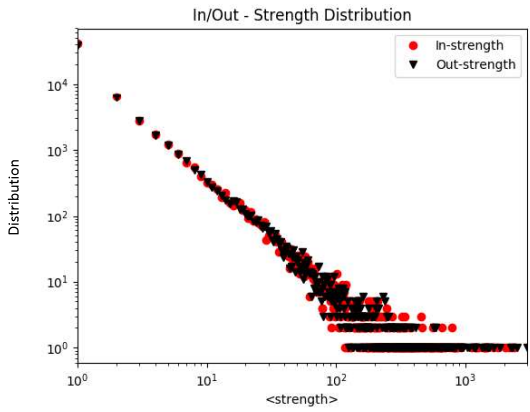
3.2.3 (c)

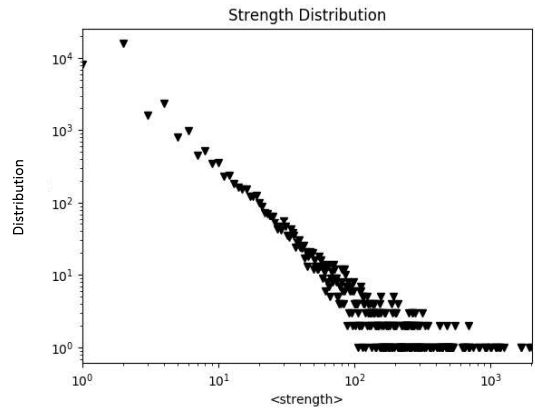
3.2.4 (a)

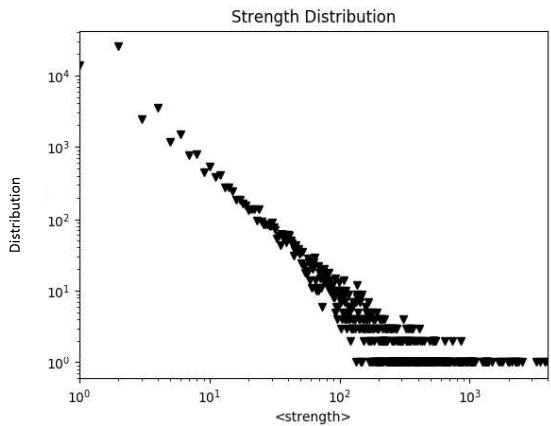
3.2.4 (b)

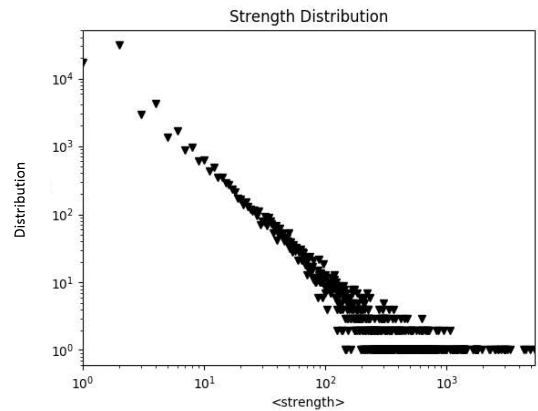
3.2.4 (c)

**Figure 3.2:** Scale-free property of Microblog WCN for Tweet Corpus of size W/4, W/2, W, respectively. 3.2.1: in-degree out-degree distribution, 3.2.2: degree distribution, 3.2.3: in-strength out-strength distribution, 3.2.4: strength distribution

**Table 3.3**: Different values of $\gamma$ for Microblog WCN for different size of corpus

| $Dataset$ | $K_{in}$ | $K_{out}$ | $K$ | $S_{in}$ | $S_{out}$ | $S$ | $E_w$ |
|---|---|---|---|---|---|---|---|
| $FSD: W/100$ | 3.320 | 2.996 | 2.852 | 3.300 | 3.010 | 2.740 | 1.361 |
| $FSD: W/20$ | 2.376, 4.141 | 2.349, 4.150 | 2.145, 3.183 | 2.309, 4.150 | 2.240, 4.154 | 2.120, 2.960 | 1.340, 4.122 |
| $FSD: W/10$ | 2.137, 3.485 | 2.108, 3.472 | 2.005, 2.950 | 2.064, 2.019 | 2.037, 3.018 | 2.028, 2.743 | 1.326, 3.548 |
| $FSD: W/4$ | 1.993, 2.828 | 2.002, 2.843 | 1.931, 2.491 | 1.991, 2.529 | 2.013, 2.550 | 1.902, 2.224 | 1.304, 3.094 |
| $FSD: W/2$ | 1.921, 2.528 | 1.906, 2.561 | 1.825, 2.166, 3.998 | 1.893, 2.293 | 1.876, 2.216 | 1.782, 2.002, 3.095 | 1.285, 3.063 |
| $FSD: W$ | 1.874, 2.401 | 1.857, 2.373 | 1.803, 2.096, 3.684 | 1.837, 2.134 | 1.827, 2.049 | 1.774, 1.930, 2.707 | 1.277, 2.948 |

Earlier, it has been observed that for traditional text, WCN follows the two-regime power law. The two-regime power law shows different scaling behaviour of WCN based on number of words in the network. For well-formed text, it is observed that the value of $\gamma$ *is less than* 1.5 for value of degree $k < 10^3$, and the value of $\gamma$ *is greater than* 2.7 for $10^3 < k < 10^5$ where $k$ is the degree. In Table 3.3, the power-law is observed using $\gamma$ for *in-degree out-degree distribution, degree distribution, in-strength out-strength distribution, and strength distribution* is analyzed. It is observed that the Microblog WCN follows scale-free property for number of nodes more than $10^2$. Thus, Microblog WCN follows scale free property for *node-degree distribution and node-strength distribution.*

**Definition 3.3: Node Degree:** The number of edges linked to any node which connect the node to other nodes in a network is called *Node Degree.*

**Definition 3.4: Degree Distribution:** The probability of node having degree $k$ in a network is called the *degree distribution*. It follows *scale-free property* for higher values of number of nodes. This is because as the size of Tweet Corpus gets increased, interlinking between the nodes is also increased. To examine in-degree and out-degree of nodes separately, in-degree out-degree distribution is observed in Figure 3.2.1.

**Definition 3.5: In-degree Out-Degree Distribution:** The probability with which in-degree or out-degree of a node in a network is $k$, is defined as *in-degree distribution* or *out-degree distribution*, respectively.

It is observed that the in-degree and out-degree distribution overlaps each other. This is because if there is path of Tweet *"A B C D E"* and another Tweet with *"A B C K L"*, then two links are created among nodes *'A and B'*, and *'B and C'* which gives edge weight as 2. For directed Microblog WCN, the links are formed *'from A to B'*, and *'from B to C'*. Thus, in-degree and out-degree is 2. This may be because both Tweets belong to same topic. Thus, there exist some latent patterns for in-degree and out-degree distribution.

**Definition 3.6: Node Strength:** The sum of edge weights of all the edges pointing to or from a node is defined as *Node Strength* or *strength.*

**Definition 3.7: Strength Distribution:** The probability of node having strength *k* is defined as *Strength Distribution*. This distribution is shown in Figure 3.2.4. The node-strength shows how strongly the node is connected to rest of the network. This is essential to examine the importance of a node. For more observations, *in-strength distribution and out-strength distribution* is evaluated for Microblog WCN.

**Definition 3.8: In-strength out-strength Distribution:** The sum of edge weights of in-coming links and out-going links are used to examine the distribution of strength of node and is defined as *in-strength distribution and out-strength distribution*, respectively. This distribution is shown in Figure 3.2.3. It is observed that in-strength and out-strength distribution shows similar behaviour as that of in-degree and out-degree distribution, respectively. As observed in Figure 3.2 and Table 3.3, both *node-degree distribution and node-strength distribution* follows scale-free property for medium to large size of the network. The edge-weight distribution and edge strength distribution are shown in Figure 3.3.1 and Figure 3.3.2, respectively.

**Definition 3.9: Edge-Weight Distribution:** The frequency of two nodes occurring together is defined as an *edge-weight*. The Microblog WCN is scaled over *edge-weight distribution* as shown in Figure 3.3.1. It shows that there are important edges in the network as they show deviation in weights than average weight.

**Definition 3.10: Edge-Strength Distribution:** The edge strength is the measure of how strongly two nodes are connected to each other as compared to other nodes in network. It is defined as shown in Equation 3.6

$$ES = \frac{2 * freq(t_i, t_j)}{freq(t_i) + freq(t_j) - freq(t_i, t_j)} \qquad 3.6$$

where $freq(t_i, t_j)$ is defined as frequency of edge occurring between $t_i$ and $t_j$ and $freq(t_i)$ or $freq(t_j)$ refers to number of times term $i$ and term $j$ occur, respectively.

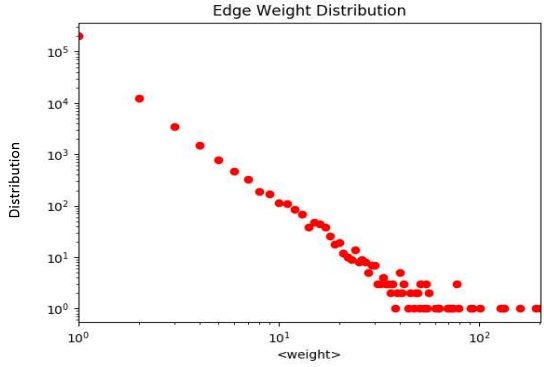

3.3.1 (a)

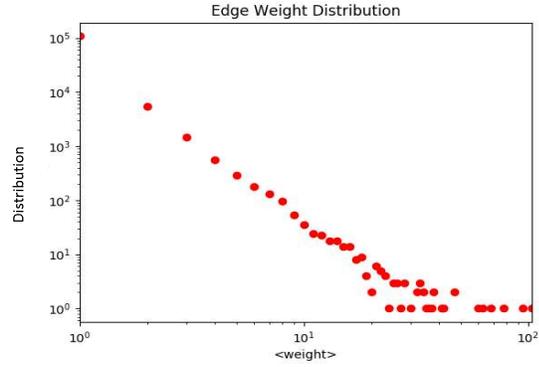

3.3.1 (b)

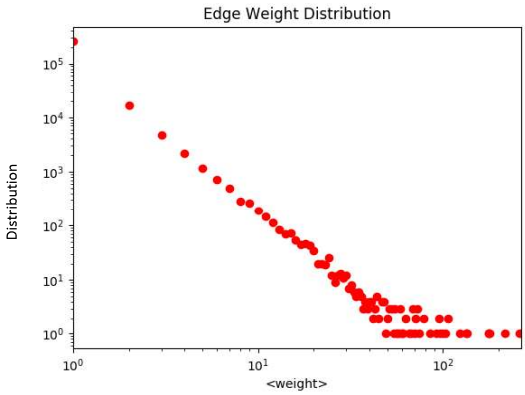

3.3.1 (c)

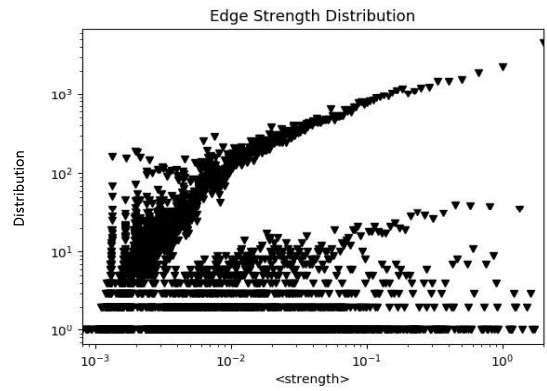

3.3.2 (a)

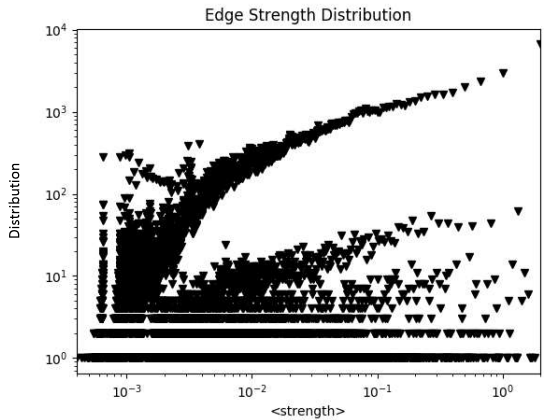

3.3.2 (b)

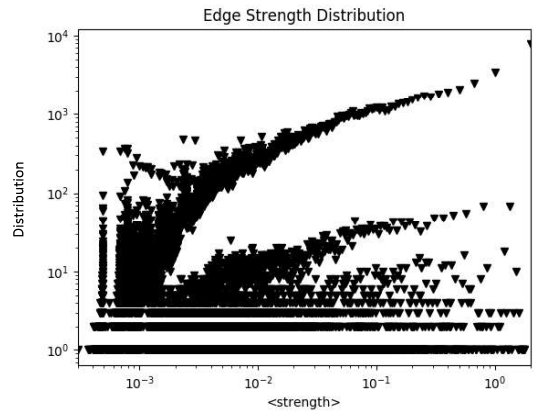

3.3.2 (c)

**Figure 3.3:** Microblog WCN for Tweet Corpus of size W/4, W/2 and W corpus. 3.3.1: Edge-weight distribution, 3.3.2: Edge-strength distribution

From Figure 3.2.2, it is observed that as the strength gets closer to 2, the probability of finding the edge with higher edge strength also increases as per logarithmic scale. The scaling of the edge strength is clearer as we increase the size of the corpus. This property is used to propose k-bridge decomposition to identify influential segments as discussed in upcoming Chapters.

Thus, the Microblog WCN follows scaling property for both node distribution and edge distribution. The dataset W/4 contains enough number of nodes for analysis of Microblog WCN as observed from Figure 3.2 and Figure 3.3. This size of Tweet Corpus can be used for finding summaries from discrete set of streaming data for SMA.

### 3.4.2. Small World feature

The small-world feature for complex networks can be identified using *Average Shortest Path Length (ASPL), and Clustering Coefficient (CC)* metrics.

**Definition 3.11: Erdos-Renyi random network:** The Erdos-Renyi (ER) random network is the network which is formed randomly using the information of *number of nodes of real – time network*, and *percentage* as the probability with which the links should be formed.

The ASPL and CC are calculated for both *Erdos-Renyi random network* and *Microblog WCN*. It is found that the *CC* of both the networks are different. The *CC* for *Microblog WCN* is much higher than *CC* for *ER random network* as observed from Table 3.4. The *ASPL* of both the networks for different sizes of Tweet Corpus follows same scale. Thus, Microblog WCN follows small-world feature.

**Table 3.4:** Small World Property measures for Microblog WCN

| Dataset | CC | $CC_r$ | Directed /Undirected | ASPL | $ASPL_r$ | ASPL (max) | $ASPL_r$ (max) |
|---|---|---|---|---|---|---|---|
| FSD: W/1000 | 0.0142 | 0 | Directed | 0.938 | 5.350 | 2.574 | 5.367 |
| | | | Undirected | 2.422 | 1.575 | 11.431 | 7.481 |
| FSD: W/100 | 0.0142 | 0 | Directed | 0.916 | 6.617 | 4.654 | 6.611 |
| | | | Undirected | 1.576 | 1.354 | 6.090 | 7.553 |
| FSD: W/20 | 0.0238 | 0.00039 | Directed | 0.710 | 6.266 | 3.956 | 6.253 |
| | | | Undirected | 1.315 | 2.355 | 4.829 | 6.431 |
| FSD: W/10 | 0.0348 | 0.00026 | Directed | 0.643 | 6.010 | 3.696 | 5.992 |
| | | | Undirected | 1.220 | 2.691 | 4.524 | 6.047 |
| FSD: W/4 | 0.0493 | 0.000135 | Directed | 0.645 | 5.714 | 3.420 | 5.703 |
| | | | Undirected | 1.258 | 3.366 | 4.240 | 5.724 |

As per the observations, the value of *ASPL* for un-directed network is closer to the value of *ASPL* for ER random network than that of directed Microblog WCN. The idea behind calculating *ASPL (max)* is to calculate the *ASPL* for the largest sub-graph of disconnected network. This is so because the *ASPL* for complete network gives the average of all the sub-graphs which might reduce due to the presence of isolated nodes and outliers in disconnected network.

### 3.4.3. Hierarchical Organization

*Hierarchical organization* is the process to check if the nature of the network is hierarchical which means that in the network, few nodes are important and are linked at equal scale, hierarchically, to make their neighbours important. The importance of the node decreases for descenders at the same scale for subsequent levels. *Hierarchical organization* is defined as the probability distribution of average clustering coefficient of all nodes with degree *k*. Hierarchically organization is defined as $C(k) \propto k^{-\beta}$ where asymptotic scaling is required for network to be hierarchically organized with linear dependency using value of $\beta$. As shown in Figure 3.6 the nodes in the network are hierarchically organized. This shows that the behaviour of the network is hierarchical in terms of importance of node. The central nodes are more important as they can communicate between other nodes in the network.

As observed in Section 3.3, the *CC* of the network is scalable over the size of corpus and thus, the network possess partial hierarchical organization for all-pair neighbouring network as observed from Figure 3.5. The network is partially hierarchically organized for all-pair neighbouring network. This is because the connections among words of one Tweet are more and thus, it is more bounded. However, the first word of the Tweet is counted as influential node which can connect other nodes, as it points to all other words of Tweet. This may or may not show any significant behaviour. Moreover, the all-pair neighbouring network is computationally expensive as more efficient systems are required which can speed up the processing.

However, for nearest neighbour network, the hierarchical organization is not scaled for number of nodes less than $10^2$. It shows anamolous behaviour and can be studied for other problem domains in near future.

Although the network is partially hierarchically organized for all-pair neighbour directed network, hierarchical property for Microblog WCN is used minimally to handle the streaming data due to high complexity.

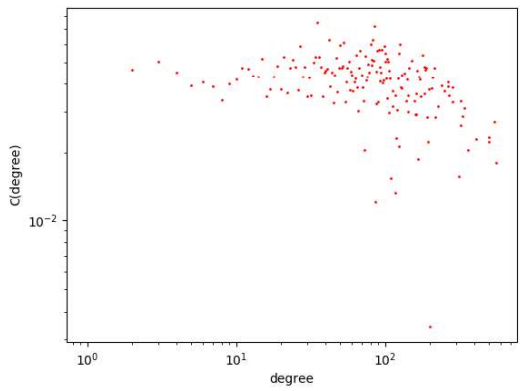
3.4.1 (a)

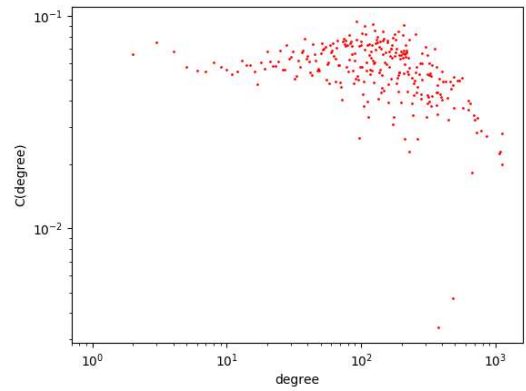
3.4.1 (b)

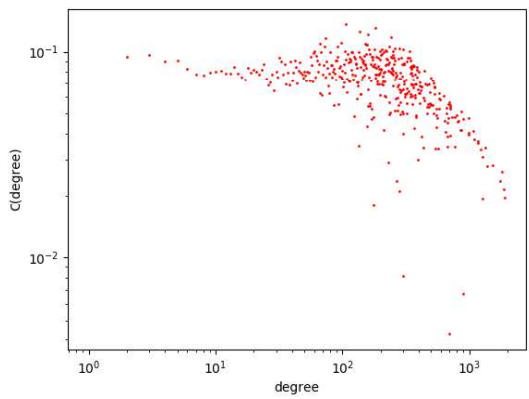
3.4.1 (c)

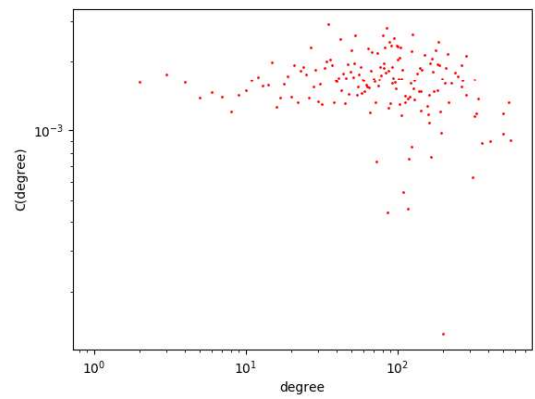
3.4.2 (a)

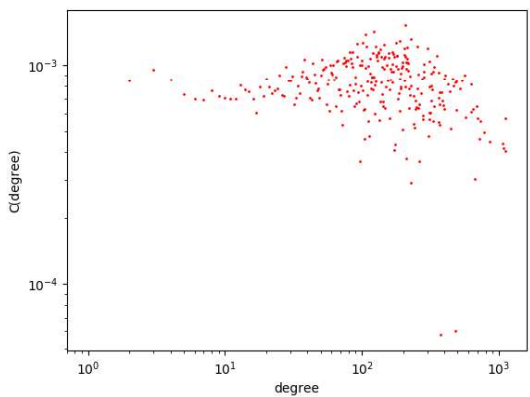
3.4.2 (b)

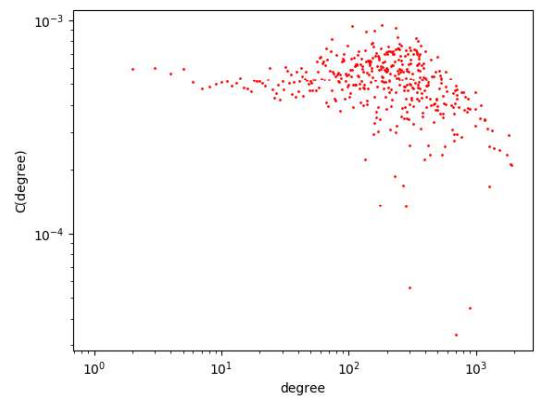
3.4.2 (c)

**Figure 3.4:** Clustering Coefficient of Tweet Corpus of size W/10, W/4, W/2 for nearest neighbour Microblog WCN. 3.4.1: Unweighted network 3.4.2: Weighted network

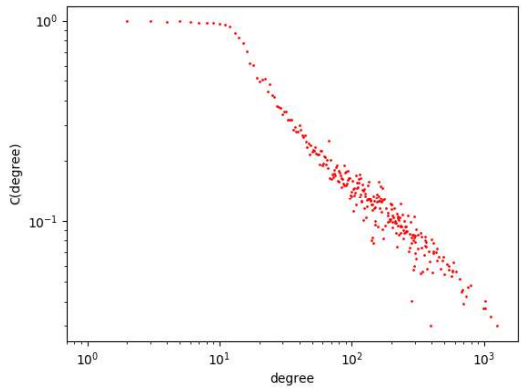

3.5.1 (a)

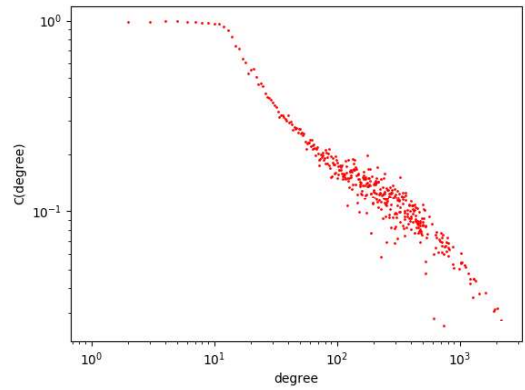

3.5.1 (b)

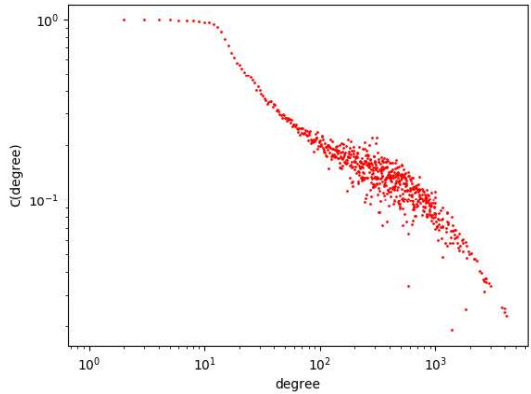

3.5.1 (c)

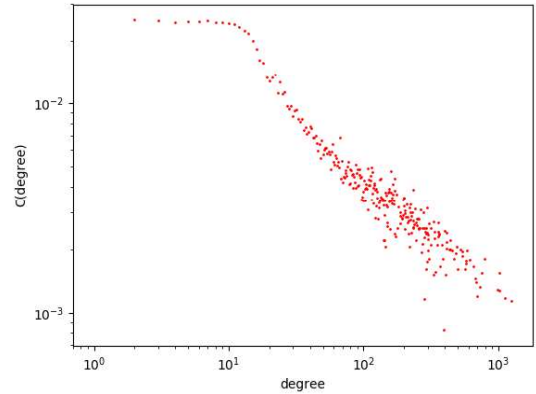

3.5.2 (a)

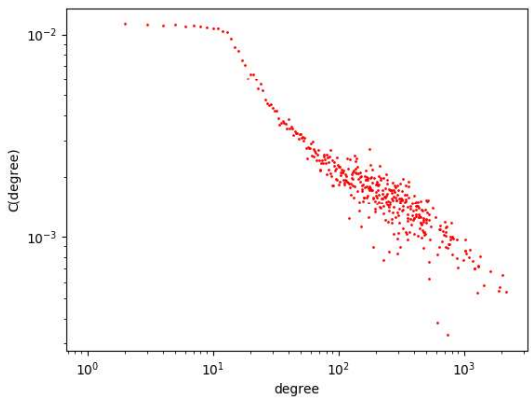

3.5.2 (b)

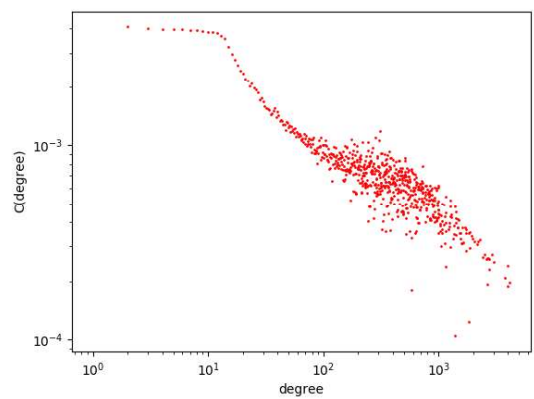

3.5.2 (c)

**Figure 3.5**: Clustering coefficient of Tweet Corpus of size W/20, W/10, W/4 for all pair neighbour Microblog WCN. 3.5.1: un-weighted network, 3.5.2: Weighted Network

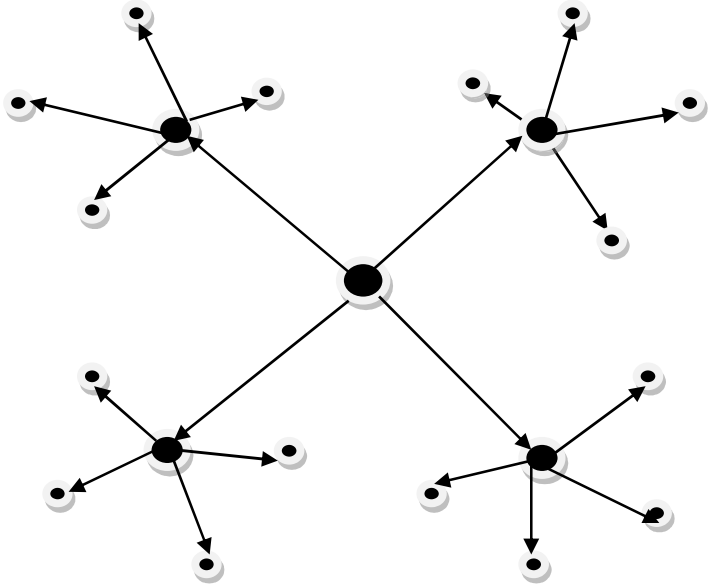

**Figure 3.6**: Hierarchical Organization of network

### 3.4.4. Assortativity

The *assortativity* is the property of the network (Newman *et al.*, 2002) to find the behaviour of nodes among each other. The *assortativity* $p(k)$ is defined as the probability that a randomly chosen node with degree $k$ is connected to the nodes with comparative degree to degree $k$ (Noldus *et al.*, 2015). The significance of the *assortativity* is that it defines the pattern of a network and is defined using *Pearson correlation coefficient* ($\tau$) as shown in Equation 3.8.

$$\tau = \frac{M^{-1}\sum_i j_i k_i - \left[M^{-1}\sum_i \frac{1}{2}(j_i + k_i)\right]^2}{M^{-1}\sum_i \frac{1}{2}(j_i^2 + k_i^2) - \left[M^{-1}\sum_i \frac{1}{2}(j_i + k_i)\right]^2} \qquad 3.8$$

where $j_i$ and $k_i$ are the degrees of the two endpoints of the $i^{th}$ edge, and $M$ is the total number of edges in the network. If $\tau > 0$, the network is said to be *assortative mixing*; while for $\tau < 0$, the network is called *disassortative mixing*. This follow the principle of *'rich get richer and poor get poorer'*. This is so because the nodes having high degree are connected to high degree neighbours, are *assortative*. If more high degree nodes are connected to low degree nodes, then the network is said to be *dis-assortative*.

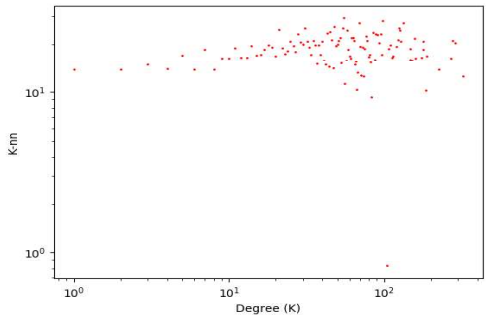
3.7.1 (a)

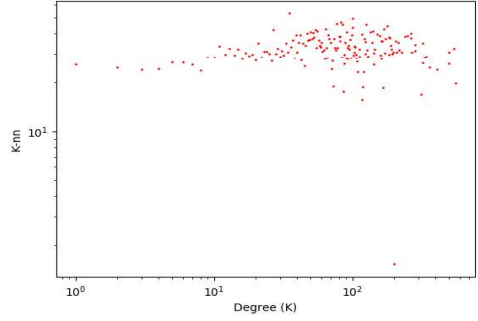
3.7.1 (b)

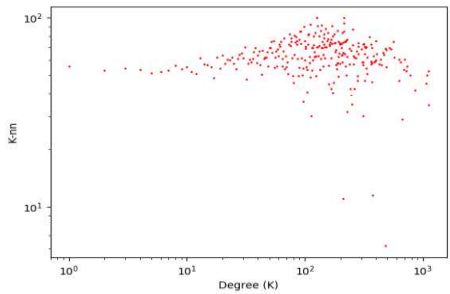
3.7.1 (c)

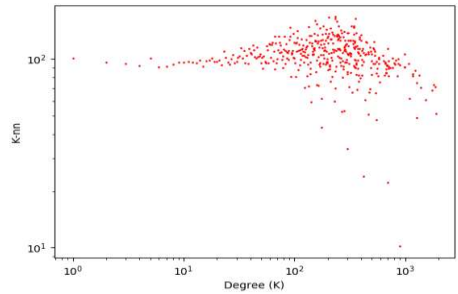
3.7.1 (d)

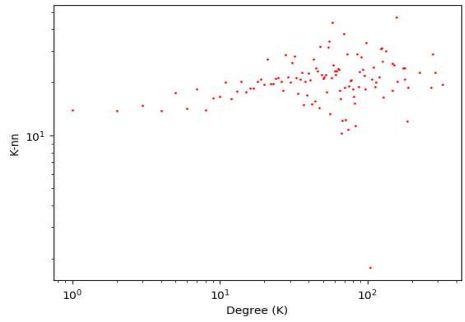
3.7.2 (a)

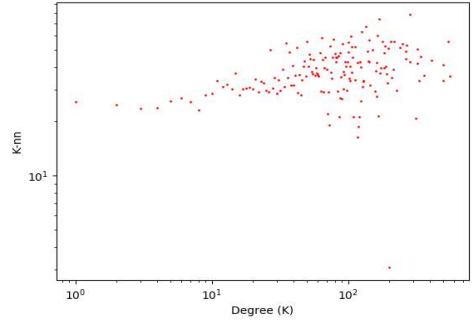
3.7.2 (b)

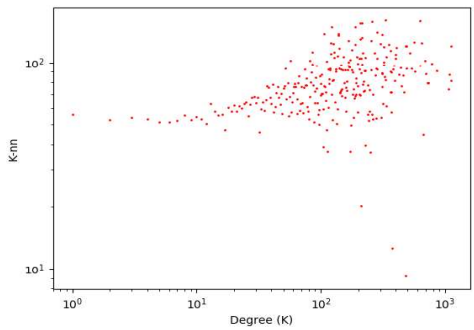
3.7.2 (c)

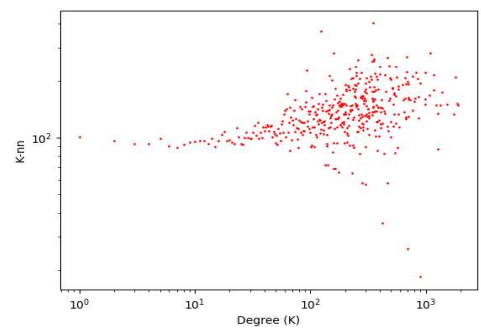
3.7.2 (d)

**Figure 3.7**: Degree correlation coefficient for Tweet Corpus of size W/20, W/10, W/4 and W/2 for Microblog WCN. 3.7.1: all pair neighbour edging. 3.7.2: nearest neighbour edging

The assortativity of the Microblog WCN depends upon the network of words evolved from Tweet Corpus. If the word is significant, it gets connected to many words which in turn gets connect to many other words. The node with high degree shows that it has high term frequency but is not always keyword. Thus, high degree nodes of network which get connected to high degree nodes give important information about Tweet Corpus. This community of high degree nodes can provide useful insights about topics, trends, and events from Microblog WCN.

Table 3.5: Assortativity measures for different corpus size of Microblog WCN

| $Dataset$ | $\tau\ (UW)$ | $\tau\ (WT)$ | $\mu\ (UW)$ | $\mu\ (WT)$ |
|---|---|---|---|---|
| $FSD: W/1000$ | $-0.385$ | $-0.421$ | $4.188$ | $4.188$ |
| $FSD: W/100$ | $-0.268$ | $-0.267$ | $3.946$ | $3.946$ |
| $FSD: W/20$ | $-0.319$ | $-0.288$ | $2.226$ | $2.409$ |
| $FSD: W/10$ | $-0.323$ | $-0.278$ | $1.846$ | $1.756$ |
| $FSD: W/4$ | $-0.329$ | $-0.265$ | $1.539$ | $1.481$ |
| $FSD: W/2$ | $-0.334$ | $-0.261$ | $1.482$ | $1.386$ |
| $FSD: W$ | $-0.334$ | $-0.256$ | $1.392, 3.724$ | $1.354, 2.615$ |

As observed from Table 3.5, the Microblog WCN follows disassortative nature of the complex networks. The observation from Figure 3.7.1 and Figure 3.7.2 shows the degree correlation coefficient for different Tweet Corpus size. It shows the disassortative nature of the Microblog WCN for both nearest neighbour and all pair neighbour network.

With increase in size of the Tweet Corpus, the number of words is increased. However, the active vocabulary of users is tentatively similar. Thus, different trends and events can be discussed using similar words which may or may not connect two type of information. For instance, there is death during two accidents and among these, one of them is during *'plane crash'* and the other one is in *'car accident'*. Both the events are described using common words like *accident, died, death* etc. This information is provided by Tweet Corpus and can show significant relations among keywords. These relations give the information about one or more events.

### 3.4.5. Spectral Analysis

The spectral analysis for Microblog WCN is obtained by using the adjacency matrix of a network. Eigenvalue is obtained from the Microblog WCN for different sizes of Tweet Corpus as observed from Table 3.6. As per the observation from this Table, the number of eigenvalues ($N_\lambda$) increases with increase in the size of the network. Thus, this information is found to be useful for identifying the centrality of the network as observed from PageRank of World Wide Web (WWW) (Mukherjee *et al.,* 2009). The values of first and second eigenvalues have significant difference. The proportion of difference between subsequent eigenvalues is reduced as size of the network gets increased. Spectrum is the set of all the eigenvalues of an adjacency matrix which is evolved from Microblog WCN.

**Table 3.6:** The eigenvalues of Microblog WCN for different corpus size

| *Dataset* | $\lambda_1$ | $\lambda_2$ | $\lambda_3$ | $\lambda_4$ | $\lambda_N$ | $N_\lambda$ |
|---|---|---|---|---|---|---|
| FSD: W/1000 | 4.169 | 4.012 | 3.667 | 3.530 | -4.134 | 513 |
| FSD: W/100 | 14.510 | 11.478 | 10.833 | 9.466 | -11.434 | 3472 |
| FSD: W/20 | 51.974 | 29.209 | 25.222 | 23.561 | -28.914 | 11056 |
| FSD: W/10 | 98.194 | 51.745 | 45.032 | 40.539 | -48.809 | 18035 |

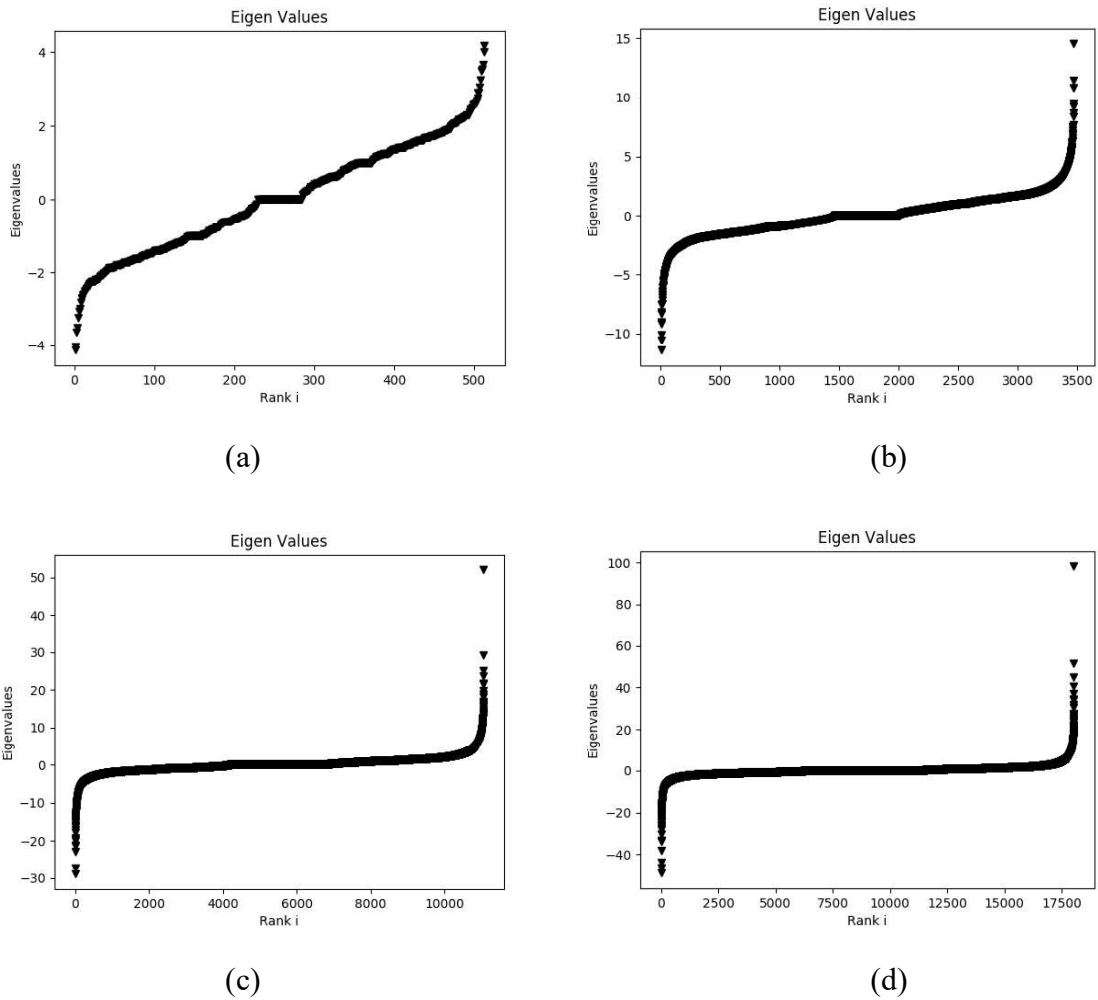

(a)          (b)

(c)          (d)

**Figure 3.8**: Eigen values of Tweet Corpus of size W/1000, W/100, W/20, and W/10, respectively for nearest neighbour Microblog WCN

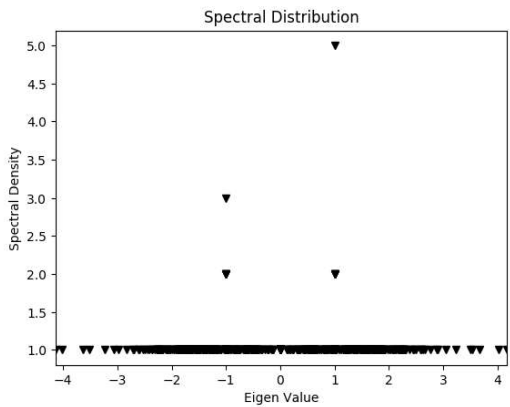
3.9 (a)

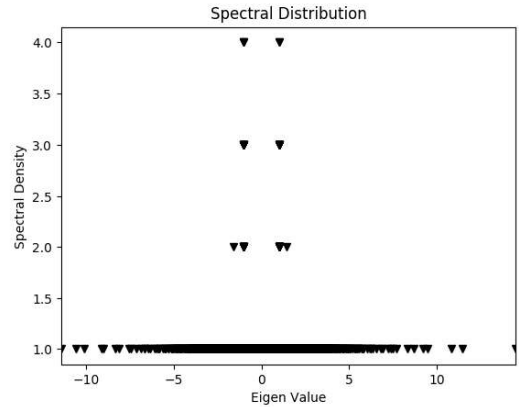
3.9 (b)

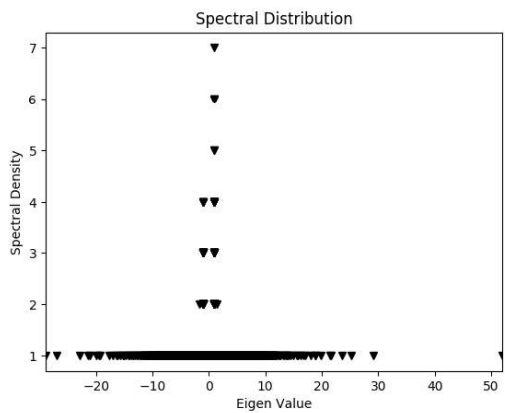
3.9 (c)

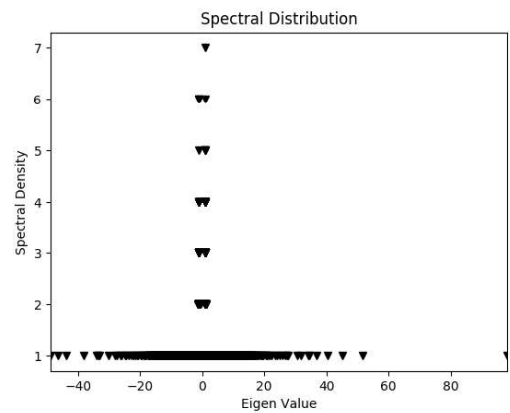
3.9 (d)

**Figure 3.9:** Spectral distribution of Tweet Corpus for nearest neighbour Microblog WCN: (a) W/1000, (b) W/100, (c) W/20, and (d) W/10

The spectral distribution of the network is shown in Figure 3.8(a), Figure 3.8(b), Figure 3.8(c), and Figure 3.8(d) for Tweet Corpus size W/1000, W/100, W/20 and W/10, respectively. The spectral density is spread near +1 and -1 according to the behaviour of the Microblog WCN. For every eigenvalue +1 or -1, the triangular patterns are observed for spectral density. Large Tweet Corpus size is computationally expensive. The area of spectral analysis of Microblog WCN is left open as new research directions in the area of NLP.

## 3.5. Concluding Remarks

In this Chapter different WCN properties are studied and the summarization of the structural analysis for Microblog WCN is described in Table 3.7. It is observed that the Microblog WCN is scale-free for degree distribution, strength distribution, and edge weight distribution. The network follows small-world property, and is disassortative. The network is partially hierarchically organized for all pair neighbour edging network and CC is scalable over increasing size of Tweet Corpus. The spectral density and eigenvalues shows the importance of graph.

**Table 3.7:** Summarization of network science properties for structural analysis of Microblog WCN

| S.N. | Key parameter | Results | Inferences |
|---|---|---|---|
| 1 | Scale-free property | $2 < \gamma < 3\ for\ k > 10^2$ | The value of $\gamma$ shows that the network is scale free for degree distribution, strength distribution and edge-weight distribution. |
| 2 | Small world feature | $\big(L\ (max)\big) \approx \big(L_r\ (max)\big)$ and $(C \gg C_r)$ | The Microblog WCN follows small-world property. |
| 3 | Hierarchical distribution | Constant magnitude for $CC(k)$ with respect to degree $k$<br><br>$CC$ remains constant with increase in number of nodes | The nearest-neighbour edging network shows anomalous behaviour and is thus, hierarchical organization not obvious.<br><br>The all-pair neighbour edging network shows partial hierarchical organization. |
| 4 | Assortativity | $\mu > 1\ for\ all\ WCN$<br><br>$\tau < 0\ for\ all\ WCN$ | The network is disassortative. |
| 5 | Spectral distribution | $\lambda_1 \gg \lambda_2\ for\ all\ WCN$<br><br>The spectral density shows triangular shape of the distribution of eigenvalues | The value of $\lambda$ shows importance of the network.<br><br>The value of highest eigenvalue is very high from second highest eigenvalue. |

There are significant patterns in Microblog WCN and thus, different structural and modular developments can be performed over the real-time Microblog WCN. The Microblog WCN is used for identifying keyphrases using the disassortative nature of the network as explained in Chapter 5. The proposed keyphrase extraction technique is used to extract different events and sub-events by finding summaries from the discrete set of streaming data as discussed in Chapter 7.

# Chapter 4.

# Comprehensive Study of Keyphrase Extraction Techniques

The words which emphasize the topic of information which is shared in Tweet are known as Keywords. Keywords can be used in combination of two or more words and are usually used as Hashtags. The Hashtag segmentation is another interesting area of research for identifying the topic of discussion. The topic describes the context in terms of set of keywords is called key-phrase as described in Chapter 2. In this research work, a novel technique is proposed for event detection from social media data using keyphrase extraction. The comprehensive study of keyphrase extraction metrics is carried out by using traditional, graphical keyphrase extraction technique over Microblog WCN.

## 4.1. Comprehensive study of keyphrase extraction techniques for Microblog WCN

There are many network science metrics which can be used for keyword extraction from WCN. In this study, the existing graph-based keyword extraction techniques are implemented over Microblog WCN for identifying the measures which are suitable for keyphrase extraction from ill-formed data.

### 4.1.1. Evolution of Keyphrase Extraction Metrics

The network science metrics, graphical and other statistical measures which are used by academic researchers for keyphrase extraction from textual documents are defined in this Section.

**A. TF-IDF**

Term Frequency – Inverse Document Frequency (TF-IDF) is introduced (Salton *et al.,* 1978) which gives the importance of the word in a document. However, it is observed that term frequency is less significant for textual documents and even less important words can occur frequently, for instance, *'hello', 'change', 'happy'* are neither stopwords nor significant with respect to topic/ trend or event detection. The document contains repetitive representation of the same word in different forms due to user-generated short text (Tweets). If a Tweet is considered as one document, then the TF-IDF can be computed as shown in Equation 4.1.

$$\text{TF} - \text{IDF}_i = \frac{\text{Number of occurrences of word } 'i'}{\text{Total number of words}} \log\left(\frac{U}{U_i}\right) \qquad 4.1$$

where $U_i$ is the total number of Tweets which contains word i, U is the total number of Tweets in the dataset. The TF-IDF measure gives information about the number of times the word occurs in Tweet Corpus, uniquely. The relation of frequently occurring terms with each other may give useful insights about the pattern of words in Microblog WCN. Some words may have high TF and low IDF which may or may not contribute towards keywords. For instance 'Happy Birthday' or 'Today I am happy' are personal wishes or greeting, respectively. But 'Happy to launch new iphone X' is something about an event "Launch of iphone X'. This word 'happy' may also relate to sentiment analysis which is out of scope of this research work. Thus, usually named entities and activity verbs are considered for analysis of WCN.

**B. KeyWorld**

The text is pre-processed and n-gram model was applied on document to count phrase frequency and list of sentences were obtained by connecting corresponding phrases. In this research work, the documents are Tweets from the Tweet Corpus and the phrase is a word. The node specific calculations are proposed as *extended path length* $L'_{G_v}$ and *extended characteristic path length* $L'_{G_v}$ which are used to calculate *contribution* $CB_v$ as shown in Equation 4.2

$$CB_v = L'_{G_v} - L'_v \qquad 4.2$$

where $L'_v$, *characteristic path length* is the length which is averaged over all pairs of nodes except node v and $L'_{G_v}$ is the *extended characteristic path length* is the length is averaged over all pairs of nodes assuming the corridor v doesn't exists (Matsuo *et al.,* 2001). However, this method is computationally expensive and difficult to apply on real time applications.

### C. LexRank

The importance of a word is calculated by using the concept of *eigenvector centrality*. The eigenvector centrality of a node A is defined by measuring the importance of the neighbouring nodes. Initially, the importance of every node is calculated using degree centrality. The redistribution of centrality to its neighbours takes place as shown in Equation 4.3.

$$p(u) = \sum_{v \in adj(u)} \frac{p(v)}{deg(v)} \qquad 4.3$$

where $adj(u)$ are the nodes adjoining node u, $deg(v)$ is the degree of adjoining node v, $p(v)$ is the degree centrality of adjoining node v. This creates matrix which satisfies the *property of stochastic matrix*.

### D. Eccentricity based keyphrase extraction

In the WCN, the eccentricity for vertex $v$ is the shortest distance or shortest path length in which vertex $v$ can reach to any other node in the network of words. *Eccentricity* defines the ease to reach node $v$ from any other node in the network. If the eccentricity of any node $v$ is low, then this means that the node $v$ is easily reachable from all other nodes and the node is marked as *keyword* in the network and if eccentricity of any node $v$ is high, it means the node is not easily reachable by other nodes. In Microblog WCN, the words may or may not get connected to many nodes as it is path based network which is evolved out of nearest-neighbour edging. Thus, the eccentricity is not significant measure to define the importance of a node.

### E. HITS: Hypertext Induced Topic Search

In HITS algorithm, the directed graph is generated from successively co-occurring words in short text documents. Each node has incoming links and outgoing links. The incoming links are called *authorities* and outgoing links are called *hubs*. The *hub score* and the *authority scores* are then calculated recursively in multiple iterations. The two types of combination of hub-authority scores are calculated, namely, the *maximum of hub and authority* as shown in Equation 4.5, and the *average of hub and authority* as shown in Equation 4.6

$$\text{HITS}(\text{Max}(H, A)) = \text{Max}\big((\text{Hub\_Score}(n)), (\text{Authority\_Score}(n))\big) \qquad 4.5$$

$$\text{HITS}(\text{Avg}(H, A)) = \frac{(\text{Hub\_Score}(n)) + (\text{Authority\_Score}(n))}{2} \qquad 4.6$$

where in-degree(n) is the *hub score* of the node $n$ and out-degree(n) is the *authority score* of the node $n$. The higher hub score than authority score shows that the position of the word is starting and higher authority score than hub score shows that the word is the ending word in most of the documents. The position of the word is significant feature while extracting keywords from Tweet Corpus (Biswas *et al.,* 2018). Thus, Hub-Authority scores are calculated for Microblog WCN evolved from Tweet Corpus.

**F. DegExt: Degree-based Extractor**

In Degree-based Extractor (DegExt), the graph is represented by assigning words of the short-text document in Tweet Corpus as vertex of WCN. Keywords are ranked on the basis of occurrence of a word in number of different Tweets. More the word occurs in number of Tweets, more salient the word is.

*Degree centrality* is calculated for each node in Microblog WCN. The top N number of vertices are extracted as keywords using *degree centrality*. In WCN, words are linked to other words. Degree centrality of the word is insignificant for identifying keywords. This is because a word may or may not get linked to many other words as it may occur as single word in a Tweet, or it may link to small number of distinct words with more frequency, or it may link to large number of words with low frequency. The word which is linked to smaller number of nodes with high edge-weight could be more important due to its occurrence in Tweet with domain-specific words and thus, this concept is used to define selectivity for SBKE.

**G. TKG: Twitter Keyword Graph**

The Twitter Keyword Graph (TKG) is the graph which is generated by using tokens (words) from Tweets as vertices and edge is created between two words if they are co-occurring within single Tweet. In this research work, the author considered three types of centrality measures namely degree centrality, closeness centrality, and eccentricity centrality. The Microblog WCN of networks is classified on the basis of type of edging, edge-weight, and edge-count. The results are thus evaluated with different combinations. The TKG is generated using *all-pair neighbouring edging* scheme with edge weighing using *inverse co-* *occurrence frequency* and *closeness centrality* outperforms in two-third of cases for identifying keywords.

**H. SBKE: Selectivity-based Keyword Extraction**

The *Selectivity – Based Keyword Extraction (SBKE)* is a graph based keyword extraction technique which is based on vertex *selectivity measure* for each vertex representing unique word. Node strength is defined as sum of weights of all links connected to node $n$. For node $n$, selectivity measure is represented as shown in Equation 4.7.

$$Selectivity(n) = \frac{Strength(n)}{Degree(n)} \qquad 4.7$$

There are many different keyword extraction techniques which are implemented on WCN evolved from well-formed documents in existing literature. In this research work, keyword extraction and keyphrase extraction techniques from existing literature are implemented for WCN evolved from Tweet Corpus.

**Table 4.1:** Keywords Obtained for different keyword extraction metrics

| SN | Topic | Top words | Bottom words |
|---|---|---|---|
| 1 | Degree | amy, winehouse, rip, amywinehouse, died, dead, sad, found, waste, news | itscarolk, karyz, nasta, adolescentquote, hell, amyamywinehouse, fucking, itamywinehouse, word, coming |
| 2 | Strength | amy, winehouse, died, rip, amywinehouse, dead, sad, found, death, cause | itscarolk, karyz, nasta, adolescentquote, hell, amyamywinehouse, fucking, itamywinehouse, word, coming |
| 3 | Betweenness Centrality | amy, winehouse, rip, amywinehouse, died, sad, dead, drugs, waste, talented | all, tweeted, dreadful, apartment, mistook, send, flatthoughts, up, cpaulhitsheet, confirmation |
| 4 | Closeness Centrality | amy, winehouse, rip, amywinehouse, died, dead, sad, found, waste, talent | sun, cher, gordon, strong, demons, smart, itamywinehouse, told, reason |
| 5 | Eigenvector Centrality | amy, winehouse, rip, died, dead, sad, amywinehouse, found, waste, talent | sun, cher, gordon, strong, demons, smart, itamywinehouse, told, reason |
| 6 | Clustering Coefficient | all, tweeted, dreadful, apartment, mistook, send, flatthoughts, up, cpaulhitsheet, confirmation | itscarolk, karyz, nasta, adolescentquote, amy, winehouse, rip, amywinehouse, died, dead |
| 8 | PageRank | amy, rip, sad, kinda, | cpaulhitsheet, xoxochristy, |
| | | amywinehouse, shocked, died, talent, ripamywinehouse, die | schofe, niallofficial, ooohhhh, marcmalkin, thisismaxonline, cticklesmcqueen, itscarolk, holymolynews |
| 9 | TF-IDF | what, wut, itscarolk, nasta, adolescentquote, winehousee, surprising, died, ripamywinehouss, seenthatcumin | house, winehouse, die, amy, died, has, amywinehouse, all, her, too |
| 10 | KeyWorld | amy, amywinehouse, unknown, appears, deadthey, ripamywinehouse, sun, heard, breaking, talent | what, bellatrix, toosoon, vegaspaul, fuck, birthday, jaidenofficial, marcmalkin, skynewsbreak, good |
| 12 | TextRank | amy, rip, sad, talent, died, amywinehouse, waste, peace, amazing, another | up, cpaulhitsheet, xoxochristy, fav, schofe, niallofficial, skynewsbreak, told, reported, ooohhhh |
| 13 | Eccentricity | ripamywinehouse, tweeted, apartment, flatthoughts, up, sorry, cpaulhitsheet, confirmation, fav, fan | amywinehouse, died, all, jimi, overdose, dreadful, mistook, young, send, finally |
| 14 | HITS (Avg (H,A)) | rip, amy, winehouse, sad, amywinehouse, died, dead, talent, drugs, waste | ripamywinehouss, cpaulhitsheet, xoxochristy, continues, hahahahahahaha, schofe, wut, bellatrix, drinks, blue |
| 15 | HITS (Max(H,A)) | amy, winehouse, rip, dead, sad, died, amywinehouse, talent, drugs, waste | ripamywinehouss, all, tweeted, dreadful, mistook, send, flatthoughts, up, cpaulhitsheet, confirmation |
| 16 | DegExt | amy, winehouse, rip, amywinehouse, died, dead, sad, found, waste, news | itscarolk, karyz, nasta, adolescentquote, hell, amyamywinehouse, fucking, itamywinehouse, word, coming |
| 17 | TKG | amy, winehouse, rip, amywinehouse, died, dead, sad, found, waste, talent | sun, gordon, strong, demons, smart, itamywinehouse, told, reason, xoxochristy |
| 18 | SBKE | mistook, unknown, mrs, lordvoldemort, weasley', rumors, sadly, holymoly | all, tweeted, dreadful, send, flatthoughts, up, cpaulhitsheet, confirmation, schofe, surprise |

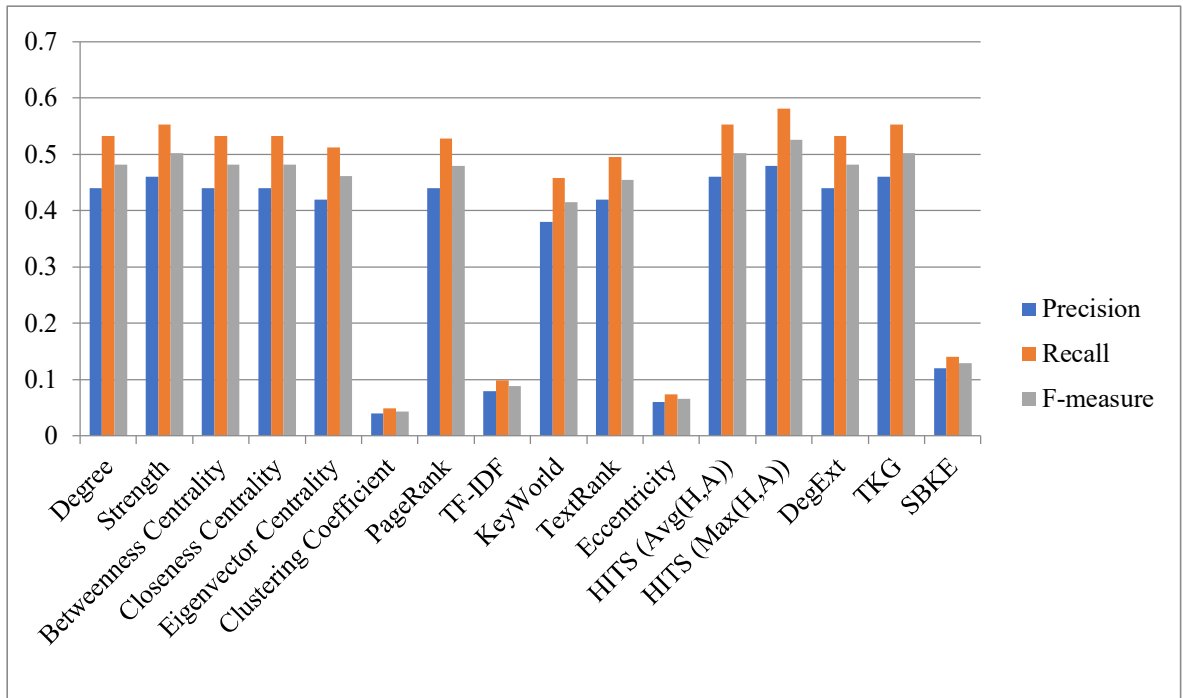

**Figure 4.1:** Graphical representation of performance evaluation of keywords extracted from domain specific Microblog WCN for N high valued keywords where N=10

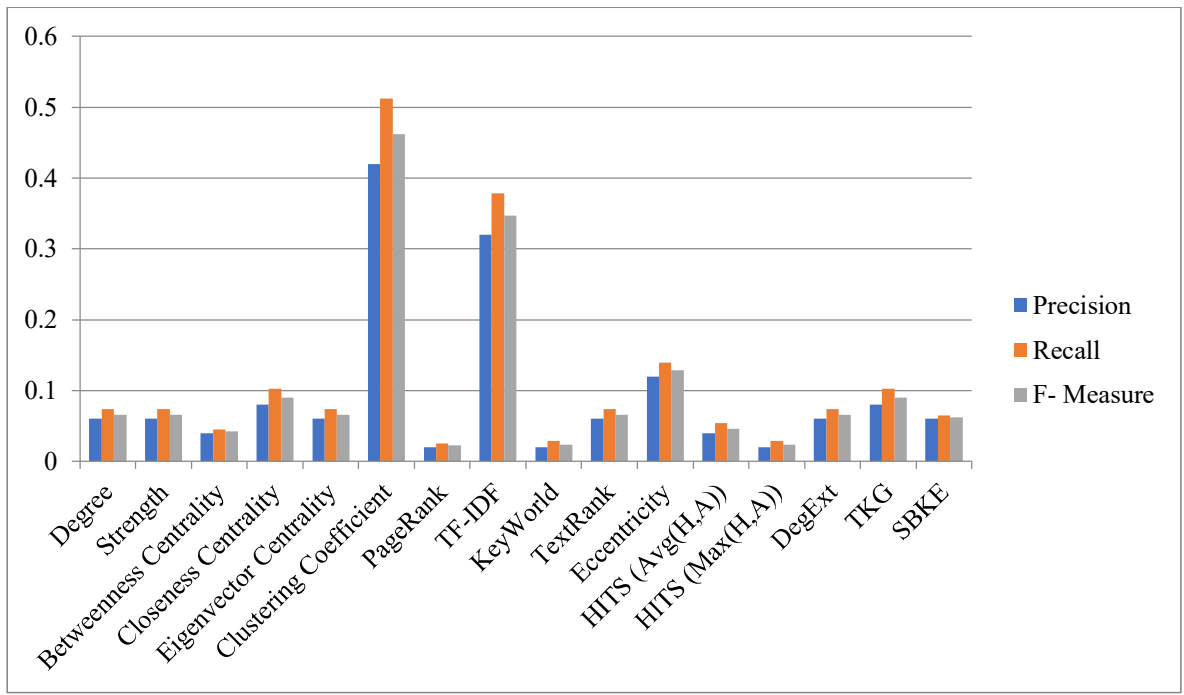

**Figure 4.2**: Graphical representation of Performance Evaluation keywords extracted from domain specific Microblog WCN for N low valued keywords where N=10

The experiments are performed to implement the existing keyphrase extraction metrics over Microblog WCN. The FSD dataset is used for identifying keywords for 27 different topic specific sets of Tweet Corpus. The results for different network metrics scores and existing keyword extraction measures, top *N* keywords (words with highest value of corresponding measure) and bottom *N* keywords (words with lowest values of corresponding measures) for $N = 10$ are collected as shown in Table 4.1.

It is observed that most of the network metrics gives important information about words in a Tweet corpus. In Topic 1, the information is related to *"Death of Amy Winehouse and rumours of Mrs. Weasley mistook her for Bellatrix"*. In this research work, the experiments are performed to solve the problem domain of keyphrase extraction from set of Tweets. The global network measures like degree, strength, *HITS*, centralities, and random-walk based measures give *keywords* for social media data as shown in Figure 4.1. It is observed that *TKG* gives better results than that of SBKE for identifying for Tweet Corpus in this research work. The results given in this Figure are evaluated automatically using recall measures. However, *SBKE* was proposed for Croatian News. This shows that the semantics of WCN for every language is different.

As observed from Figure 4.2, the network metrics like those of clustering coefficient, TF-IDF, and eccentricity gives less or non-significant set of keywords for Tweet Corpus. The CC is evaluated using the measure of the extent of connectivity of neighbouring words of word *w* with each other. Consider a set of Tweets as *A* and another set of Tweets as *B*. Both set A and set B are used to generate a Microblog WCN. Both set A and setB are associated with different topics. One of the Tweet from Set A is *"Flight crashed in Malaysia"* and another Tweet from Set B is *"Stock market economy crashed"*. Now if Microblog WCN is evolved out of these two Tweets, after pre-processing, the Microblog WCN is generated as shown in Figure 4.3, the words *'Flight'* and *'Malaysia'* have tendency to connect with each other in corresponding set of Tweets.

Similarly, the words *'Stock', 'Market', and 'Economy'* have more tendency to connect with each other. Likewise, if a word like *'Crashed'* gets connected to multiple topics, then the value of Clustering Coefficient will be low. Also, if any unimportant word like *'Happy'* gets connected to many Tweets which may or may not get connected to each other, then the degree is high but CC is low. Thus, CC does not play any significant role in identifying keywords in Microblog WCN. But important node may get connected to multiple important nodes which may give the community of keywords.

Another reason could be that the words can be inter-related using general information, for instance, *'good morning', 'happy to share', 'happy in morning', 'sharing is good'* are all inter-related but it does not give any topic specific meaningful information for the set of words extracted using high CC metrics. Similarly, the most frequently occurred words may or may not be useful, for instance, *'good morning'*, *'happy to share'*. This gives high term frequency but IDF is calculated by identifying unique words in the corpus. The unique words in the corpus may or may not be identified as users use different slangs, abbreviations, and jargons in their Tweets.

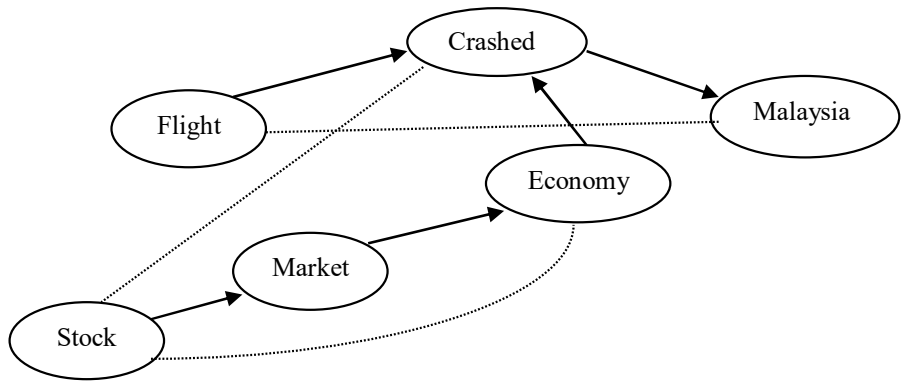

**Figure 4.3:** Microblog WCN from two Tweets of different topics

### 4.1.2. Observations and Discussions

As observed from Figure 4.2, based on the bottom N values of keyword extraction techniques, for $N = 10$, the CC gives good results. This could be due to the fact that Microblog WCN was generated from topic specific set of Tweets only. Thus, the words are related to one topic and hence, are connected to each other.

The keywords are connected to keywords and this forms the chain of words. This chain of words is connected through high weighted edges. This shows that edge weights are important than node degree.

Another important keyword extraction metrics which is observed as significant metric is HITS (Max(H,A)). This is based on *HITS algorithms* where maximum value is chosen out of hub and authority scores. The hub score denotes the degree of last position of the word and authority score defined first position of the word. Higher the hub score more is the probability of a word to be the last word. Higher the authority score more is the probability of a word to be the first word in Tweets. The position of the word is very important and researchers have considered this measure for PositionRank, a keyphrase extraction approach (Florescu *et al.*, 2017).

Other important measures are HITS (Avg (H,A)), Strength, and TKG. The HITS(Avg(H,A)) denotes implementation of HITS algorithm to find average of hub score and authority score of a word in Microblog WCN. This is high for important words due to the fact that more important words are well connected in the network. The HITS(Avg(H,A)) and Strength were earlier used for well-formed data and complex networks.

It is observed that other than TKG, HITS, CC, and Strength, two other important measures are Betweenness Centrality and PageRank. The betweenness centrality measure is significant in terms of importance of a node or word which links other words. When Tweet is pre-processed and tokenized into a set of words in a sequence, it is termed as a *path*. Every word is marked as a node and the connection between adjacent words as an edge. High betweenness centrality shows that the word is more important and that it comes in between the path of most of the connections in the network. Although, this measure is suitable as per the semantics of a network, it is observed that words at the first and the last position are important and thus, should be given high score (Biswas *et al.,* 2018) which is usually not satisfied by betweenness centrality.

The PageRank algorithm is used to define the importance of word by voting system. Initially, all words are allocated some score randomly. One agent starts from one node and moves within the network randomly. It undergoes thousands of iteration before it converges. The damping factor is used to induce the importance of the vote of a node who votes for other nodes. Usually, damping factor is considered as 0.85. The PageRank algorithm is used for complex networks using *nodes-scores* and *edge-score* for different ranking measures like LexRank, TextRank, TopicRank, and PositionRank. These ranking measures are also used to collect the top ranked words and identify the keyphrase.

The summarized inferences for different keyphrase extraction metrics are mentioned in Table 4.2. It is observed that centrality gives the importance of word in terms of its connectivity among other words in the network. Selectivity measure is used in SBKE and shows that the importance of node is measured using its frequency of occurrence with words. Clustering coefficient defines the connectivity of words among each other which is not good measure for Microblog WCN as the network is evolved using chain of words occurring as a path of nearest-neighbour edging network. PageRank, and TextRank are the ranking measures which identify the keywords and rank them using different measures. These techniques are based on random-walk and voting of neighbouring nodes. Eccentricity measures the ease of approaching a word by all other words in the network. However, eccentricity is logically not significant as the Microblog WCN is based on the path network.

**Table 4.2:** Inference for different keyword extraction measures for Microblog WCN

| Keyword Extraction Algorithm | Inference |
|---|---|
| Degree | Gives the number of words with which word w occurs. Measure of degree centrality for each word. |
| DegExt | |
| SBKE | Measures the repeated occurrence of word *w* with its neighbouring words with respect to number of word it is co-occurring with. |
| Strength | Calculated word w frequency. |
| Betweenness Centrality | Centrality calculates the measure of number of paths a word w have for word to word connectivity. |
| Closeness Centrality | |
| Eigenvector Centrality | |
| TKG | Calculates the significance of word *w* in Tweets when WCN is generated. Performs best by using all pair neighbour edging network, weighted edges using inverse co-occurrence frequency and closeness centrality. |
| Clustering Coefficient | Measure the extent of similarity among neighbours of word w. |
| HITS (Max(H,A)) | |
| HITS (Avg (H,A)) | |
| TF-IDF | Statistical measure of importance of word *w*. |
| KeyWorld | Based on the average of shortest path measure between two words. Used small world phenomenon. |
| PageRank | Measures the impact of neighbouring words co-occurring with words on word *w* using their votes. It is based on random walk based measure in Microblog WCN. |
| TextRank | |
| Eccentricity | Measures reciprocal of number of co-occurrences of pair of words in given feeds. Zero for isolated node. |

Thus, the keyphrase extraction metrics which can be used for further studies of Microblog WCN are strength, betweeenness centrality, clustering coefficient, PageRank, HITS, and TKG. The semantics of the Microblog WCN for existing keyword extraction algorithms is shown in Table 4.3.

**Table 4.3:** Nature of the Microblog WCN for different keyword extraction techniques

| Keyword Extraction Algorithm | F-measure | | Weighted/ Non-weighted metrics | All pair Neighbour/ Nearest neighbour | Directed/ Undirected |
|---|---|---|---|---|---|
| | Top | Bottom | | | |
| Strength | 0.502074 | 0.066096 | Non-weighted | All pair Neighbour | Un-Directed |
| Betweenness centrality | 0.481900 | 0.042353 | Weighted | All pair Neighbour | Un-Directed |
| Clustering coefficient | 0.043871 | 0.461710 | Non-weighted | All pair Neighbour | Un-Directed |
| PageRank | 0.479843 | 0.022222 | Non-weighted | Nearest neighbour | Directed |
| HITS (Avg (H,A)) | 0.502074 | 0.045802 | Non-weighted | Nearest neighbour | Directed |
| HITS (Max(H,A)) | 0.525771 | 0.023529 | Non-weighted | Nearest neighbour | Directed |
| TKG | 0.502074 | 0.089725 | Non-weighted | All pair Neighbour | Un-Directed |

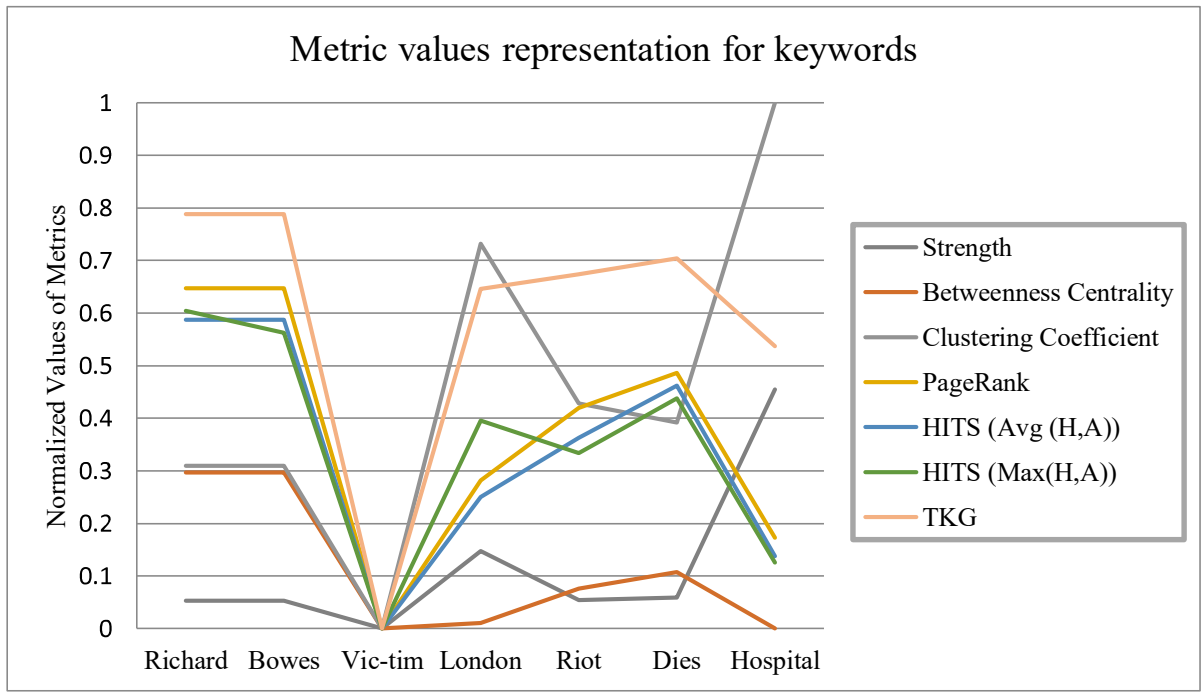

**Figure 4.4:** Values for selected keyword extraction techniques for Microblog WCN evolved from *topic "Richard Bowes, Victim of London Riot, Dies in Hospital" of FSD* dataset

The normalized values of selected metrics are obtained for the topic *"Richard Bowes, Victim of London Riot, Dies in Hospital"* of FSD dataset. The values for keyword and keyphrase extraction metrics are obtained as shown in Figure 4.4. The words on x-axis are obtained after pre-processing the text of the topic description and by removing stop-words.

It is observed that the discussion is about *Richard Bowes* and thus, the two words *'Richard' and 'Bowes'* appears side by side. This appearance is validated by similar values of most of the keyword extraction metrics as shown in Figure 4.4. It appears as if the *'victim'* word is not mentioned by any of the user as it is discussed in traditional news that the person has suffered some mishap. Thus, in active vocabulary of users, the words like *'fire', 'mishap', and 'accident'* are more common than that of *'victim'*. The keywords in ground truth topic may or may not exist in active vocabulary of users. The word *'Hospital'* marks the end of the post and thus, the *HITS (Max(H,A))* gives high value like that of *CC*. The *CC* for *'London'* and *'Hospital'* are high which ensures that the words which are used by users in Tweet are more connected to these two words. It can be concluded that the many network metrics shows useful patterns in Microblog WCN. To examine the effectiveness of random-walk based keyphrase extraction measures over Microblog WCN, the performance is evaluated using ROUGE scores.

## 4.2. Performance evaluation of existing keyphrase extraction techniques

The diversity of the topics and words which are used by users to post Tweet is much more on the internet than that of offline social community because the scope of discussion about wide range of topics increases over internet. The semantics of Microblog WCN differs from those of well-formed WCN. In this research work, existing keyphrase extraction techniques are implemented on Twitter FSD dataset and performance is evaluated.

The keywords from Tweet Corpus are ranked and top $n$ keywords are obtained from Microblog WCN. These top keywords may or may not connect with each other using edges in the network. The community of connected keywords create sub-graphs. Each sub-graph gives set of keywords which are known as *keyphrases*. The random walk based PageRank algorithms used *node-score* and *edge-score* for different keyphrase extraction techniques which were evolved in literature.

The resulting keyphrases for a Topic "Richard Bowes, victim of London riots, dies in hospital" of FSD dataset is shown in Figure 4.4. This topic is not specified by top $N$ keyphrases having $N = 1$ or even for $N = 5$ by existing techniques. The results are obtained for *TextRank, TopicRank, NErank, and PositionRank* which gives keywords and keyphrases in Microblog WCN. The TextRank gives information about those words which are related to *'injury' and 'hospital'*. Thus, important contextual information is obtained using *TextRank*. In *TopicRank*, the information is given about the person, place and activity. This gives more specific information and closer snippets to that of ground truth of a topic which is mentioned in the dataset. The NErank was proposed for Twitter feeds and gives information about something happened but less information about the actual verbs. Similarly, the position of a word can describe keywords and relevant information but no information is obtained with chain of keywords in precise lexical sequence. Although, TopicRank outperforms all the existing techniques by providing many important keyphrases but the information is not structured.

Table 4.4: Keyphrases extracted using different techniques for topic of FSD dataset

| Techniques | Keyphrase@1 | Keyphrase@5 |
| --- | --- | --- |
| TextRank | Injured | injured, injured assaulted, needless, hospital, midnight |
| TopicRank | Riot | riot, rip richard mannington, monda man, stamp fire, west london die |
| NErank | Bowes Man | bowes man, tried stamp, richard bowes, london, Cameron |
| PositionRank | Richard Mannington Bowes | Richard Mannington bowes, old Richard Mannington, depressing Richard Mannington, Richard Mannington, death Richard bowes |

The dataset used for experimental results and evaluation is FSD dataset. The performance evaluation metrics used for experiments is ROUGE score (Recall-Oriented Understudy for Gisting Evaluation) as shown in Table 4.5. The ROUGE-N score gives the common n-grams over total number of n-grams as obtained from the lexical sequence of a topic. This metrics is used to measure recall for keyphrase extraction and text summarization (Erkan *et al.,* 2004) problem. The duplicates obtained during evaluation are removed before intersection of extracted keyphrase and reference text. The ROUGE-N score is defined as shown in Equation 4.8.

$$ROUGE-N = \frac{(n-grams\ in\ extracted\ keyphrase) \cap (n-grams\ in\ reference\ text)}{Total\ number\ of\ n-grams\ in\ reference\ text} \quad 4.8$$

Thus, ROUGE-1 and ROUGE-2 are used to calculate the performance for unigram and bi-gram of keyphrases, respectively. The ROUGE-L is the measure of Longest Common Subsequence (LCS) which measures the LCS between the automatically extracted keyphrases and the reference text. The highest value of common sub-sequence is returned by ROUGE-L of TopicRank.

**Table 4.5:** Performance evaluation of keyphrases as obtained using different keyphrase extraction metrics for FSD dataset

| Algorithm | ROUGE – 1 | | | ROUGE – L | | | ROUGE – 2 | | |
|---|---|---|---|---|---|---|---|---|---|
| | N=1 | N=3 | N=5 | N=1 | N=3 | N=5 | N=1 | N=3 | N=5 |
| TextRank | 19.13 | 24.07 | 29.16 | 0.67 | 0.77 | 1.00 | 5.55 | 5.55 | 5.55 |
| TopicRank | 26.54 | **54.93** | **56.32** | 0.88 | **2.30** | **2.56** | 11.11 | 27.82 | 27.82 |
| NErank | 13.73 | 53.21 | 54.16 | 0.77 | 1.67 | 2.11 | 1.38 | 11.88 | 11.88 |
| PositionRank | **38.11** | 48.19 | 49.22 | **1.44** | 2.11 | 2.33 | **23.54** | **31.56** | **31.56** |

The PositionRank gives the best performance in terms of ROUGE-2 score. In this Table, N describes number of top keyphrases. Thus, for values of N as 1, 3, and 5, among all other keyphrases top 1, top 3 and top 5 keyphrases are obtained, respectively. As per analysis, it is observed that if $N = 1$, the PositionRank outperforms all the existing techniques in terms of ROUGE – 1, ROUGE – 2, and ROUGE – L score. The TopicRank gives best performance for higher values of N in ROUGE-1 and ROUGE-L score. The NErank gives good performance after TopicRank for ROUGE-1 and ROUGE-2 scores. This is because the ROUGE score for identifying keyphrases from social media data is high but it does not provide keyphrases in ordered lexical sequence of words. The ROUGE-L score for NErank is lower than both PositionRank and TopicRank.

All the existing techniques are performed for social media data. The Tweet Corpus which is used for evaluations used a single Topic in every experiment. Thus, the words are interrelated and relevant data is extracted. The keyphrases are obtained from set of Tweets which are similar to each other and thus, this process can also be known as Tweet-summarization.

## 4.3. Concluding Remarks

In this Chapter, existing keyword and keyphrase extraction techniques are studied from literature. It is observed that the network science metrics shows important insights on social media data. Similar observations are used in existing literature to identify keywords from social media data using Microblog WCN for Twitter data (Abilhoa *et al.,* 2014; Biswas *et al.,* 2018) and Croatian News (Beliga *et al.,* 2016). Also, many keyphrase extraction techniques have used WCN (Mihalcea *et al.,* 2004; Wan *et al.,* 2008; Florescu *et al.,* 2017) and Microblog WCN (Bellaachia *et al.,* 2012) to identify keyphrases from social media data. Thus, these observations have opened up new research directions for social media data analysis. In upcoming Chapters, network models and network science properties are used to identify keyphrases and thus, events.

# Chapter 5.
# Identifying Influential Segments from Social Media Data

Influential segments are the set of words which gives important information about the short-text. The important information can be a the set of words which are relevant to a topic, the headlines to define the topic, and the brief information about a topic. The network science properties which are used for keyphrase extraction from social media data are degree distribution (in-degree out-degree distribution), and strength distribution (in-strength out-strength distribution) as studied in Chapter 3. It is observed that the edge-weight plays significant role to extract important bi-grams. The words in a Tweet are tokenized and placed in Microblog WCN as nodes. The lexical sequence of a Tweet is maintained using directed WCN. Thus, the values of in-coming edge-weight and out-going edge-weight are comparable for any word in Microblog WCN. The edge weight gives importance of co-occurring words in a network. The network model is proposed using edge-weight which is used to extract keyphrases from Microblog WCN.

## 5.1. Introduction

Initially, the weighted edges of Microblog WCN are sorted in descending order. Heuristic value gives the number of high weighted edges which are kept in resulting disconnected graph to obtain sub-graphs for keyphrase extraction. Heuristic measures are used to find sub-graphs from Microblog WCN. In this research work, high weighted edges are preserved in the network and remaining edges are removed. This results into disconnected sub-graphs. These sub-graphs are used to extract meaningful keyphrases in ordered lexical sequence of words.

The heuristic measures are based on logical human interpretations which are abstractive. To remove this limitation, the low weighted edges are removed iteratively from the graph and sub-graphs so obtained. The key idea behind this approach is based on the network model k-core decomposition which is used for decomposition of complex networks.

K-core decomposition (Alvarez-Hamelin *et al.,* 2006) is the recursive process of removing all nodes with degree less than $k$ iteratively from the network. The edge-weight based decomposition of Microblog WCN is used to obtain the sub-graphs which are used for keyphrase extraction. To reduce the number of iterations, an approximate keyphrase extraction technique is proposed using different static controlling parameters for threshold based approach. The parameter dependency for termination is removed when a keyphrase extraction technique, BArank is proposed using k-bridge decomposition and assortativity property for Microblog WCN.

## 5.2. Decomposition of Microblog WCN

The complex networks are those networks which evolve with time and are dynamic in nature. The structure and dynamics of these complex networks are studied (Boccaletti *et al.,* 2006) for networks which are created using links among different user profiles. These links are either two – way communication as in Facebook using undirected edges, or one – way communication as in Twitter using directed edges. The network evolves dynamically by insertion and deletion of links based on friendship network and/ or follower-followee network. These networks follow core-periphery structure (Borgatti *et al.,* 2000).

**Definition 5.1: Core-Periphery Structure:** The core-periphery structure is the structure of complex networks in which some nodes are core nodes which are usually those nodes which have high degree, and other are periphery nodes which have low degree but are connected to core nodes.

This core-periphery structure shows that there are some profiles which are more connected in the social community and others have fewer connections. More connectivity in complex networks may define important and significant nodes for Microblog WCN. The degree – distribution and strength-distribution of nodes in Microblog WCN follows scale free property. The in-degree out-degree distribution and in-strength out strength distribution are shown in Figure 5.1 (a) and Figure 5.1(b), respectively. The idea behind decomposition of Microblog WCN is that the in-coming edges and out-going edges have equivalent values due to path based network which is created using words of Microblogs. Each node in the network can be classified into:

- **Centre Element:** The node having similar values of in-coming strength and out-going strength is called *centre element* because the word which occurs in between the Tweet will have predecessor or successor. Thus, the word will have equal in-strength and out-strength.

- **First Element:** The word from which the Tweet often starts is called *First Element*. The word is the first element if it has higher out-strength of node than that of in-strength of node. The probability of node being the first element depends upon the fraction of Tweets in which the word occurs as the first element after pre-processing the text.

- **Last Element:** The last position of the word in many Tweets of Tweet Corpus is a *Last Element* if in-strength of node is higher than that of out-strength of node.

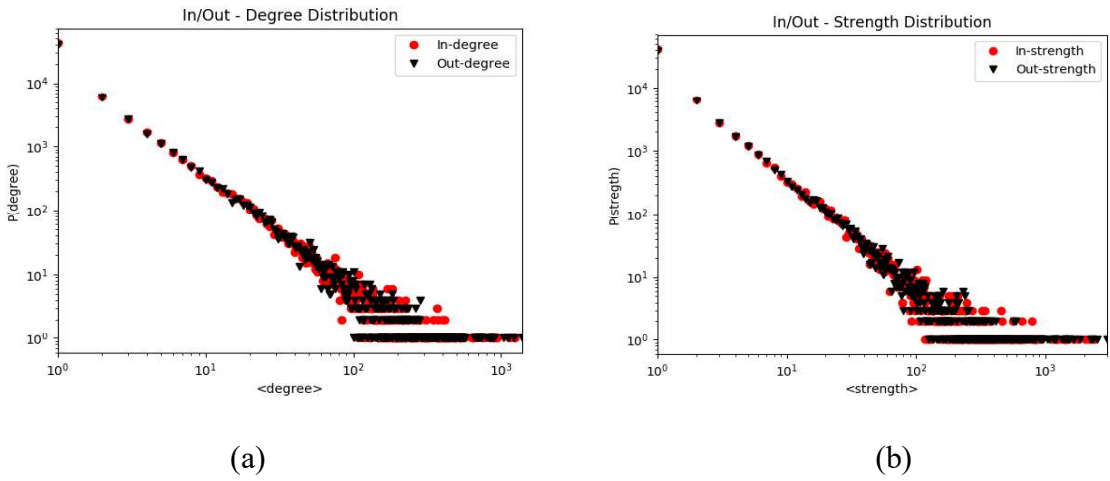

(a)            (b)

**Figure 5.1:** (a) In-degree out-degree distribution of Microblog WCN. (b) In-strength out-strength distribution of Microblog WCN

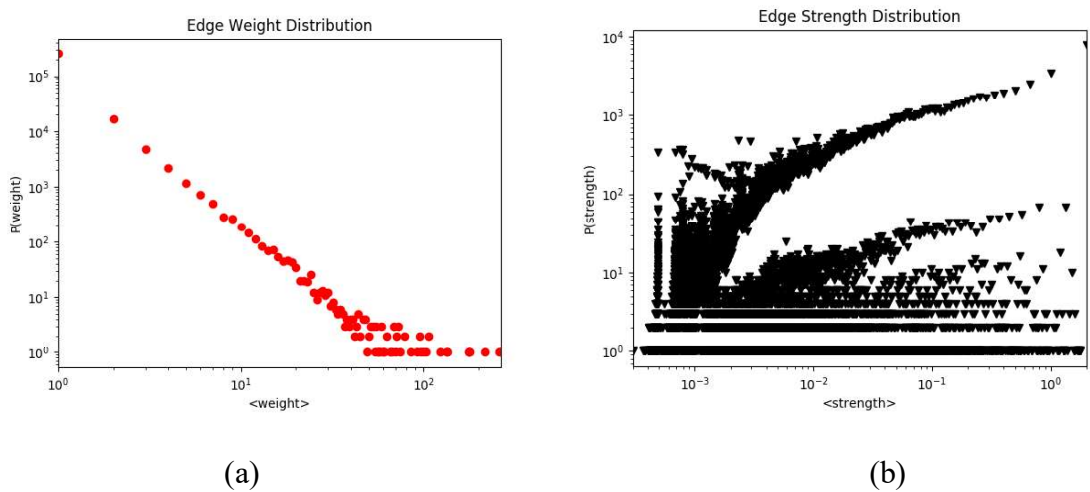

(a)            (b)

**Figure 5.2:** (a) Edge weight distribution of Microblog WCN (b) Edge strength distribution of Microblog WCN

The edge-weight distribution and the edge-strength distribution for Microblog WCN is as shown in Figure 5.2(a) and Figure 5.2(b), respectively, the edge-weight shows the scale-free property for Microblog WCN which shows that range of the weight of edges varies. The high weighted edges give important bi-grams. The connected bi-grams are the sub-graphs from which a keyphrase is obtained. Thus, high weighted edges are considered to be important bi-grams for keyphrase extraction.

### 5.2.1. Heuristic approach

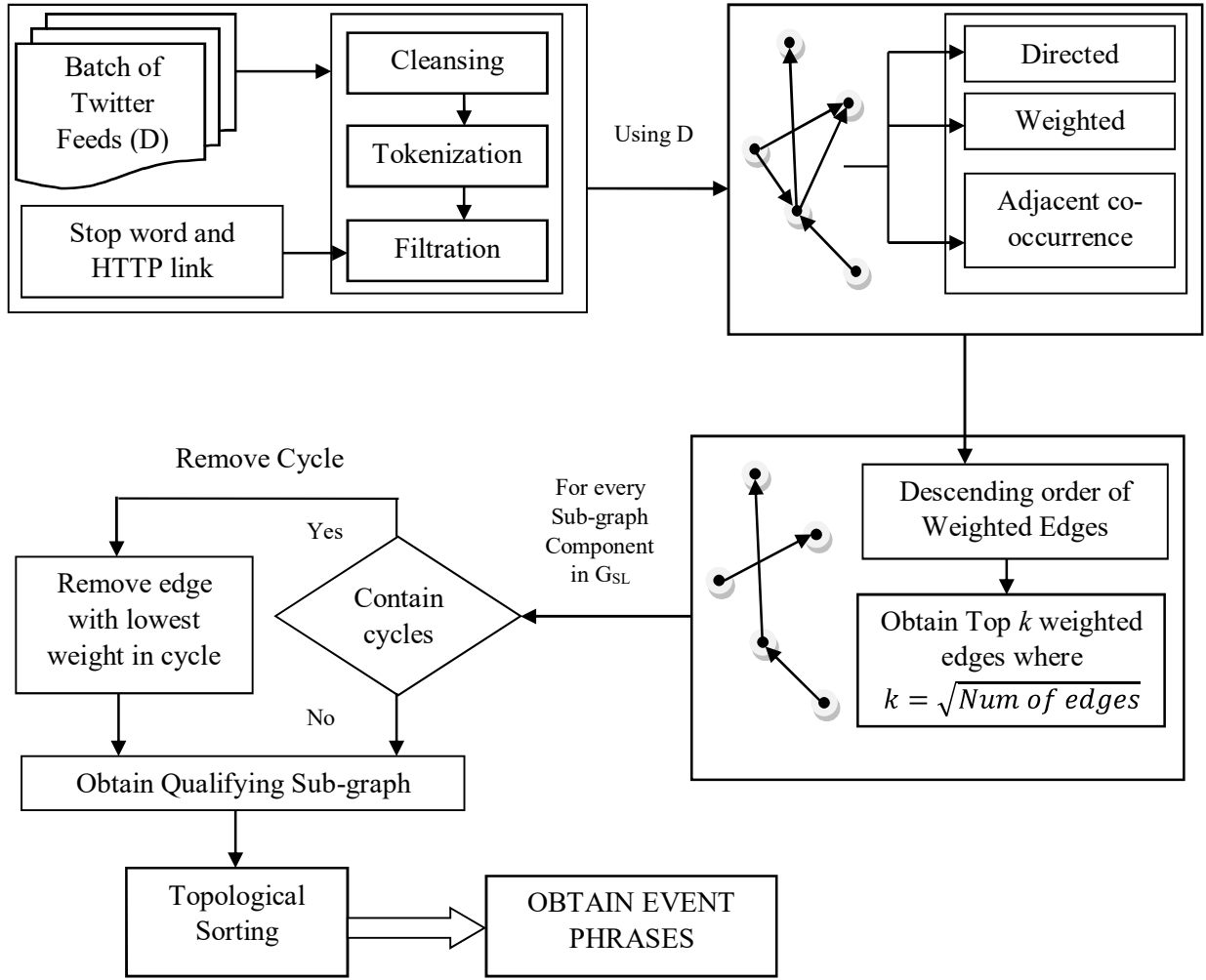

**Figure 5.3:** Heuristic based keyphrase extraction approach for Microblog WCN

The Heuristic approach is proposed for keyphrase extraction from social media data as shown in Figure 5.3. The social media data is used to create Microblog WCN ($G_D$). Consider the set of words w ϵ D, where $w = \{w_1, w_2, ... w_t\}$ is collection of words in Tweets. Here $w_i$ is each word which is considered as a node $n_i$ in graph $G_D$. Every word $w_i$ in Tweet $d_k$ is connected to adjacent word $w_j$ in Tweet $d_k$. This generates directed graph $G_D$ for words occurring within same Tweet $d_k$ and every node $w_i$ points to next adjacent node $w_j$. This directed graph of all Tweets $D = \{d_1, d_2, ... d_n\}$ is called Word Co-occurrence Network ($WCN$). The frequency of co-occurring words is considered as an edge-weight which represents number of times the edge is connecting two words as an adjacent pair in Tweet Corpus.

The edge weights of Microblog WCN are sorted in a list in descending order. Top $k$ weighted edges are preserved in the Microblog WCN among total number of edges. Heuristic values are used to identify the top $k$ number of the highest edge-weights which are preserved in graph to obtain appropriate sub-graphs with popular content in its vicinity. Different heuristic values which are used for this research work are:

- **Root Two:** This gives the square-root of the total number of edges in Microblog WCN which gives the measure of number of edges preserved in Microblog WCN as shown in Equation 5.1

$$k = \sqrt{Total\ Number\ of\ Edges} \qquad 5.1$$

- **Divided by Two:** This gives top $k$ weighted edges as one-half of total number of edges as shown in Equation 5.2.

$$k = \frac{Total\ Number\ of\ Edges}{2} \qquad 5.2$$

- **Divided by Three:** This gives top $k$ weighted edges as one-third of total number of edges as shown in Equation 5.3.

$$k = \frac{Total\ Number\ of\ Edges}{3} \qquad 5.3$$

- **Log Method:** This is given by k as log of total number of edges as shown in Equation 5.4.

$$k = \log(Total\ number\ of\ edges) \qquad 5.4$$

The *Root-Two* measure out-performed all the other heuristic measures namely *Divided by Two, Divided by Three, and Log Method*. This is because *divided by two* or *divided by three* does not give all highly weighted edges. The number of edges which are extracted from Microblog WCN using *divided by 2* and *divided by 3* methods are increased. Hence, there are some edges which are low weighted. The log method is not scalable over wide range of Tweet corpus size. Thus, root two method was found to be suitable heuristic approach to obtain required sub-graphs from Microblog WCN. This study is performed as pilot study which acts as preamble to identify keyphrases by obtaining disconnected sub-graphs with high weighted edges.

### 5.2.2. K-Bridge Decomposition

The Root Two method is a heuristic approach which is based on human interpretations and human judgements. It gives abstract results. To remove this limitation of heuristic approach, a decomposition based keyphrase extraction technique is proposed. As the Microblog WCN is scalable over edge-weight distribution, it contains some high weighted edges. A topic of FSD dataset is used to find the plot of edge-weight with respect to frequency of edge-weights as shown in Figure 5.4.

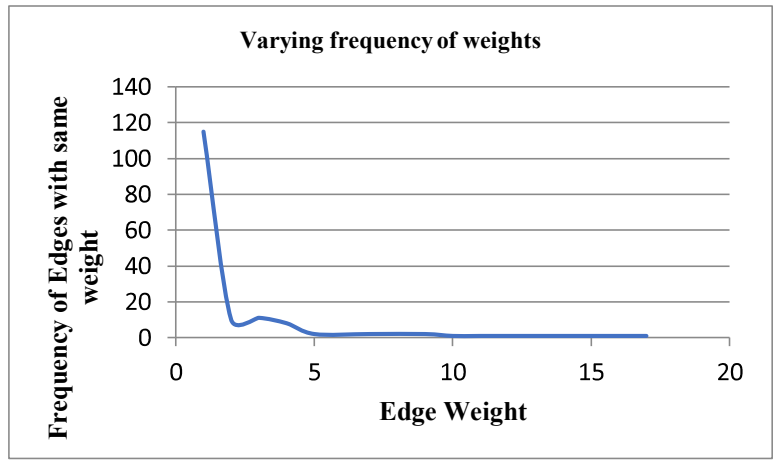

**Figure 5.4:** Varying Frequency of Weights from Microblog WCN

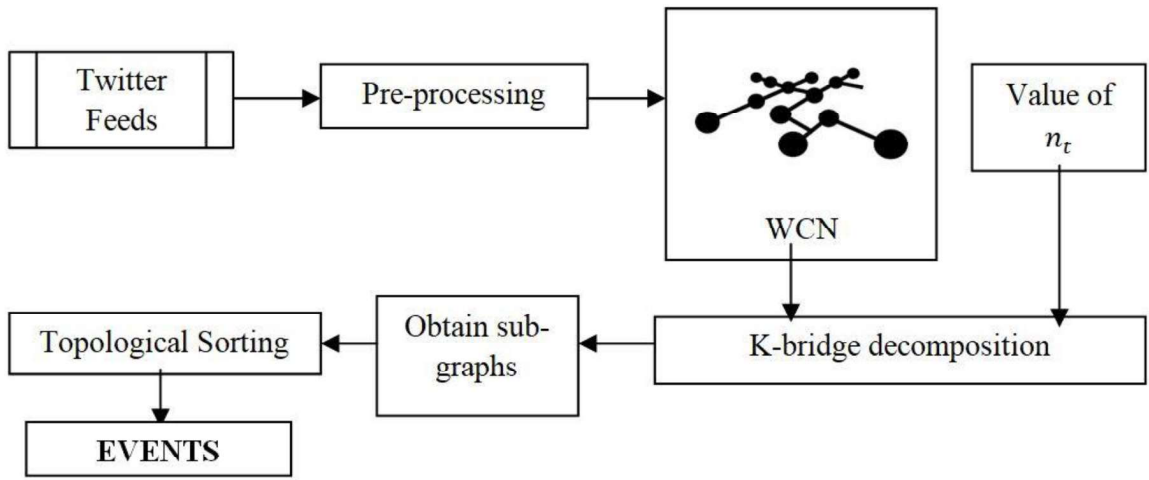

**Figure 5.5:** Architecture of K-bridge decomposition from Social Media Data

The key idea is that in every iteration, the minimum weighted edge is removed which may or may not result into disconnected graph. By iteratively removing an edge from each graph or sub-graph, it will give connected high weighted edges only. The k-bridge decomposition algorithm is shown in Algorithm 5.1. As observed, the Microblog WCN is used to decompose iteratively till each sub-graph contains less than $n_t$ where $n_t$ is the maximum number of nodes in any component or sub-graph of a disconnected network. Thus, $n_t$ is the controlling parameter to terminate the iterative process of decomposition in Microblog WCN. The architecture for k-bridge decomposition is shown in Figure 5.6. The Tweet Corpus is pre-processed and is tokenized to create Microblog WCN. In this study, the Tweets for single topic of FSD dataset are extracted. The Microblog data is user-generated and contains many slangs, abbreviations and other redundant information like those of smiley and jargons. A unique word is marked as a node and co-occurring words are represented by an edge between the words. While k-bridge decomposition, several sub-graphs are obtained.

**Algorithm 5.1**: The k bridge decomposition (G, t=1)

1. Consider as the list of edges as $L(e_i) = G.edges()$
2. For each sub-graph with number of nodes $< n_t$
$$e_{i+1} = \begin{cases} 0, & if\ e_i = 1 \\ e_i - 1, & if\ e_i > 1 \end{cases}$$
3. Edges with *zero edge weight* are removed from the graph.
4. Nodes with *zero degree* are removed from the graph.

### 5.2.3. Threshold based Decomposition

In k-bridge decomposition, the controlling parameter which is used to terminate the iterative process of decomposition is $n_t$. This parameter is used to limit the number of nodes in every sub-graph which is being decomposed using k-bridge decomposition. However, to process huge amount of social media data, the iterative decomposition is computationally expensive and slow-down the process. Thus, to speed up the processing, a threshold $t$ is defined as a boundary. The value of threshold is used to remove all the edges from Microblog WCN having edge-weight less than or equal to $t$ as shown in Algorithm 5.2. The k in k-bridge decomposition is defined as threshold value of edge-weight $t$.

**Algorithm 5.2:** The k bridge decomposition (G, t): Threshold based Approach

1. Consider as the list of edges as $L(e_i) = G.edges()$
2. All edges having edge-weight below t are removed from graph and edges having edge-weight above threshold t are computed as
$$e_{i+1} = \begin{cases} 0, & if\ e_i \leq t \\ e_i - t, & if\ e_i > t \end{cases}$$
3. Edges with zero edge-weight are removed from the graph.
4. Nodes with zero degree are removed from the graph.
5. Return G

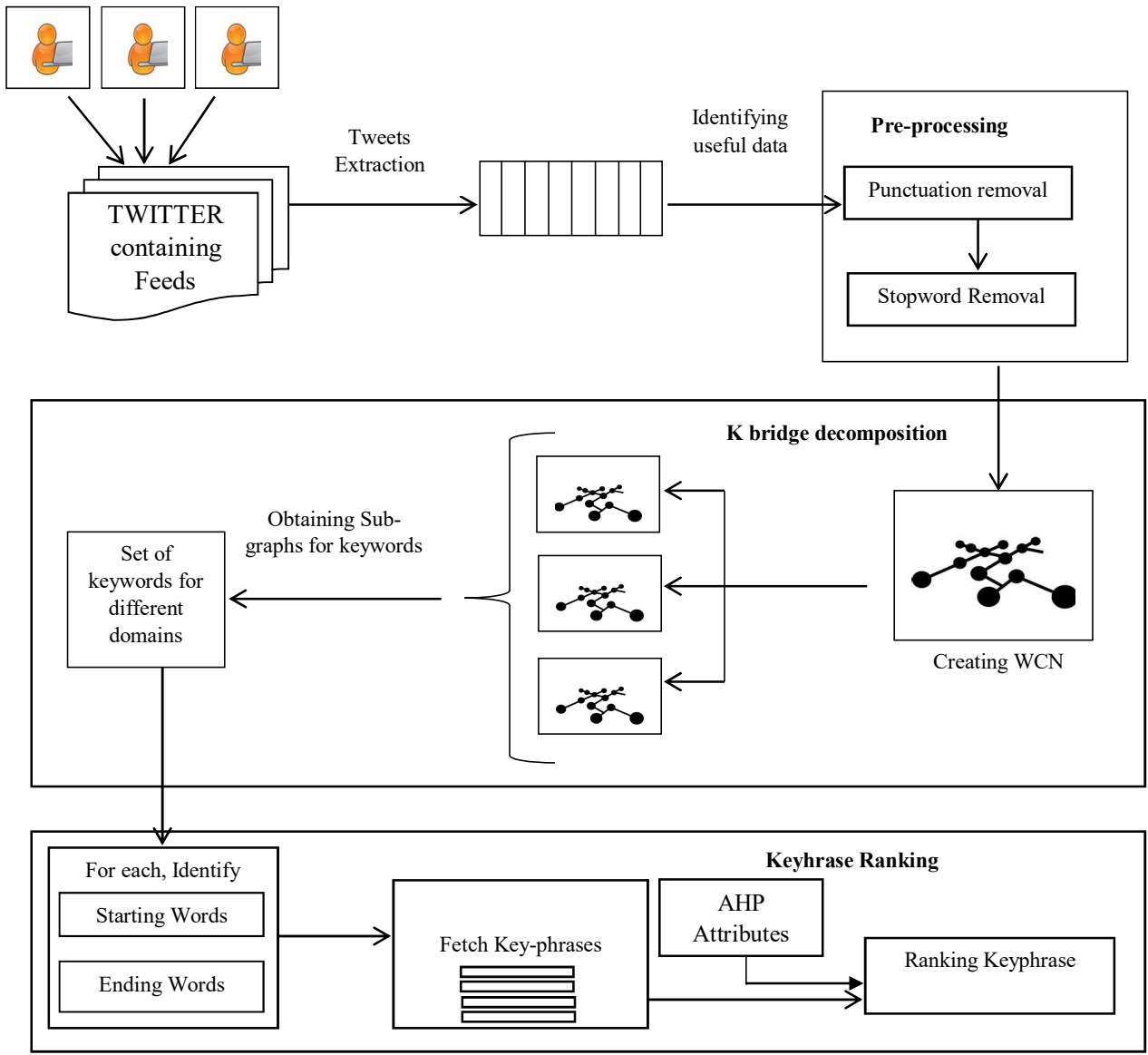

**Figure 5.6:** Architecture of keyphrase extraction using approximation approach of k-bridge decomposition

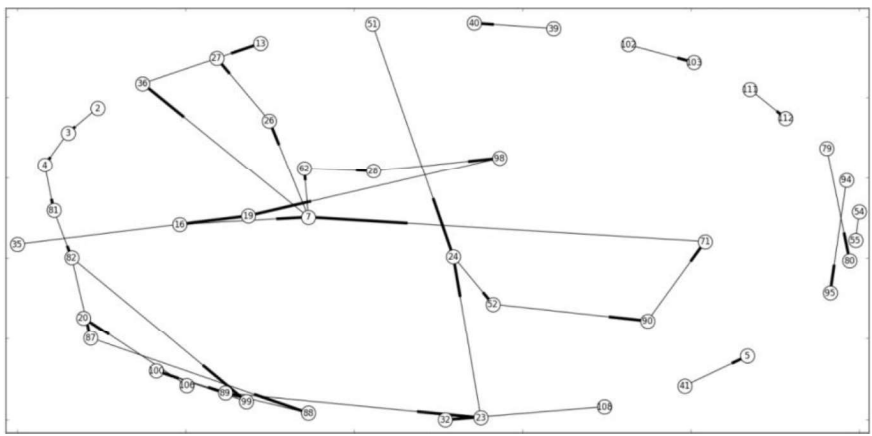

**Figure 5.7:** Microblog WCN after k-bridge decomposition

For edge weights more than $t$, the new edge-weight will be calculated as difference between current edge-weight and $t$. The parameter $t$ is defined as the value at $p$ index in the *list of edge-weights* which is sorted in descending order. The significance of $t$ is to keep top $p$ edges in the network so as to obtain the useful keyphrases information in vicinity. The parameter $p$ is set as the value which specifies the number of high weighted edges which should be kept in resulting graph.

For each sub-graph, a new threshold $t$ is calculated using value $p$ which is static and fixed. This algorithm is also defined as an approximation algorithm of k-bridge decomposition as it reduces the complexity of existing k-bridge decomposition by reducing number of iterations in Microblog WCN.

The architecture of approximation approach for k-bridge decomposition is given in Figure 5.6. The WCN is created using Tweet Corpus and an approximation approach of k-bridge decomposition is used to obtain sub-graphs which are well connected with each other. The connections are in the form of directed edges as observed from Figure 5.7. The directed edges are used to obtain ordered words in a keyphrase to preserve lexical sequence. After decomposition, many components which contain single word and are considered as unimportant components due to no connection. Thus, sub-graphs having single node is removed from the graph.

## 5.3. Identifying influential segments from social media data

The sub-graph gives highly important chains of words in vicinity. These words are important and combined together to generate keyphrases. In existing literature, the top $k$ keywords which are obtained from graph-based keyphrase extraction techniques and if they are connected together in a graph, then these connect words are obtained as keyphrase. The existing techniques do not maintain lexical sequence of words. In this research work, to obtain keyphrases from the sub-graphs which are obtained after decomposition, two different approaches are used namely Multiple Lexical Sequences (MLS) and topological sorting.

### 5.3.1. Multiple Lexical Sequences (MLS)

The Multiple Lexical Sequences (MLS) are the multiple keyphrases which are obtained from same sub-graph using different paths from the *first element* to the *last element*. Based on the position of words in sub-graph, each word can be classified into *first element, last element or* *centre element* as discussed in Section 5.2. The *first element* and *last element* can be identified as the node with incoming degree 'zero' and out-going degree 'zero', respectively as shown in Algorithm 5.3. The list of *first element* and *last element* is named as *Ls and Le*, respectively. From each element in list *Ls* to each element in list *Le*, all the paths are obtained in proper lexical sequence using directed graph. These multiple lexical sequence (MLS) are different sequence of words in sub-graph which are used to obtain different context from a Microblog WCN.

**Algorithm 5.3:** Keyphrase Extraction using Multiple Lexical Sequence (MLS)

1. G=*k bridge decomposition (k)*
2. Identify list First Element ($Ls$)

$$\forall I_{score}(n) = 0$$

3. Identify list Last Element ($Le$)

$$\forall O_{score}(n) = 0$$

4. For *eachStart* in *Ls*:
5.     For e *eachEnd* in *Le*:
6.         if $\exists path\ (eachStart, eachEnd)$:
7.             $(keyphrase = find\_path(eachStart, eachEnd)$
8. Return Keyphrase so obtained

The frequency of occurrence of text or high edge-weights does not necessarily make keyphrases important. Thus, the contextual information which is obtained using MLS is ranked using Analytical Hierarchical Process (AHP) as discussed in Chapter 6.

### 5.3.2. Topological Sorting

The topological sorting is the linear ordering of any sub-graph of directed Microblog WCN. For every edge $w_1w_2$, the words are ordered such that $w_1$ comes before $w_2$. This approach is used to extract at-least one keyphrase from each sub-graph having two or more than two nodes. This approach is based on Directed Acyclic Graph (DAG). Thus, the sub-graph which is obtained after k-bridge decomposition is converted into acyclic graph by removing the lowest edge weight from each cycle in a graph. Topological sorting is used to describe a set of words as keyphrase in ordered lexical sequence.

## 5.4. Experimental results and evaluations

The experimental results and evaluations are performed over Tweet Corpus using FSD dataset which contains Tweets related to multiple events. The k-bridge decomposition is used to extract keywords and keyphrases from Tweet Corpus. It is used for identifying different events or sub-events using longest MLS from Tweet Corpus. A set of Tweets is obtained for each topic and keyphrases are extracted from each set. The resulting keyphrases are obtained for different topics using threshold based approach for k-bridge decomposition.

### 5.4.1. Experimental results for k-bridge decomposition

In this research work, the experimental results are obtained for identification of keywords and keyphrases from Tweet Corpus. This approach depends on $n_t$ which act as controlling parameter to terminate the iteration of decomposition for every sub-graph. Thus, for varying value of $n_t$, which is taken as 10, 20, and 30, the results are obtained as shown in Table 5.1, Table **5.2**, and Table **5.3**, respectively. It can be concluded that the value of $n_t < 10$, gives better results than that of higher values of $n_t$. This is because the results for multiple events may or may not get separated due to the use of common vocabulary when higher threshold value is used. The keyphrase is obtained as the longest path found in MLS after decomposition. The difference between KeyWord Based Results (KWBR) and KeyPhrase Based Results (KPBR) is that KWBR are all the words from sub-graph but KPBR are the lexical sequence of set of words which are there in longest chain from first element to last element. In KPBR, there may be some words which are missed and are not included in that chain. But it gives high K-PREC than that of KWBR for all values of $n_t$ as observed in Table 5.4.

**Table 5.1:** Experimental results for keyword and keyphrase extraction for multiple topics using Tweet Corpus for $n_t < 10$

| S. No. | Keywords obtained | Keyphrase obtained | Actual Topics |
|---|---|---|---|
| 1 | rating agency debt bill downgrades first house credit | credit rating debt first | S&P downgrades US credit rating; |
| 2 | amy winehouse died dead found rip | rip amy winehouse found dead | Death of Amy Winehouse |
| 3 | man ealing riot attacked, richard mannington | richard mannington bowes, man attacked ealing | Richard Bowes, victim of London riots, dies in hospital |
| | bowes | | |
| 4 | carrying crashed russia plane crash khl hockey team lokomotiv | plane carrying team | Plane carrying Russian hockey team Lokomotiv crashes, 44 dead |
| 5 | space shuttle atlantis nasa | | Space shuttle Atlantis lands safely, ending NASA's space shuttle program |
| 6 | norway youth summer dressed oslo camp | dressed camp norway | Gunman opens fire in children's camp on Utoya island, Norway, |
| 7 | delhi high court blast outside explosion | blast outside delhi high court | Terrorist attack in Delhi |
| 8 | earthquake virginia ago felt aug mineral | earthquake virginia aug ago mineral | Earthquake in Virginia |

**Table 5.2:** Experimental results for keyword and keyphrase extraction for multiple topics using Tweet Corpus for $n_t < 20$

| S. No. | Keywords obtained | Keyphrase obtained | Actual Topics |
|---|---|---|---|
| 1 | rating agency debt bill downgrades passes first downgrade downgraded aa+ breakingnews house via loses aaa credit | breakingnews debt aaa credit rating agency downgrades first downgrade | S&P downgrades US credit rating; |
| 2 | amy winehouse died dead found rip | rip amy winehouse found dead | Death of Amy Winehouse |
| 3 | man ealing riot attacked, richard mannington bowes | richard mannington bowes, man attacked ealing | Richard Bowes, victim of London riots, dies in hospital |
| 4 | reported terrible kills khl carrying russian crashed russia plane crash lokomotiv members ice hockey team crashes | terrible plane russian carrying hockey team lokomotiv crashed | Plane carrying Russian hockey team Lokomotiv crashes, 44 dead |
| 5 | kennedy center orbits earth journey miles year space shuttle atlantis around final nasa florida lands nasas landed' | | Space shuttle Atlantis lands safely, ending NASA's space shuttle program |
| 6 | norway gunman delhi high court blast outside explosion summer dressed youth oslo childrens camp | Children camp norway | Gunman opens fire in children's camp on Utoya island, Norway; Terrorist attack in Delhi |
| 7 | earthquake virginia ago felt aug mineral | earthquake virginia aug ago mineral | Earthquake in Virginia |

Table 5.3: Experimental results for keyword and keyphrase extraction for multiple topics using Tweet Corpus for $n_t < 30$

| S. No. | Keywords obtained | Keyphrase obtained | Actual Topics |
|---|---|---|---|
| 1 | news amy winehouse died time rip breaking loses aaa credit rating passes debt ceiling bill downgrades dead downgraded aa+ found house representatives debtceiling agency giffords first downgrade via breakingnews | breaking credit downgraded rating agency downgrades aaa first debt house representatives | S&P downgrades US credit rating; |
| | | | Death of Amy Winehouse |
| 2 | man ealing riot attacked, richard mannington bowes | richard mannington bowes, man attacked ealing | Richard Bowes, victim of London riots, dies in hospital |
| 3 | reported terrible kills khl carrying russian crashed russia plane crash lokomotiv members ice hockey team crashes | terrible plane russian carrying hockey team lokomotiv crashed | Plane carrying Russian hockey team Lokomotiv crashes, 44 dead |
| 4 | kennedy center orbits earth journey miles year space shuttle atlantis around final nasa florida lands nasas landed | | Space shuttle Atlantis lands safely, ending NASA's space shuttle program |
| 5 | norway gunman delhi high court blast outside explosion summer dressed youth oslo childrens camp | Children camp norway | Gunman opens fire in children's camp on Utoya island, Norway; Terrorist attack in Delhi |
| 6 | earthquake virginia ago felt aug mineral | earthquake virginia aug ago mineral | Earthquake in Virginia |

Table 5.4: Performance evaluation for keyphrase extraction technique using varying values of $n_t$

| Value of $n_t$ | K-PREC | | T-REC | | K-REC | |
|---|---|---|---|---|---|---|
| | KWBR | KPBR | KWBR | KPBR | KWBR | KPBR |
| 10 | 0.488 | 0.527 | 80% | 70% | 0.629 | 0.534 |
| 20 | 0.410 | 0.532 | 80% | 60% | 0.628 | 0.693 |
| 30 | 0.381 | 0.507 | 80% | 50% | 0.682 | 0.532 |

For varying values of $n_t$, the value of $n_t < 20$ gives best K-REC and K-PREC among all other experiments. This is because neither too high nor too low value of $n_t$ is suitable for identifying KWBR and KPBR. For low value i.e. $n_t < 10$, T-REC is observed to be highest for KPBR whereas it remains constant for KWBR. This is due to the fact that as the value of threshold parameter is increased, the maximum number of words in the sub-graph which is obtained after k bridge decomposition, is increased. Thus, more number of words are extracted which may be insignificant. This may also connect keywords of two or more topics together in one sub-graph. For low value of $n_t$, less number of words are obtained in resulting sub-graphs and thus, both K-REC and K-PREC are reduced. The reason behind low K-PREC for all values of $n_t$ is that the keywords which are present in annotated dataset are lesser than resulting keywords as the information given in the topic is limited. As per manual evaluation, the keywords are observed as significant and relevant to the topic.

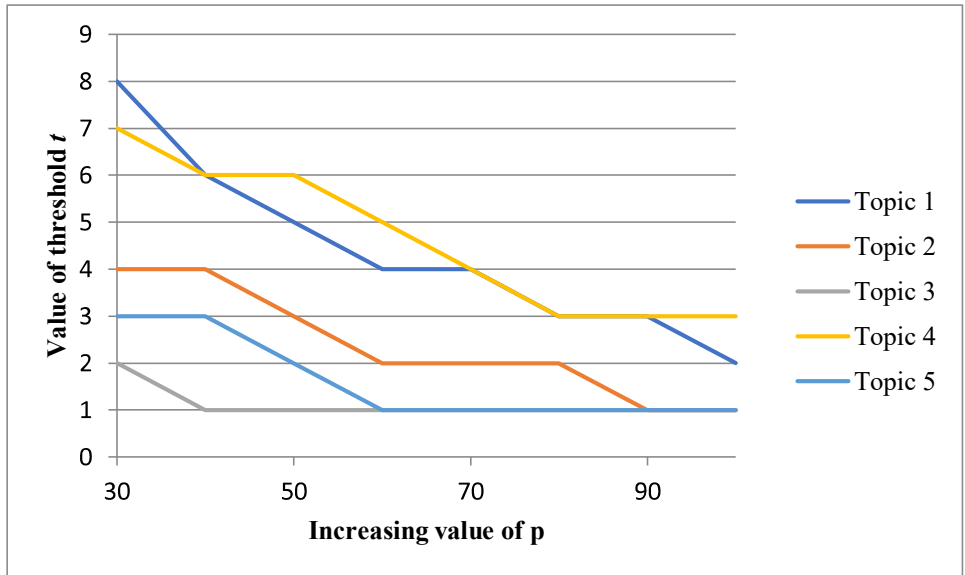

**Figure 5.8:** Varying value of threshold t with respect to p

**Table 5.5:** Experimental results for topic related Tweets using Threshold based approach for k-bridge decomposition

| Topic | Value of p | Value of t | Keyphrase Obtained |
|---|---|---|---|
| Death of Amy Winehouse | 30 | 8 | rip amy winehouse found dead |
| | 40 | 6 | rip amy winehouse found dead her |
| | 50 | 5 | lordvoldemort amywinehouse has died the cause death unknown but there are rumors mrs weasley mistook her for bellatrix |
| | 60 | 4 | |
| | 70 | | |
| | 80 | 3 | |
| | 90 | | |
| | 100 | 2 | holymoly sadly true amy winehouse dies from overdose amywinehouse has died the cause death unknown but there are rumors mrs weasley mistook her for bellatrix |
| Space shuttle Atlantis lands safely, ending NASA's | 30 | 4 | kennedy space shuttle atlantis |
| | 40 | 4 | kennedy space shuttle atlantis |
| | 50 | 3 | nasas year shuttle atlantis landed nasa kennedy space center after orbits around earth and journey miles sts |
| | 60 | 2 | Bringing nasas year shuttle atlantis landed nasa kennedy |
| space shuttle program | 70 80 | | space center after orbits around earth and journey miles sts |
| | 90 | 1 | First picture atlantis touching down for final mission bringing nasas year shuttle atlantis lands kennedy space center after orbits around earth and journey miles sts |
| Richard Bowes, victim of London riots, dies in hospital | 30 | 2 | man attacked ealing riot dies |
| | 40 50 60 | 1 | richard mannington bowes who was critically injured tried stamp out fire during riots ealing west london |
| Plane carrying Russian hockey team Lokomotiv crashes, 44 dead | 30 | 7 | russian plane carrying khl team lokomotiv |
| | 40 | 6 | |
| | 50 | 6 | |
| | 60 | 5 | russian plane carrying khl team lokomotiv, charter plane carrying khl team lokomotiv |
| | 70 | 4 | russian plane carrying khl team lokomotiv yaroslavl |
| | 80 90 100 | 3 | dchesnokov terrible tragedy russia charter plane carrying khl team lokomotiv crashed only one person survived |
| Gunman opens fire in children's camp on Utoya island, Norway | 30 | 3 | summer camp |
| | 40 | 3 | Summer camp |
| | 50 | 2 | labour party youth camp norway |
| | 60 | 1 | skynewsbreak sky sources widespread reports attack childrens summer camp island just outside oslo norway gunman dressed police officer |

### 5.4.2. Experimental results for Threshold based Approach

The results are obtained for threshold based approach for k-bridge decomposition as discussed in Section 5.2.3. The FSD dataset is used for the experimental results and evaluations. The influential segments are extracted using MLS based keyphrase extraction approach. Many different keyphrases are obtained and experimental results are observed using different values of controlling parameters $p$ and $t$ as shown in Figure 5.8.

The value of p is chosen from the index of the descending ordered list of edge-weights which describes the total number of highest weighted edges which should be considered for extracting keyphrases. It is observed that more keywords and more information is extracted as the value of $p$ is increased. The value of $p$ describes the number of top edges which should be kept in the resulting network. The value of $t$ is the edge-weight at index $p$ is the list. The value of $t$ may have same value at index $p - 1$ or at $p + 1$. As the number of edges which should be preserved in the network is increased, the value of $t$ is decreased and thus, for low threshold value of edge-weight $t$, less number of edges are removed. This gives more number of words in the resulting keyphrase. Different values of $t$ and $p$ can be used and these may vary for different scale and different types of Tweet Corpus. In this approach, experiments are performed to calculate the keyphrases using various values of $p$ and $t$ as shown in Table 5.5.

For different topics, the keyphrases are obtained with high K-PREC for different values of $p$ and $t$. For instance, for the topic *"Death of Amy Winehouse"*, the resulting keyphrase have high K-REC and high K-PREC for value of $t = 5$ which is obtained at $p = 50$ than that of value at $t = 8$ for $p = 30$. For the topic *"Plane carrying Russian hockey team Lokomotiv crashes, 44 dead"*, the keywords in keyphrase are increased as we increase the value of $p$ and more information is obtained at $p = 30$ than that of $p = 80$. The number of keywords obtained in the resulting keyphrase is increased and contextual information varies. The context of the keyphrase is ranked using *Analytical Hierarchical Process (AHP)* which is explained in Chapter 6.

## 5.5. BArank: A keyphrase extraction technique

BArank is an extension of the previous studies: identifying influential segment using k – bridge decomposition over Microblog WCN. In existing research work, the termination of decomposition depends on controlling parameters for keyphrase extraction from social media data. The static parameters $p$ and $t$ were used for Microblog WCN to terminate the iterative process of removing lower weighted edges. The Microblog WCN is disassortative as observed in Chapter 3. The disassortativity of the network ensures that all high degree words are not connected to high degree nodes only. The idea behind the proposed keyphrase extraction technique is to identify the break-even point during iterative k-bridge decomposition to check the *assortativity property* of Microblog WCN. This break-event point is used for automatic termination to deduce important sub-graphs from Microblog WCN.

**Definition 5.2: Break-Even Point:** The point where the sub-graph gets in a disassortative network of Microblog WCN converted into non-disassortative graph from disassortative graph, is called break-even point.

### 5.5.1. Proposed Technique: BArank

In this research work, a non-parametric keyphrase extraction technique is proposed for Microblog WCN. This means that there is no static controlling parameter to terminate iterative process in Microblog WCN. The BArank keyphrase extraction technique uses k-**B**ridge decomposition, **A**ssortativity and **rank**ing measure for identifying top-n keyphrases. The contributions which are made for keyphrase extraction techniques are given below.

- **Terminal State:** Earlier, the static parameters were used to terminate the decomposition in Microblog WCN such that either the maximum number of nodes is fixed for every sub-graph as $n_t$, or $p\ and\ t$ threshold are used for decomposition. BArank does not need any static controlling parameter and executes automatic termination using break-event point. The edge-based decomposition of the network is performed till the network gets non-disassortative as shown in Figure 5.9. Resulting sub-graph contains the most connected keywords.

- **Topological Sorting:** The keyphrases were extracted using MLS keyphrase extraction technique from sub-graphs in threshold based approach of k-bridge decomposition. The BArank uses topological sorting for extracting keyphrase from directed graph which gives single keyphrase. This keyphrase represents the complete information in the sub-graph.

- **Ranking:** In this research work, many different sub-graphs were obtained after k – bridge decomposition of domain specific Microblog WCN. Using topological sorting, single keyphrase is obtained for every sub-graph. Thus, keyphrases are ranked using the density of the sub-graph. The density (Angel *et al.*, 2012) is measured as the ratio of the sum of edge weights in the sub-graph to the total number of edges.

The pseudo-code for BArank is shown in Algorithm 5.4. The k – bridge decomposition is the edge based network reduction which provides a probe to study the hierarchical properties of large scale networks, focussing on the network's regions of increasing centrality and connectedness properties (Alvarez-Hamelin *et al.*, 2006). The network is disassortative and the complex network can be reduced by removing low edge-weights as shown in Algorithm 5.4.

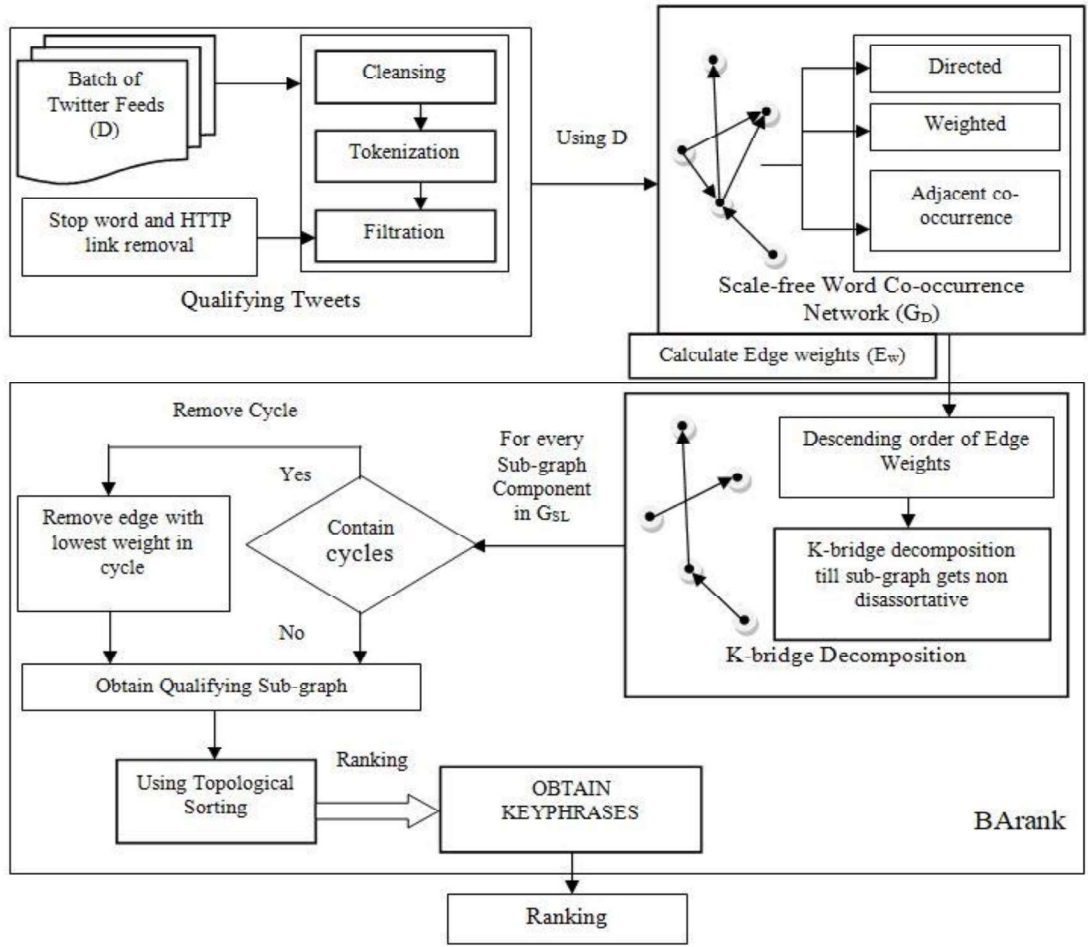

**Figure 5.9:** Architecture of BArank for identifying keyphrases from social media data

---

**Algorithm 5.4:** BArank:A Keyphrase Extraction Technique

---

1. Given: Set of Tweets $D$
2. For each document $d$ in $D$:
3.     $T_d = Preprocess(d)$
4.     For each word $w_i$ in $T_d$:
5.         $w_i = Preprocess(w_i)$
6. $G = CreateWeightedDiGraph(D)$
7. $If\ G\ is\ Disassortative()$:
8.     For $eachSubgraph\ in\ G.subgraph$:
9.         $L_e = Ascending(G.subgraph.edges())$
10.         $i = 0$
11.         While $eachSubgraph! = Assortative()$:
12.             Remove $L_e[i]$
13.             $i = i + 1$
14. For $eachSubgraph\ in\ G.subgraph$:
15.     $keyphrase = topologicalSorting(eachSubgraph)$
16. $RankingKeyphrases(G)$

The mathematical validation for assortativity analysis of Microblog WCN is as follows:

**Step 1:** The assortativity is defined as shown in Equation 5.5 and is denoted by $\tau$.

$$\tau = \frac{M^{-1}\sum_i j_i k_i - \left[M^{-1}\sum_i \frac{1}{2}(j_i + k_i)\right]^2}{M^{-1}\sum_i \frac{1}{2}(j_i^2 + k_i^2) - \left[M^{-1}\sum_i \frac{1}{2}(j_i + k_i)\right]^2} \qquad 5.5$$

where $j_i$ and $k_i$ are the degrees of the two endpoints of the $i^{th}$ edge, and $M$ is the total number of edges in the network. If $\tau > 0$, the network is said to have *assortative mixing*; while if $\tau < 0$, the network is said to have *disassortative mixing*.

**Step 2:** This *assortative mixing coefficient* ($\tau$) is then equivalent to Equation 5.6

$$\tau = \frac{\sum_i j_i k_i - M^{-1}\left[\sum_i \frac{1}{2}(j_i + k_i)\right]^2}{\sum_i \frac{1}{2}(j_i^2 + k_i^2) - M^{-1}\left[\sum_i \frac{1}{2}(j_i + k_i)\right]^2} \qquad 5.6$$

**Step 3:** For the negative value of $\tau$, the *disassortative mixing* Equation is obtained as shown in Equation 5.7

$$\frac{\sum_i j_i k_i - M^{-1}\left[\sum_i \frac{1}{2}(j_i + k_i)\right]^2}{\sum_i \frac{1}{2}(j_i^2 + k_i^2) - M^{-1}\left[\sum_i \frac{1}{2}(j_i + k_i)\right]^2} < 0 \qquad 5.7$$

**Step 4:** This implies that for *disassortative mixing*, the resulting coefficients should follow Equation 5.8.

$$either\ (A < B < C)\ or\ (C < B < A), \qquad 5.8$$

where $A = \sum_i j_i k_i$, $B = M^{-1}\left[\sum_i \frac{1}{2}(j_i + k_i)\right]^2$ and $C = \sum_i \frac{1}{2}(j_i^2 + k_i^2)$. This implies that $B$ should lie in between $A$ and $C$.

**Step 5:** As per analysis, the resulting coefficients for assortative mixing should follow Equation 5.9

$$either\ (a > b\ and\ c > b)\ or\ (b < a\ and\ b < c), \qquad 5.9$$

where $A = \sum_i j_i k_i$, $B = M^{-1}\left[\sum_i \frac{1}{2}(j_i + k_i)\right]^2$ and $C = \sum_i \frac{1}{2}(j_i^2 + k_i^2)$. This implies that either $B$ is smaller than both $A\ and\ C$, or $B$ is greater than both $A\ and\ C$. Thus, the break-even point for disassortative network is when A = B or C = B as observed in Figure 5.10.

The assortativity mixing coefficient for different parameters A, B, and C is observed for iterative process of k-bridge decomposition for Microblog WCN.

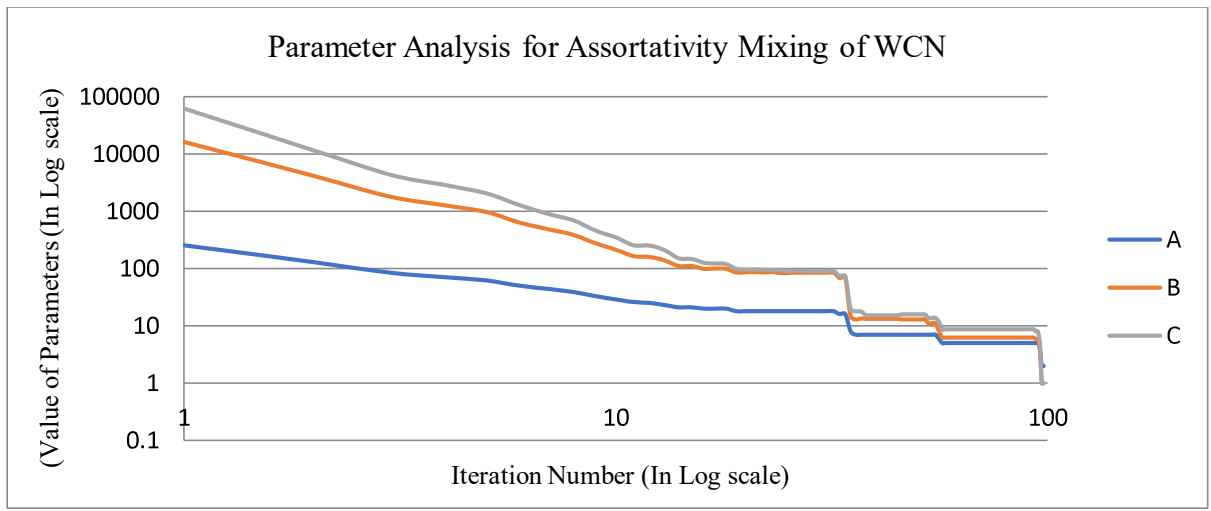

**Figure 5.10:** Parameter Analysis of assortativity mixing component for Microblog WCN during decomposition

The network is decomposed by removing lowest weighted edges from each sub-graph and high weighted edges are preserved. Initially the parameter $B$ lies in between $A$ and $C$ and shows disassortativity as mentioned in Equation 5.8, till the value gets equal to 4 where the break-even point is obtained. The results are obtained for the graph having number of nodes 4 and topological sorting gives the keyphrase as *"rip amy dead winehouse"* for Tweet Corpus of Topic *"Death of Amy Winehouse"* from FSD dataset.

Ranking keyphrases is used to obtain keyphrases as event-phrases. The most-important *event-phrases* and *sub-event phrases* depend upon the occurrence of nodes and the extent of closely knitted nodes in the network. Thus, the keyphrases obtained by topological sorting are ranked using the density ($Den$) of a sub-graph as defined in Equation 5.10.

$$Den\ (S) = \frac{\sum_{(i,j)\epsilon M} w_{i,j}}{M} \qquad 5.10$$

where M is the total number of edges and $w_{i,j}$ is the weight of an edge connecting nodes $i\ and\ j$ in the sub-graph ($S$). The graph based keyphrase extraction techniques namely TextRank, TopicRank, and NErank are used as baseline measures to compare and validate the BArank.

## 5.5.2. Experimental results and evaluation

The Microblog WCN is stable as observed in Chapter 3, and shows small-world property. The evaluations are performed for different topic of FSD dataset. The Microblog WCN follows scale-free property for degree distribution and edge distribution. The edge distribution shows important bi-grams with high co-occurring frequency of words among each other. Thus, the edge-based decomposition has proved to be better and effective for keyphrase extraction. The NErank used node-scores and edge-scores for random-walk based keyphrase extraction algorithm using Twitter data.

The TextRank, TopicRank, NErank and BArank keyphrase extraction algorithms are implemented over FSD dataset. The implementations are made using Python 2.7.11. The observations are obtained using the performance measure ROUGE score, K-REC, K-PREC, and F-measure as shown in Table 5.6. The BArank outperforms all the existing graphical keyphrase extraction techniques.

**Table 5.6:** Performance analysis of TextRank, TopicRank, NErank, and BArank for Microblog WCN

| $Algorithm$ | $ROUGE-1$ | $ROUGE-L$ | $ROUGE-2$ | $K-REC$ | $K-PREC$ | $F-measure$ |
|---|---|---|---|---|---|---|
| $TextRank$ | 19.13 | 0.67 | 5.55 | 19.13 | 8.61 | 11.87 |
| $NErank$ | 13.73 | 0.77 | 1.38 | 13.73 | $25.76$ | 17.91 |
| $TopicRank$ | 26.54 | 0.88 | 11.11 | $26.54$ | 18.11 | $21.53$ |
| $BArank$ | **37.80** | **1.62** | **12.66** | **37.80** | **39.95** | **38.84** |

**Table 5.7:** Keyphrase obtained using TextRank, TopicRank, NErank, and BArank for Microblog WCN

| Technique\Topic | Plane carrying Russian hockey team Lokomotiv crashes, 44 dead | Terrorist Attack in Delhi | Betty Ford dies |
|---|---|---|---|
| TextRank | Breaking news | Ox blood ruffin | Isolabella former |
| TopicRank | Plane crash | High court delhi | Betty Ford |
| NErank | Russia Lokomotive hockey | Reuters India | Isolabella former |
| BArank | Plane crash Hockey team | Blast outside Delhi high court | RIP lady Betty Ford |

The K-REC, K-PREC, and F-measure is obtained using automatic evaluation. The BArank outperforms the existing techniques with 38.84 F-measure, and shows an edge over K-REC and K-PREC measures with 37.80 score and 39.95 score, respectively. TopicRank and NErank gives second best performance with value of K-REC as 26.54, and the value of K-PREC as 25.76, respectively. BArank outperforms the existing techniques by identifying important and relevant words which describes the topic as shown in Table 5.7.

Keyphrases obtained from TextRank are less likely to have keywords and have more repetitive and popular terms. The NErank gives good performance over selected topics by using nodes-score and edge-weight score. The resulting keyphrases by TopicRank does not give much of the words which signify the event specific information and thus, shows low K-PREC. The BArank gives improved results for F-measure among all the existing techniques. This is because it gives those nodes in the output which are connected to high degree nodes and thus, gives keyphrase which represents non-disassortative nature of Microblog WCN.

## 5.6. Concluding Remarks

In this Chapter, the decomposition based keyphrase extraction technique is proposed. The results are obtained for KWBR and KPBR for k-bridge decomposition in terms of K-PREC, K-REC, and T-REC. To reduce the number of iterations, the threshold based k-bridge decomposition is proposed and resulting keyphrases are obtained using MLS and topological sorting. The parameter-free keyphrase extraction approach is proposed as BArank using k-bridge decomposition and assortativity network science property. It is observed that BArank outperforms all the existing keyphrase extraction techniques. The contextual information in keyphrase is ranked to obtain event-phrase or sub-event phrase using AHP optimization technique as explained in Chapter 6.

# Chapter 6.
# Ranking of Contextual Information using AHP

In existing graphical keyphrase extraction techniques, the keyphrases are usually ranked on the basis of its frequency or degree of words in WCN. The keyphrase which is obtained using k-bridge decomposition give set of words in an orderly fashion. This orderly fashion is due to the directed Microblog WCN in which lexical sequence of words is maintained. Different events and sub-events are described by one or more keyphrases as obtained from sub-graphs.

Keyphrase contains the contextual information. This contextual information should be ranked to identify top keyphrases. The context of the information obtained using MLS may or may not be repetitive due to multiple keyphrases extraction. The contextual information should be ranked and classified to identify event and sub-event phrases. Based on the structure of Microblog WCN, four attributes are proposed for AHP optimization technique.

## 6.1. Analytical Hierarchical Process (AHP) for contextual ranking

The Twitter dataset which is used for keyphrase extraction is FSD dataset. For each topic, Tweets are extracted to generate Tweet Corpus. For this Tweet Corpus, different keyphrases are obtained in a list. For ranking, different methods have been applied in the literature, namely, Analytic Hierarchy Process (AHP), fuzzy Multiple Attribute Decision- Making (MADM) model, linear 0-1 integer programming, weighted average method, and genetic algorithms. However, the keyphrases are ranked using MADM based AHP optimization technique (Saaty *et al.,* 1990). This approach is used to rank the contextual information among alternative keyphrases. The essential components of AHP are

- **The alternatives:** Different keyphrases which are obtained using MLS.
- **Attributes:** The features of keyphrases as observed from the study of the structure of Microblog WCN.

- **The main-goal:** To obtain the contextual information so as to obtain the most significant keyphrases as top ranked alternatives which are recommended as *event-phrases*.

Various steps are followed for AHP approach which are discussed as follows:

**Step 1:** The attributes of AHP are the contributing features to rank the alternatives. In this research work, the *alternatives* term is used for different keyphrases which should be ranked. To rank the keyphrases, *the sum of node-degree of keywords in keyphrase* was the key factor in existing random-walk based graphical keyphrase extraction technique. However, in this research work, instead of popularity of any word, the influence of the keyphrase, as a whole, is analysed. To find the influence of a keyphrase, four attributes are proposed as discussed in Section 6.2 using the semantics of Microblog WCN.

**Step 2:** The values for these attributes are obtained in each keyphrase. The matrix (M) for attributes is obtained as shown in Equation 6.1. In this matrix, $V_{ij}$ denotes the value of attribute in $i^{th}$ row and $j^{th}$ column. The row denotes the $i^{th}$ alternative and column denotes the $j^{th}$ attribute.

$$M = \begin{bmatrix} V_{11} & V_{12} & V_{13} \\ V_{21} & V_{22} & V_{23} \\ \vdots & \vdots & \vdots \\ V_{n1} & V_{n2} & V_{n3} \end{bmatrix} \quad\quad 6.1$$

**Step 3:** The matrix M which is obtained in Step 2 is normalized to obtain normalized decision matrix. The normalization for each attribute is performed on the basis of *beneficiary and non-beneficiary* attribute. The *beneficiary* attributes are those attributes which have high significance of high values and low significance of low values. Thus, for beneficiary attributes, each element of the column is normalized using maximum value in the column as shown in Equation 6.2. The normalized element of the column $j$ is obtained as $NV_{ij}$ as shown in Equation 6.3.

$$Max_j = Max(V_{ij}) \quad\quad 6.2$$

$$NV_{ij} = \frac{V_{ij}}{Max_j} \quad\quad 6.3$$

The *non-beneficiary* attributes are those attributes which have low significance for high values and high significance for low values. For non-beneficiary attributes, the minimum value is obtained from all the elements of column $j$ as shown in Equation 6.4. The normalized element of the column $j$ is obtained as $NV_{ij}$ as shown in Equation 6.5.

$$Min_j = Min(V_{ij}) \qquad 6.4$$

$$NV_{ij} = \frac{Min_j}{V_{ij}} \qquad 6.5$$

The resulting normalized decision matrix for three attributes is obtained for calculating the score with weighted matrix. This normalized decision matrix is denoted by $N_m$ as shown in Equation 6.6.

$$N_m = \begin{bmatrix} NV_{11} & NV_{12} & NV_{13} \\ NV_{21} & NV_{22} & NV_{23} \\ \vdots & \vdots & \vdots \\ NV_{n1} & NV_{n2} & NV_{n3} \end{bmatrix} \qquad 6.6$$

**Step 4:** The relative importance of attributes is analyzed and pair-wise comparison matrix is calculated using *Saaty 1-9 preference scale* as shown in Table 6.1. The attribute compared with it-self is always assigned value 1. The numbers from 1-9 gives verbal judgements for comparative importance of proposed attributes as compared to each other. If any attribute $a_{ij}$ is compared with other attribute $a_{ji}$, then the values are reciprocal of each other.

**Table 6.1:** Saaty's pairwise comparison scale

| Scale | Compare factor I and J |
|---|---|
| 1 | Equally important |
| 3 | Weakly important |
| 5 | Strongly important |
| 7 | Very strongly important |
| 9 | Extremely important |
| 2,4,6,8 | Intermediate values between adjacent scales |

The pair-wise comparison matrix is required to obtain Consistency Ratio (CR). The Pair-wise Comparison Matrix ($A_1$) is obtained for three attributes as shown in Equation 6.7 and *Geometrics Mean (GM)* is calculated for three attributes as shown in Equation 6.8.

$$A_1 = \begin{bmatrix} a_{11} & a_{12} & a_{13} \\ a_{21} & a_{22} & a_{23} \\ a_{31} & a_{32} & a_{33} \end{bmatrix} \qquad 6.7$$

$$GM = (a_{11} * a_{12} * a_{13})^{1/3} + (a_{21} * a_{22} * a_{23})^{1/3} + (a_{31} * a_{32} * a_{33})^{1/3} \qquad 6.8$$

GM is obtained as the sum of three components for three attributes. The weights are calculated as matrix $A_2$ having $n$ rows and 1 column where $n$ is number of attributes. The matrix $A_2$ is obtained as shown in Equation 6.9

$$A_2 = \begin{bmatrix} \dfrac{(a_{11} * a_{12} * a_{13})^{\frac{1}{3}}}{GM} \\ \dfrac{(a_{21} * a_{22} * a_{23})^{\frac{1}{3}}}{GM} \\ \dfrac{(a_{31} * a_{32} * a_{33})^{\frac{1}{3}}}{GM} \end{bmatrix} \qquad 6.9$$

The GM method of AHP is used to determine the relative normalized weights of the attributes because of its simplicity and easy determination of maximum eigenvalue. This helps in reduction in inconsistency of judgements.

**Step 5:** A new matrix $A_3$ is obtained as the product of two matrices $A_1$ and $A_2$. Finally $A_4$ is calculated as $A_3/A_2$. The average value of $A_4$ is obtained as maximum eigenvalue $\lambda_{max}$. The Consistency Index is denoted by CI and is calculated as shown in Equation 6.10.

$$CI = \frac{\lambda_{max} - m}{m - 1} \qquad 6.10$$

where, m is the number of attributes. The smaller the value of CI, smaller is the deviation from consistency.

**Step 6:** The Consistency Ratio (CR) is calculated using CI and Random Index (RI). RI gives the random degree of consistency which is independent of attribute values and varies for different number of attributes (Saaty, 1990). The values of RI for small number of attribute based problems are given in Table 6.2. Relative weights are assigned for attributes in $A_1$ and *consistency ratio (CR)* is obtained as given in Equation 6.11.

$$CR = \frac{CI}{RI}, where\ CR < 0.1 \qquad 6.11$$

where CR is Consistency Ratio, CI is Consistency Index which gives the deviation or degree of consistency, and RI is Random Index which gives random degree of consistency. The CR is better for lower values and thus, 10% deviation is allowed to accept the consistency (Saaty, 1990). The CR is calculated using RI as 0.90 and it is found to be consistent with if $CR < 0.1$.

**Table 6.2:** Values of the Random Index (RI) for small problems

| M | 2 | 3 | 4 | 5 | 6 | 7 | 8 | 9 | 10 |
|---|---|---|---|---|---|---|---|---|---|
| RI | 0 | 0.58 | 0.90 | 1.12 | 1.24 | 1.32 | 1.41 | 1.45 | 1.51 |

**Step 7:** If the *Pairwise Comparison Matrix* is found to be consistent, the overall score for each alternative (keyphrase) is calculated as $N_m * A_2$. Each alternative will get the corresponding score. These scores will be ranked in descending order and corresponding alternatives will be ranked accordingly.

The AHP is used in relative mode where decision makers have prior knowledge of the attributes for different alternatives to be used or when objective data of the attributes for different alternatives is not available. The major contributions of this research work are: the attributes have been proposed for AHP using the semantics of the structure of Microblog WCN as shown in Section 6.2, and the pair-wise comparison matrix is introduced which is used to obtain consistent results for contextual ranking of keyphrases as shown in Section 6.3.

## 6.2. Attributes of AHP

The attributes are proposed as feature set of alternatives (keyphrases) from Microblog WCN. These attributes are based on the structure of Microblog WCN. The Microblog WCN is scale-free for node-degree distribution and edge-weight degree distribution. The influence factors for any keyphrase, which is obtained from a sub-graph is determined as:

- *The normalized distribution of co-occurrences:* The normalized distribution of keyphrases is useful because the Microblog WCN is a path based network which is obtained using the chain of words in lexical sequence as it appears in a Microblog. The contribution of this factor is given as *Edge Strength Density.*
- *The variance of frequency:* The edge-strength density is variable among the centre elements of the keyphrase and boundary elements of the keyphrase. The rate of change of edge strength over boundary elements for convergence/ divergence is comparable to the edge-strength of centre elements. The strength difference gives the variance of frequency of co-occurrences of words.
- *Degree of keyphrase:* The degree of keyphrase determines the frequency of occurrence of keywords in a keyphrase. This measure is used as the only measure for ranking of keyphrases in existing graphical random-walk based keyphrase extraction techniques. The *phrase degree* is proposed to measure the degree of keyphrase.

- *The deviation of phrase degree:* The deviation of phrase degree of centre elements should be obtained. This is because more is the deviation of centre elements with other elements in the network, the lower ranking should be given to such keyphrase.

Different attributes are proposed for contextual ranking which are described as follows:

### 6.2.1. Edge Strength Density

The *edge-strength density* gives the importance of a keyphrase by measuring the co-occurring frequency of all the adjacent bi-grams in the keyphrase per unit edge. The number of edges in the keyphrase as obtained from unique path is equal to the total number of words minus one. The edge-strength density is measured as the ratio of the sum of edge weights in a keyphrase to the total number of edges in a keyphrase as shown in Equation 6.12.

$$Edge\ Strength\ density\ (ESD) = \frac{\sum Edge\ weight}{Number\ of\ edges} \qquad 6.12$$

The edge-weight distribution is scaled over the Microblog WCN, it is observed that edge-weight plays significant role in defining important bi-grams in the network. Thus, this attribute is used as a beneficiary attribute.

### 6.2.2. Strength Difference

If there is no other link of words of a keyphrase with other words in the network, then this keyphrase is more important than those which have links with other words in neighbourhood. The set of words which are inter-related as a path and have high edge-strength in the path is defined as influential segment.

The Strength Difference (SD) is defined as the difference between edge-strength density and the average of in-strength and out-strength in the keyphrase as shown in Equation 6.13. The in-strength is calculated for the *first element* and out-strength is calculated for the *last element*.

$$Strength\ Difference\ (SD) = ESD - Avg(In - strength, out - strength) \qquad 6.13$$

It measures the convergence or divergence of a keyphrase. If a keyphrase have high external links with *centre elements*, then the keyphrase is less significant because it may not provide complete contextual information. This is non-beneficiary attribute. If the SD is less, it defines the consistent frequency of set of words in the keyphrase of a network.

### 6.2.3. Phrase Degree

The Phrase Degree (PD) gives information about all the nodes of a keyphrase which is considered as one unit. The degree of this unit is calculated as the total number of incoming and out-going links from this unit as shown in Equation 6.14.

$$Phrase\ Degree\ (PD) = \sum_i \deg(n_i) - (2n - 2) \qquad 6.14$$

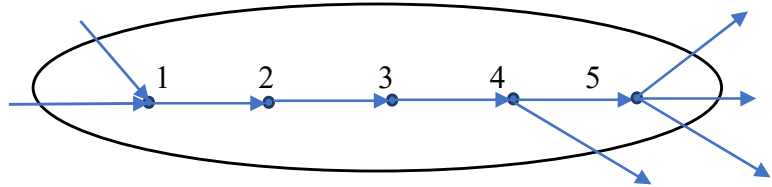

**Figure 6.1:** Phrase degree for Keyphrase extracted using k-bridge decomposition from Microblog WCN

The PD for keyphrases is shown in Figure 6.1, every word in the keyphrase is marked with index from 1 to 5. It is observed that the words from 1 to 5 commute better than the phrase from nodes 2 to 4. From 1 to 5, the edges are connected and any deviation is made at either node 1 or node 5. But for keyphrase 2 to 4, less deviation is made at 2 and 4. If node 2 and node 3 are co-occurring in $X$ number of Tweets, then it is very likely that node 1 and node 2 will also co-occur along with node 2 and node 3. The deviations from nodes usually define the *first element* and the *last element*. The phrase degree is a beneficiary attribute, and has more significance for high values of edge-weights and less deviation.

### 6.2.4. Degree Computation

The Degree Computation (DC) attribute is defined for AHP as a fraction of the sum of $in-degree\ and\ out-degree$ with respect to Phrase Degree of a phrase P as shown in Equation 6.15.

$$Degree\ Computation\ (DC) = \frac{In-degree + Out-degree}{PD\ (P)} \qquad 6.15$$

where $PD\ (P)$ represents *Phrase Degree* of a keyphrase P. The DC measures the deviation of in-degree and out-degree links for boundary nodes. It means that varying deviation among centre nodes is less significant than deviation at boundary nodes. The boundary nodes are the *first element* or *last elements*. This attribute is beneficiary attribute. If there is minimal degree deviation at centre node, the information is the most significant with the value of DC close to 1. This is because in *phrase degree,* most of the degree is contributed by boundary nodes. With minimal deviation at centre node, it shows that all the words occur together and create more meaningful context than their occurrence with other nodes in the network.

## 6.3. Pairwise Comparison Matrix and Consistency Ratio

The *Pair-wise Comparison Matrix (PCM)* is the relativity measure which is used to create the importance based comparison matrix. The *Pairwise Comparison Matrix* which has been proposed for Microblog WCN is shown in Table 6.3.

Table 6.3: Pairwise Comparison Matrix A for proposed attributes

| A | ESD | SD | PD | DC |
|---|---|---|---|---|
| **ESD** | 1.00 | 5.00 | 3.00 | 5.00 |
| **SD** | 0.20 | 1.00 | 0.33 | 1.00 |
| **PD** | 0.33 | 3.00 | 1.00 | 3.00 |
| **DC** | 0.20 | 1.00 | 0.33 | 1.00 |

As the most important factor is the density of a keyphrase which is defined by using the normalized distribution of co-occurrences in a keyphrase as Edge-strength density (ESD). ESD and SD are proposed using edge-weight distribution and among these ESD is strongly important than SD. This comparison is because density (Angel *et al.,* 2012) is considered as important parameter to calculate the importance of a weighted sub-graph. Also, the variation of edge-weight may also depend upon many other factors like position of co-occurring words (Biswas *et al.,* 2018). The PD and DC is proposed on the basis of node-degree distribution. However, PD was found to be weakly important than DC because PD is an important ranking measure for graphical keyphrase extraction techniques, namely, TextRank and TopicRank. Since, SD and DC both defines the variation in edge-weights and deviation of node-degree, respectively, they are considered to be equally important. GM is calculated using Pairwise Comparison Matrix. The geometric mean is obtained as 5.268 and $A_2$ matrix is obtained as shown in Equation 6.16.

$$A_2 = \begin{bmatrix} 2.942/5.268 \\ 0.506/5.268 \\ 1.313/5.268 \\ 0.506/5.268 \end{bmatrix} \quad 6.16$$

After calculating $A_3$ and $A_4$, the CI is obtained as 0.01446 and RI is considered as 0.90 for four attributes. The CR is calculated as 0.0162 which is less than 0.1. Thus, this Pairwise Comparison Matrix is consistent. The $A_2$ matrix is used as *weighted matrix* for experimental results and evaluation.

## 6.4. Experimental results and evaluation

The experimental results and evaluations are performed over FSD dataset. The number of keyphrase so obtained is 17 for the topic *"Richard Bowes, victim of London riots, dies in Hospital"* from FSD dataset and each of these keyphrase is considered as alternatives. For each alternative, different attributes are calculated as shown in Table 6.4. The PCM is obtained and is used for ranking contextual information using AHP optimization mechanism. The results are evaluated for the topic *"Richard Bowes, victim of London riots, dies in Hospital"*, the resulting keyphrases are ranked using AHP as shown in Table 6.4. The values of each attribute are normalized and *normalized decision matrix* is calculated. The product of *normalized decision matrix* and weighted matrix gives the total score for each keyphrase. The keyphrases are ranked in descending order using these scores.

**Table 6.4:** Values of AHP attributes for different keyphrases indexed as alternative phrase number using topic "Richard Bowes, victim of London riots, dies in Hospital" of FSD dataset

| Alternative Phrase Num | ESD | SD | PD | DC |
|---|---|---|---|---|
| 1 | 4.73 | 0.76 | 55 | 0.2 |
| 2 | 4.53 | 0.96 | 54 | 0.203 |
| 3 | 4.86 | 0.13 | 48 | 0.208 |
| 4 | 4.67 | 0.33 | 47 | 0.21 |
| 5 | 8.2 | 4.7 | 37 | 0.19 |
| 6 | 8.6 | 5.6 | 30 | 0.2 |
| 7 | 1.0 | 0.5 | 1 | 1.0 |
| 8 | 3.0 | 0.0 | 9 | 0.67 |
| 9 | 3.71 | 1.21 | 34 | 0.147 |
| 10 | 4.0 | 2.0 | 27 | 0.15 |
| 11 | 1.0 | 0.5 | 1 | 1.0 |
| 12 | 1.0 | 0.5 | 1 | 1.0 |
| 13 | 1.0 | 0.5 | 1 | 1.0 |
| 14 | 1.0 | 0.5 | 1 | 1.0 |
| 15 | 1.0 | 2.0 | 10 | 0.6 |
| 16 | 1.0 | 0.5 | 4 | 0.75 |
| 17 | 1.0 | 0.5 | 1 | 1.0 |

**Table 6.5:** Contextual ranking of influential segments as obtained for topic "Richard Bowes, victim of London riots, dies in Hospital" of FSD dataset

| Rank | Phrase | Number of Words | Inference |
|---|---|---|---|
| 1 | the man attacked ealing has died | 6 | Important headlines |
| 2 | the man attacked ealing west london | 6 | |

| | | | obtained |
|---|---|---|---|
| 3 | richard mannington bowes who was critically injured tried stamp out fire during riots ealing has died | 16 | Detailed description of topic has been obtained |
| 4 | richard mannington bowes who was set upon tried stamp out fire during riots ealing has died | 16 | |
| 5 | richard mannington bowes who was critically injured tried stamp out fire during riots ealing west london | 16 | |
| 6 | richard mannington bowes who was set upon tried stamp out fire during riots ealing west london | 16 | |
| 7 | put out fire during riots ealing has died | 8 | |
| 8 | put out fire during riots ealing west london | 8 | |
| 9 | very sad | 2 | Most frequently used phrase have been obtained. |
| 10 | assaulted rioters | 2 | |
| 11 | camerons respons | 2 | |
| 12 | monday confirmed | 2 | |
| 13 | ukriots seanboscott | 2 | |
| 14 | are murderers | 2 | |
| 15 | stop looting | 2 | |
| 16 | link please | 2 | |
| 17 | world news | 2 | |

**Table 6.6:** Contextual ranking of influential segments as obtained using FA cup dataset

| Rank | Influential Segment | Number of Words | Inferences |
|---|---|---|---|
| 1 | chelseafc final congratulations chelsea liverpool win the facupfinal | 8 | Important headlines obtained |
| 2 | chelseafc final whistle gone chelsea liverpool cfcwembley facupfinal | 8 | |
| 3 | chelseafc final congratulations chelsea win the facupfinal | 8 | |
| 4 | chelseafc final congratulations chelsea liverpool win the cup champions | 9 | |
| 5 | come liverpool damnitstrue pick one chelsea goal ramires facup cfc lfc | 11 | Detailed description of topic has been obtained |
| 6 | chelseafc final whistle gone chelsea liverpool win the cup champions | 10 | |
| 11 | come liverpool damnitstrue pick one chelsea goal cfcwembley facupfinal cfc lfc | 11 | |
| 12 | come liverpool damnitstrue pick one chelsea goal ramires facup cfc lfc | 11 | |
| 13 | come liverpool damnitstrue pick one chelsea goal cfcwembley facupfinal cfc lfc | 11 | |

| 14 | winning the winner | 3 | Most frequently used phrase have been obtained. |
| 15 | are the winner | 3 | |
| 16 | blue today | 2 | |
| 17 | winning the cup champions | 4 | |
| 18 | are the cup champions | 4 | |

Table 6.7: Contextual ranking of influential segments as obtained using US Elections dataset

| Rank | Influential Segment | Inferences |
|---|---|---|
| 1 | for the president obama election | Important headlines obtained |
| 2 | for the white house | |
| 3 | warm congratulations barackobama look forward continuing work together | Detailed description of topic has been obtained |
| 4 | news? local | |
| 5 | warm congratulations president barackobama look forward continuing work together | |
| 6 | barack obama election | Most frequently used phrase have been obtained |
| 7 | tulang yakusa | |
| 8 | congrats barackobama look forward continuing work together | |
| 9 | warm congratulations president obama election | |
| 10 | and congressman | |

**Definition 6.1: Rule of Three Slots:** The rule of three slots shows that the contextual information is classified into three slots namely headlines, description, and relevant phrases.

It is observed that the contextual information is segregated in three slots where the first slot presents the topic headlines, second slot provides the brief information about the topic, and the third one gives relevant phrases. Based on number of words and variation between total number of words in the ranked keyphrase, the contextual information can be classified.

More observations for two other datasets, namely FA cup datset, and US Elections dataset are made for this research work as shown in Table 6.6 and Table 6.7, respectively. It is observed that in all the three datasets, the ranking of contextual information in keyphrase follows the rule of three slots with some minor deviations.

As observed from Table 6.5, Table 6.6, and Table 6.7, the contextual information is ranked such that it can be easily classified into headlines, details and relevant phrases. The headlines are denoted by top rankers which give information about the topic and medium length of the keyphrase. These topics are marked as *event-phrases and sub-event phrases*. The detailed information is obtained from mid rankers which have long length of keyphrases and describes the topic. The relevant phrase is the shortest phrase or short segments which are related to the topic but can not describe the complete topic. Although there are some deviations but majority of the information is classified and thus, event-phrases are obtained as top rankers, as obtained using AHP optimization mechanism. As observed from Table 6.4, the PD contributes the highest variation of value among different alternatives followed by ESD. The importance of ESD is more than that of PD as observed from PCM. On the other hand, DC shows minimal variation which is equally important as that of SD but less important than ESD and PD. The proposed keyphrase extraction and ranking technique is compared with different baseline techniques.

## 6.5. Comparison with Baseline Techniques

In existing literature, different keyphrase extraction techniques were proposed. The baseline measures for keyphrase extraction techniques are TextRank, Topicrank, NErank, and PositionRank. Using the k-bridge decomposition keyphrase extraction technique and contextual ranking of keyphrases using AHP (kBD_AHP), different keyphrases are obtained for each topic of FSD dataset as shown in Table 6.8. It is observed that the information obtained in the form of headlines of keyphrases have high K-REC and K-PREC values as compared to that of detailed information. However, the detailed information gives qualitative event-phrases with well-formed lexical sequence.

**Table 6.8:** Resulting headlines and detailed description of topic specific Tweet Corpus

| Topic | Keyphrase Obtained |
|---|---|
| Death of Amy Winehouse | rip amy winehouse dies from |
| | rip amy winehouse has died the cause death unknown but there are rumors mrs weasley mistook her for bellatrix |
| Space shuttle Atlantis lands safely, ending NASA's space shuttle program | Breakingnews last space shuttle atlantis touching makes final landing |
| | Breakingnews last space shuttle atlantis touching down after orbits around earth and journey miles sts |
| Richard Bowes, victim of London riots, dies in hospital | the man attacked ealing has died |
| | richard mannington bowes who was critically injured tried stamp out fire during riots ealing has died |
| Plane carrying Russian hockey | the plane crash kills |

| team Lokomotiv crashes, 44 dead | the plane carrying khls lokomotiv crashed was taking members hockey team crashes reported dead |
| --- | --- |
| Gunman opens fire in children's camp on Utoya island, Norway | breakingnews gunman dressed policeman has attacked labour party youth camp island just outside oslo norway the wake |
| | skynewsbreak sky sources widespread reports attack childrens summer camp island just outside oslo norway gunman dressed police officer |

Table 6.9: Keyphrases extracted using different keyphrase extraction techniques for the topic of FSD dataset

| *Techniques* | Keyphrase@1 | Keyphrase@5 |
| --- | --- | --- |
| *TextRank* | Injured | injured, injured assaulted, needless, hospital, midnight |
| *TopicRank* | Riot | riot, rip richard mannington, monda man, stamp fire, west london die |
| *NErank* | Bowes Man | bowes man, tried stamp, richard bowes, london, cameron |
| *kBD_AHP* | Richard Mannington Bowes set upon tried stamp fire riots | Richard mannington bowes set upon tried stamp fire riots, ealing died, Richard Mannington Bowes set upon tried stamp fire riots, ealing west London, critically injured tried stamp fire riots ealing west london |

Table 6.10: Performance evaluation obtained for topics using different keyphrase extraction metrics for FSD dataset

| *Algorithm* | *ROUGE − 1* | | | *ROUGE − L* | | | *ROUGE − 2* | | |
| --- | --- | --- | --- | --- | --- | --- | --- | --- | --- |
| | $N=1$ | $N=3$ | $N=5$ | $N=1$ | $N=3$ | $N=5$ | $N=1$ | $N=3$ | $N=5$ |
| *TextRank* | 19.13 | 24.07 | 29.16 | 0.67 | 0.77 | 1.00 | 5.55 | 5.55 | 5.55 |
| *TopicRank* | 26.54 | **54.93** | **56.32** | 0.88 | **2.30** | **2.56** | 11.11 | **27.82** | **27.82** |
| *NErank* | 13.73 | 53.21 | 54.16 | 0.77 | 1.67 | 2.11 | 1.38 | 11.88 | 11.88 |
| *kBD AHP* | **35.49** | 50.46 | 55.40 | **1.77** | 2.11 | 2.33 | **25.39** | 26.63 | 26.63 |

It is observed in Table 6.10, the kBD_AHP outperforms all the existing techniques in terms of ROUGE scores for N=1. For high values of N, the TopicRank outperforms all the existing techniques in terms of ROUGE scores. Among all the keyphrase extraction techniques as evaluated for FSD dataset, for high values of N, kBD_AHP gives comparable and second best results after TopicRank. The contextual information as obtained using

kBD_AHP is in orderly fashion of lexical sequence of words in the keyphrase as shown Table 6.8.

## 6.6. Concluding Remarks

In this Chapter, the contextual information, as obtained in keyphrases, is ranked using AHP optimization mechanism. The experimental results and evaluation shows that the ranked keyphrases can be classified and it follows the rule of three slots. The contextual information is classified as headlines, detailed description, and relevant keyphrases. Using k-bridge decomposition and AHP contextual ranking mechanism, the keyphrase extraction and ranking techniques is proposed as kBD_AHP and compared with existing techniques. It is observed that the kBD_AHP outperforms existing keyphrase extraction techniques in terms of ROUGE scores for top keyphrases as obtained using automatic evaluation process. Also, the top ranked keyphrases can be marked as event or event-phrase.

# Chapter 7.
# Event Detection from Social Media Data

Event detection technique is the process of identifying events from social media data. Earlier, identification of event was performed using topic modeling (Ferrari *et al.,* 2011; Zhou *et al.,* 2014), finding similar items using Locality Sensitive Hashing (LSH) (Kaleel *et al.,* 2015; Petrovic *et al.,* 2010), using named entities and temporal clustering (Li C *et al.,* 2012, Li R *et al.,* 2012) and identifying bursty keywords (Fung *et. al.* 2005; He, 2007; Mathioudakis *et al.,* 2010). In this research work, event detection technique is proposed using cognitive patterns in Microblog WCN evolved from social media data

## 7.1. Twitter Word Co-occurrence Model: Heuristic Approach

The TWCM is proposed using Heuristic approach of decomposition of Microblog WCN as described in Chapter 5. The pseudo-code for the proposed technique Twitter Word Co-occurrence Model (TWCM) is described in Algorithm 7.1. The set of Tweets is collected using Tweet-id and is considered as one document $d \in D$, where $D = \{d_1, d_2, \dots d_n\}$ is the collection of Tweets where $d_k$ represents each Tweet. The Tweet Corpus D is pre-processed by *automatic tokenization* and is given as an input to TWCM. The directed Microblog WCN is created using Tweet Corpus D. The Microblog WCN ($G_D$) has following properties:

**Property 1**: *Directed graph to preserve the ordered lexical sequence of Tweets*.

**Property 2**: *Weighted graph to record word to word co-occurrence frequency*.

**Property 3**: *Nearest-neighbour edging because in all-pair neighbour edging word to word co-occurrence is computationally inexpensive.*

After creating $G_D$, the edges having higher edge weights are identified and are used for identifying the most frequent bi-grams. These bi-grams are arranged in descending order on the basis of edge-weights. Top k number of edges are obtained with the highest edge-weights where k is given in Equation 7.1.

$$k = \sqrt{n_e} \quad \text{where } n_e \text{ is total number edges in } G_D \qquad 7.1$$

**Definition 7.1: Edge Sub-List $E_{sl}$:** The list of top k number of edges having highest edge-weight is termed as *edge sub-list* and is denoted by $E_{sl}$.

The *Edge sub-list* represents the set of most frequently co-occurring words in Tweet Corpus. As each Tweet contains not more than 140 characters, they contain important words and bi-grams because repetitive information is posted in limited space. The edges which are not in edge sub-list are removed from the graph $G_D$.

**Definition 7.2: Directed Sub-graph Component $G_{sl}$:** The resulting graph obtained by removing edges from $G_D$ which are not in *edge sub-list* ($E_{sl}$) is termed as *directed sub-graph component* ($G_{sl}$).

Any cycle in any *directed sub-graph component* is identified and removed by removing cycles in sub-graph. The cycles in sub-graph are removed by removing the least weighted edge in the cycle. From sub-graph components obtained in $G_{sl}$, the topological sorting is used to obtain different phrases from $DAG$ which are called keyphrases.

**Definition 7.3: Event-Phrase:** The ordered lexical sequence of keywords obtained from directed sub-graph component which gives contextual information about a topic is termed as keyphrases. Only single keyphrase is obtained from each sub-graph of Microblog WCN evolved from single topic-specific Tweets. This keyphrase is considered as event-phrase for TWCM.

---

**Algorithm 7.1:** Twitter Word Co-occurrence Model

---

1. **Initialize**: *Set of Twitter Feeds* $\rightarrow D$
2. For each tweet $D_i$ in $D$:
3.     Pre-processing of $D_i$
4. Create directed Word Co-occurrence Network ($G_D$) of Twitter Feeds D
5. $n_e \leftarrow$ *Total number of edges in* $G_D$
6. Threshold limit $k \leftarrow \sqrt{n_e}$
7. $E_l \leftarrow$ *Descending order of weighted edges of* $G_D$
8. $E_{sl} \leftarrow$ *Top k weighted edges in* $G_D$
9. Create Word Co-occurrence Network ($G_{sl}$) from Edges Sub-list $E_{sl}$
10. For each $S_{sl}$ sub-graph component obtained in $G_{sl}$:
11. *Phrase* $\leftarrow$ *Topological Sorting or* $S_{sl}$
12.     Add *Phrase* in *List of Events* $L_e$
14. Return *List of Events* $L_e$

---

**Table 7.1:** Event phrases obtained using different techniques for Topic "Plane carrying Russian hockey team Lokomotiv crashes, only one survived" from FSD dataset

| Topic →<br>Technique ↓ | **Plane carrying Russian hockey team Lokomotiv crashes, only one survived** | Recall (%) | Redundancy (%) |
|---|---|---|---|
| **FPM-DSV** | Breakinngnews crash hockey ice lokomotive majority members plane, Russia hockey team urt victims | 60 | 33.33 |
| **SFPM** | Plane crash Russia hockey lokomotiv Russian carrying khl crashed dead | 60 | 50 |
| **HUPC** | Plane team crash, hockey team crashed, lokomotiv plane, hockey plane team | 50 | 80 |
| **TRCM** | #Breakingnews #hockey #majority# #khl #lokomotiv | 20 | 0 |
| **TWCM** | Ice charter urt Russian crash kills hockey terrible tragedy Russia plane carrying khl team lokomotiv crashed only one person survived crashes reported dead | 100 | 20 |

**Table 7.2:** Event phrases obtained using different techniques for a Topic "Space shuttle Atlantis lands safely, ending NASA's space shuttle program" from FSD dataset

| Topic →<br>Technique ↓ | **Space shuttle Atlantis lands safely, ending NASA's space shuttle program** | Recall (%) | Redundancy (%) |
|---|---|---|---|
| **FPM-DSV** | Atlantis centerkennedy space urt, lands shuttle space | 50 | 25 |
| **SFPM** | Shuttle atlantis space year NASA program sts NASAs lands kennedy final center home florida mission earth NASA2019s around miles programme | 75 | 33.33 |
| **HUPC** | Shuttle, atlantis, shuttle space, atlantis shuttle, atlantis urt, atlantis shuttle space, space urt, atlantis space, kennedy space, atlantis NASA shuttle, program shuttle space, NASA space | 75 | 83.3 |
| **TRCM** | Shuttle NASA sts Atlantis NASAs, NASA Atlantis, atlantis shuttle, NASAs Atlantis Shuttle | 37.5 | 100 |
| **TWCM** | Landed NASAUS NASA year kennedy space center orbits shuttle atlantis programme, around earth journey miles | 75 | 16 |

To validate and compare the proposed technique with baseline methods, namely, *FPM-DSV, SFPM, HUPC and TRCM*, the FSD dataset is used. There are total 27 topics which contain information about variety of events namely *bomb explosion, organ transplant, plane crashes, riots, terrorist attack, death, earthquake, peace pact, credit rating and many more*. The results for topic *"Plane carrying Russian hockey team Lokomotiv crashes, only one survived"* and *"Space shuttle Atlantis lands safely, ending NASA's space shuttle program"* are shown in Table 7.1 and Table 7.2, respectively. It is observed that the existing technique *FPM-DSV* records the frequent patterns as keyphrases and the patterns which are more than 75% similar to each other are extracted as single pattern. The frequent patterns of words in keyphrases are obtained using *FPM-DSV*. The redundancy is increased due to overlapping. The multiple frequent patterns are obtained using *HUPC*. *HUPC* has improved recall due to structured algorithm introduced for handling frequent patterns.

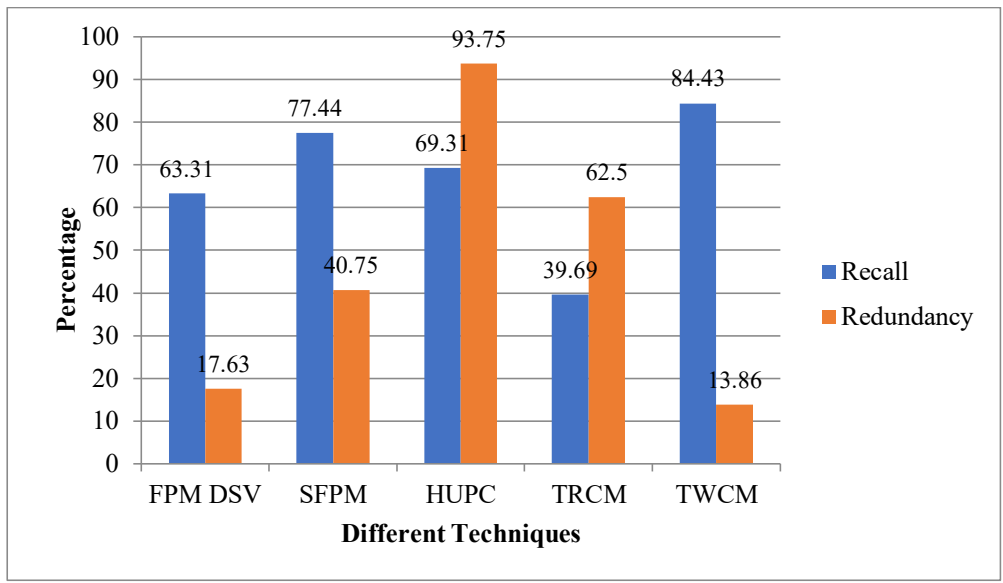

**Figure 7.1:** Recall (%) and Redundancy (%) achieved by different techniques

The *TRCM* approach uses Hashtags for rule mining. Hashtags do not always describe events and gives low recall value for *TRCM*. The repetitive Hashtags and combination of Hashtags in resulting key-phrase description shows higher redundancy rate. *SFPM* outperforms *TRCM*, *FPM-DSV*, and *HUPC* in terms of recall and redundancy measures. The top *k* terms and initial elements in set S (final list of keywords) are defined by user in *SFPM*. The reference corpus is required for term selection in *SFPM*. To remove this limitation of initialization with static parameters, a graphical event detection technique *TWCM* is proposed which act as preamble for this research work. *TWCM* outperforms all the existing event detection techniques for social media data

The number of words in existing and proposed techniques is different due to varying and changing parameters. The precision is the ratio of the number of relevant records retrieved to the total number of records retrieved. The relevance of words as obtained from keyphrase extraction cannot be compared by words of topic because the ground truth topic does not contain complete information. The keyphrases which have been obtained contains redundant frequent pattern. Thus, precision performance measure is found to be an inappropriate measure for comparison and validation of TWCM with existing techniques. It is observed that the TWCM outperforms all the existing techniques using performance measures *recall (%) and redundancy (%)* as shown in Figure 7.1.

It is observed that the existing techniques have higher redundancy rate. Redundancy rate is an important performance measure because when results are obtained, the repetitive information is provided about same topic. This information makes the process computationally expensive and thus, lowers the speed. TWCM achieves the maximum recall rate with 84.43% and the minimum redundancy rate with 13.86% among all existing techniques. On the contrary, the highest redundancy rate is given by HUPC 93.75% and lowest recall rate is given by TRCM which is probably due to use of the Hashtag in TRCM. The TWCM is the preamble for event detection technique which is proposed using cognitive patterns in Microblog WCN.

## 7.2. An Event Detection Technique for social media data

A graph-based event detection technique is proposed for processing discrete set of Tweets from streaming social media data. The idea is to use cognitive patterns in Microblog WCN to extract sub-graphs in which each sub-graph represents contextual information in the form of keyphrase. These keyphrases are ranked to obtain top $N$ keyphrases which are termed as *event-phrase*.

The architecture for an event detection technique for social media data is shown in Figure 7.2. The streaming social media data gives discrete set of Tweets from which events can be detected. To extract the keywords with high frequency and keyphrases about peak events, summaries should be obtained from discrete set of Tweets. This discrete set of Tweet is used to create Microblog WCN.

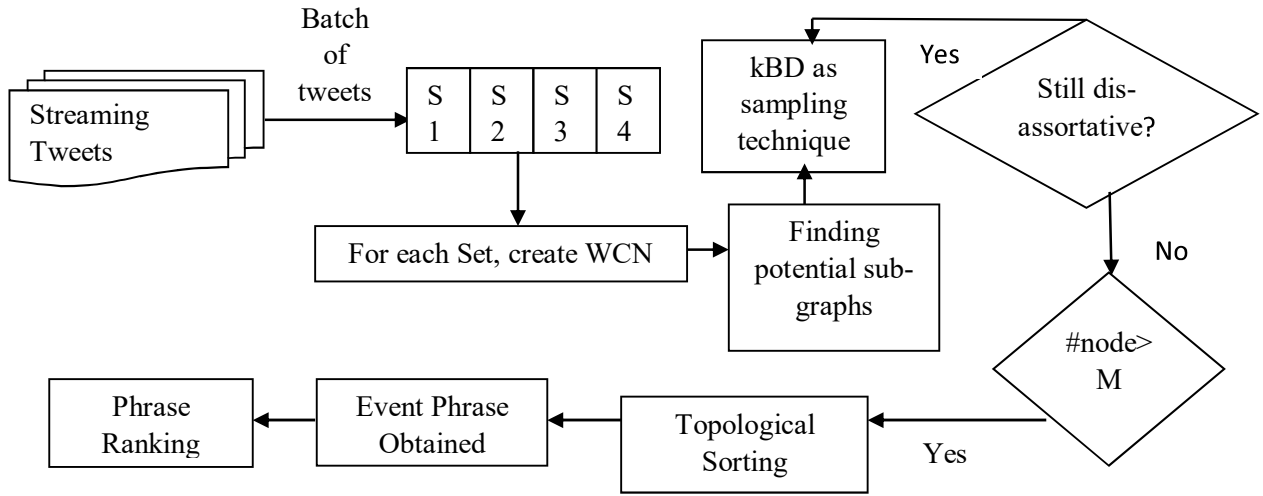

**Figure 7.2:** Architecture for event detection technique for social media data

### 7.2.1. Workflow for an event detection technique for social media data

The probability of distribution of edge strength in Microblog WCN shows significant patterns as observed in Chapter 3. This edge-strength distribution shows that beyond some threshold, which is observed as $10^2$, the distribution follows linear dependency. It is observed that the in-degree distribution and out-degree distribution over Microblog WCN evolved from the discrete set of Tweets shows significant and similar patterns. The edge-weight distribution follows scale-free property. These network science properties are used to propose an event detection technique using k-bridge decomposition and assortativity property for Microblog WCN. The sub-events which are not posted at equally high frequency may or may not be obtained as an event phrase. If low weighted edges are not placed adjacent to high weight edges, then after k-bridge decomposition, even less discussed or topics can be identified.

**Table 7.3:** Notations used for event detection technique

| Notation | Name | Meaning |
|---|---|---|
| $t_i$ | $i^{th}$ Tweet | Tweet having the value of index as $i$ |
| $\Delta T$ | Window size | Number of tweets in a batch |
| $d_i$ | Tokenized Tweet $t_i$ | Number of words in a tweet $t_i$ |
| $w_n$ | Tokenized word | $n$ number of tokenized words in a tweet |
| $C_k$ | Collection of Unique words | Total $k$ unique words in a discrete set of tweets which represents number of nodes in WCN |
| $t_s$ | Domain stop-words | Most frequent $t$ words which are considered as domain specific stop-words to detect sub-events |
| $G_{WCN}$ | Microblog WCN | The WCN evolved from Tweet in discrete set |
| $G_{WCN_i}$ | Sub-component of $G_{WCN}$ | $i^{th}$ sub-graph component in Microblog WCN |
| $M_q$ | The Parameter to Qualify | Minimum number of nodes to qualify graph |

|  | Graph |  |
| --- | --- | --- |
| $m$ | Multiplicative factor | Multiplicative factor to calculate $M_q$ |
| $M_G$ | The Parameter for Event Phrase extraction | Minimum number of nodes in graph for extracting event phrases |
| $G_Q$ | Qualified Graph | A graph which is qualified for processing. It is based on minimum number of nodes required in a graph |
| $E_p$ | Event Phrase | Set of words which represent an event |
| $S_{E_p}$ | Set of Event Phrase | The set of events obtained from discrete set |

### 7.2.1.1. Discrete Social Media Data Stream

As the size of Tweet Corpus is large, the discrete batches of Tweets are considered for finding summaries. The number of Tweets in a batch of discrete social media data stream is denoted by the window size ($\Delta T$).

**Definition 7.4: Sliding Window:** The window keeps on moving to collect discrete set of social media streams as a batch of Tweets. This window is called *Sliding Window*.

---

**Algorithm 7.2:** Iterative Processing of Discrete Set: Event_Detection($\Delta T, M_q, M_G, G_Q$)

---

**Input**: $\Delta T, M_G, G_Q, t_s, \mathrm{m}$
**Output**: $S_{E_p}$
$M_q = m * M_G$
**While** ($\Delta t\ != Null$)
    **For** $t_i$ in $\Delta t$:
        $d_i = Pre - process\ and\ tokenize\ (t_i, t_s)$
        $C_k = unique\_words()$
        $G_{WCN} = Construct\ microblog\_WCN(nodes = C_k, edges = path(d_i))$
        **For** $G_{WCN_i}$ in $G_{WCN}$:
            **if** $len(G_{WCN_i}.nodes()) > M_q$
                $G_Q = G_{WCN_i}$
                $E_p = ObtainEvents(M_G, G_Q)$
                $S_{E_p}.append(E_p)$
            **End if**
        **End for**
    **End for**
    **Return** $S_{E_p}$
**End while**

---

**Definition 7.5: Discrete Set:** The batch of Tweets which is obtained from discrete social media stream is considered as temporary Tweet Corpus which is called *Discrete Set*.

Consider each Tweet, $t_i$ where $i^{th}$ Tweet denotes the indexing of Tweet. The set of Tweets is extracted as discrete set using window size ($\Delta T$). To capture $\Delta T$, set of Tweets is pre-processed. The window is moved from one batch to another after processing. The $\Delta T$ parameter can be used for divergence of time in seconds, minutes, and hours. Formally, let $d_i$ denotes the set of words in a Tweet $t_i = \{w_1, w_2, w_3 \dots w_n\}$ where $n$ is the number of words in any Tweet $t_i$. The collection of all the unique words from Tweets in discrete set is denoted by $C_k$ where $k$ represents the number of unique words in the discrete set.

### 7.2.1.2. Identifying Events

The key idea of the proposed approach is to identify events as topic of discussion in discrete set of Tweets. All the events should be extracted irrespective of the frequency of discussion. To solve this problem, decomposition based keyphrase extarction approach is used for Microblog WCN. The WCN component should be validated as a qualified graph forn decomposition.

**Algorithm 7.3**: Identifying Events from Qualified Graph: $\text{ObtainEvents}(M_G, G_Q)$

***Input***: $M_G, G_Q, E_p = [\ ]$
***Output***: $E_p$
***While*** $(G_Q \text{ is } dis - assortative)$
　　　$K - bridge\_decomposition(G_Q)$
***End while***
***If*** $len(G_Q.nodes()) > M_G$
　　　$E_p.append(Topological\_sorting(G\_Q))$
*end if*
***Return*** $E_p$

***Definition 7.6: Qualified Graph:*** The Microblog WCN evolved from discrete set of Tweets may or may not contain disconnected sub-graphs. The disconnected graph having minimum number of nodes ($M_q$) gets qualified for further processing and is called *Qualified Graph* as shown in Algorithm 7.3.

### 7.2.1.3. K-Bridge Decomposition: Assortativity based termination

The qualified graph is taken as an input for k-bridge decomposition. In this module, the edges having lowest edge-weights are removed from the graph iteratively. After each iteration the assortativity is checked for each component of the resulting graph.

The nature of Microblog WCN is disassortative. Disassortative property shows that the nodes in the sub-graph are connected to each other in such a way that they do not follow "Rich get Richer and poor get poorer" phenomenon. This shows that higher degree nodes are connected to lower degree nodes and lower degree nodes may be connected to higher degree nodes. Thus, the k-bridge decomposition is performed till the break-event point obtained and network gets non-disassortative. This is because the resulting non-disassortative network shows high cohesion among nodes of the sub-graph. It is also observed that the resulting sub-graph may contain any number of nodes. However, presence of one or two nodes in the resulting sub-graph may not give useful results. Thus, the minimum number of nodes in resulting sub-graph $M_G$ should be given as input by the user.

**Definition 7.7: Event Graph:** The resulting non-disassortative graph is said to be an *Event Graph* as it identifies the set of event related keywords in the resulting sub-graph.

**Definition 7.8: Qualified Event Graph**: If the total number of nodes in Event Graphs is more than the threshold $M_G$, the resulting graph qualifies for event-phrase extraction which is called *Qualified Event Graph*.

**Definition 7.9: Event Phrase:** The phrases or set of words in a segment which describes the title or topic of an event to provide contextual information about an event or sub-event is defined as an *Event Phrase*.

Thus, the resulting set of nodes in directed event graphs undergo topological sorting. The key-phrases are obtained which are taken as alternatives for ranking the contextual information. Top ranked keyphrases are marked as *event phrases*, as discussed earlier.

### 7.2.1.4. Ranking Event Phrases

It is observed that the Event Phrases which are obtained in the results may or may not show redundant information or sub-events. The contextual information in key-phrases is ranked using AHP to obtain as top ranked key-phrases. The different attributes are proposed for AHP as discussed in Chapter 6. The reason behind not using DF-IDF (Aeillo *et al.,* 2013) and other keyword ranking based techniques like Twitter Keyword Graph (TKG) (Abilhoa *et al.,* 2014), Selectivity Based Keyword Extraction (SBKE) (Beliga *et al.,* 2016), and Keyword Extraction using Collective Node Weight (KECNW) (Biswas *et al.,* 2018) was to rank key-phrases based on semantics of cognitive patterns rather than frequency of occurrence of a keyword.

### 7.2.2. Experimental results and evaluation

The experimental results are performed over the 72 Twitter TDT dataset. The Microblog WCN follows cognitive patterns. A novel event detection technique is performed over discrete sets of Tweets to extract summaries during iterative processing.

### A. Experimental setup

The experiments are performed for 72-Twitter-TDT dataset FA Cup. The FA Cup is the Football Association challenge Cup which is world-wide competition since 1871. In 2012, Tweets on final match are extracted and thirteen one-minute windows which were highly informative (Aiello *et al.*, 2013) are marked with annotated dataset. These topics which are marked are the sub-events of the major event namely "Football Match of FA Cup". The ground truth is marked with keywords and the story is extracted using the keywords from annotated dataset to evaluate the performance.

To evaluate and validate the performance with existing techniques, it is compared with BNgram and Nguyen's Method (Nguyen *et al.*, 2017). In this research work, the experiments are carried out and values of parameters are set as $\Delta T = 200, m = 2, M_q = 2, and\ t_s = 0.01$. These parameters are tuned for different types of data. The performance measures used for experiments and evaluation are T-REC, K-PREC, and K-REC.

### B. Experimental Results: FA Cup Dataset

As observed from existing techniques (Nguyen *et al.*, 2017), the T-REC for *Nguyen method* is same as that for BNgram. However, Nguyen method outperformed BNgram method in terms of K-PREC but BNgrams gives better result for K-REC. In

Table **7.4**, event-phrases are obtained for FA cup dataset using proposed event detection technique. Event phrases are mapped with ground truth keyword specific information in annotated dataset. The values of T-REC, K-PREC, and K-REC for different event detection techniques show that our method gives best performance in terms of K-PREC and K-REC as shown in Table 7.5. It is observed that K-PREC and K-REC are improved upto 30% without affecting T-REC. Moreover, the performance is analyzed further for $N = 10$ and T-REC, K-PREC, and K-REC is obtained as 0.846, 0.7718, and 0.7807, respectively.

**Table 7.4:** Event Phrases obtained for FA Cup dataset using proposed event detection technique for social media data

| File | Events Obtained | Ground Truth |
|---|---|---|
| 16_16 | finalfacup kick-off cfc ktbffh | [kick kicked]   [0-0 0 0];start;off |
| 16_26 | goal cfcwembley @chelsea | [ramires chelsea];[goal 1-0 1 0] score;yes |
| 16_41 | great mazy run kalou box gets ambushed liverpool defence | [salomon kalou]; run; box mazy |

**Table 7.5:** Performance evaluation for comparison of event detection techniques

| Dataset | Technique | T-REC | K-REC | K-PREC |
|---|---|---|---|---|
| | LDA | 0.6923 | 0.682 | 0.163 |
| | SFPM | 0.6154 | 0.657 | 0.234 |
| FA Cup | BNgram | **0.7692** | 0.587 | 0.355 |
| | Nguyen *et al.* | **0.7692** | 0.548 | 0.453 |
| | Our Method | **0.7692** | **0.717** | **0.719** |

### 7.2.3. Effectiveness of different parameters

The parameters used for evaluating Event Phrases for FA cup are tuned manually. For increase in the size of window $i.e. \Delta T > 200$, the K-PREC begins to deteriorate. This could be due to the fact that the information obtained from Tweets is user-generated and thus, resulting event phrases may contain ill-formed text. The ill-formed text may contain redundant terms which may or may not be counted as single node in Microblog WCN. Similarly, reducing the size might miss important segments due to heated discussion about random events.

Another important factor which is observed as critical is Domain stop-words ($t_s$). Increasing the domain specific stop-words may remove important terms like 'gggggooooollll' and many other. Decreasing this parameter may result in extracting common terms which are related to events rather than sub-events. Another important parameter is the parameter to qualify graph ($M_q$) which defines minimum number of words in the resulting sub-graph. It is not necessary that multiple keywords should be combined to tell a story. Even the small set of words or even two words can mark a story, for instance, 'Chelsea goal'. Such phrases are re-tweeted multiple times and are important.

## 7.3. Concluding Remarks

An event detection technique is proposed for social media data. The proposed technique outperforms existing techniques because it follows cognitive patterns among words in a Microblog WCN and the network science property *assortativity* measure is used to identify the association of terms with each other. More the words co-occur with each other, better are the cognitive patterns extracted. The resulting sub-graph gets non-disassortative. Thus, the nodes connect with each other and forms dense sub-graph which gives important key-phrase. This is further ranked using AHP optimization technique to obtain top ranked keyphrases as event phrases.

# Chapter 8.
# Conclusion and Future Work

In a nutshell, various conclusions are made based on different keyphrase extraction and event detection techniques. Initially, the structure of Microblog WCN is analyzed using network metrics, network models, and network science properties. Different observations are made and decomposition based keyphrase extraction technique, BArank, is proposed based on these observations. The BArank is used to extract keyphrases and event-phrases are identified using contextual ranking of information in *keyphrases,* which are considered as alternatives for AHP. An event detection technique is proposed to extract event-phrase using BArank by finding summaries from discrete set of streaming Twitter data. This research work is concluded and summarized in this Section. Also, the scope of future work and possible extensions is discussed in this Section.

## 8.1. Conclusion

The existing graphical keyword extraction and keyphrase extraction techniques of literature are studied and analyzed over Tweet Corpus. It is observed that the value of different network science metrics plays a significant role in identifying keyword and relevant information about keywords. Thus, there exists the need to study network science properties for Microblog WCN. In this research work, a novel approach of event detection is proposed for social media data by identifying and using cognitive patterns from Microblog WCN.

The Tweet Corpus is used to generate Microblog WCN. The structure of Microblog WCN is analyzed for different network science properties. It gives useful insights about the patterns among words which are used in Tweets. It is observed that the structure of Microblog WCN is different than that of traditional WCN. The networks of words give information about different patterns due to use of active language of users. The Microblog WCN is scale-free for the *node-degree distribution*, the *node-strength distribution*, and *edge-weight distribution*. The Microblog WCN follows *small-world property* and is *disassortative*.

It shows partial hierarchical organization for *all-pair neighbour edging* network and spectral behavior of Microblog WCN is discussed using *eigenvalues*.

It is observed that the Microblog WCN is scale-free for *node-degree* and *edge-degree* distribution. Thus, decomposition based keyphrase extraction technique is proposed. The traditional graph-based keyphrase extraction techniques are applied over social media data, and it is observed that due to presence of ill-formed text, the performance is deteriorated. The cognitive patterns of Microblog WCN are used to propose the k-bridge decomposition which decomposes a network and retains high weighted edges only.

A pilot study is carried out for identifying key-phrases using heuristic values for decomposition of Microblog WCN. It is observed that the *Root Two* measure gives best results among all the other heuristic measures, namely, *divided by 2, divided by 3, and Log Method*. The heuristic approach is based on human intelligence and is abstractive. Thus, *k-core decomposition* based network model is used to propose *k-bridge decomposition* using edge-weight. The termination of decomposition of Microblog WCN depends upon the static controlling parameter which is the maximum number of nodes for each resulting sub-graph. Due to iterative removal of the lowest edge-weight, the proposed technique is computationally expensive. Thus, a threshold value $t$ is defined for sub-graphs to remove all the lower weighted edges. The proposed keyphrase extraction technique outperforms all the existing keyphrase extraction techniques for the K-PREC performance measure. A novel keyphrase extraction technique, BArank is proposed to extract keyphrases using *k-bridge decomposition* and *assortativity* for automatic termination of decomposition in Microblog WCN.

The keyphrases give the contextual information about an event or sub-event. The contextual information ranking is performed using an optimization mechanism, Analytical Hierarchical Process (AHP). Some network science based attributes are proposed to extract top key-phrases which represent event-phrases. These attributes are used for contextual ranking and as a result, the information is segregated into three slots namely headline, description, and relevant phrases. This classification of data into three parts is called *the rule of three slots*. The top-ranked key-phrases are the headlines are marked as sub-event phrases.

An event detection technique for social media data is proposed using the BArank. The discrete set of Tweets is obtained from streaming Tweets using Tweepy API. Keyphrases are obtained by finding summaries from discrete sets. The key-phrases are ranked using AHP to obtain top N event-phrases. Based on the experiments and evaluations for FA cup dataset, the proposed method outperforms existing techniques for K-REC and K-PREC without affecting T-REC.

## 8.2. Future Work

In future, an event detection technique can be proposed using different cognitive patterns of Microblog WCN for following areas:

- **Line Networks:** The networks of co-occurring words which are considered as edges in Microblog WCN are used as nodes and nodes of Microblog WCN are used as edges. For instance, in a Tweet *"Amy Winehouse Died",* the edge will be *'Winehouse'*. But the nodes will be *(Amy, Winehouse), (Winehouse, Died)*. The conversion of nodes and edges of Microblog WCN into edges and nodes, respectively, forms a new network called *line network*. The structure and dynamics for line networks can give useful insights for SMA.
- **Streaming Data:** The event detection for streaming data can be analyzed using the dynamics of Microblog WCN. For every new Tweet, the nodes and edges are added and removed from the Microblog WCN. The community detection is performed for dense sub-graphs as obtained using high edge-weights on streaming data.
- **Heterogeneous Information Networks (HIN):** This approach is used for social media data in recent years by academic researchers. The HIN has different types of nodes and edges which denote entities and the relationship between these entities, respectively. The nodes and relations can be used for identifying meta-paths in the network. The HIN for social media data may use *entities* like location, temporal data, user id, tweet id and *relations* like follower-followee, re-tweeted by, at-mentioned by and many more. These patterns can be used for identifying similar Tweets and finding clusters of Tweets.

This research work can be expanded on multi-channel and multi-modal data for extracting events from heterogeneous type of data, for instance, textual data, images, videos, and other cross-media data. The patterns of the network can be analyzed for dissemination of the outbreak which can be used as an added feature for event detection. The proposed event detection model can be used as an application to traffic congestion detection and management, awareness programs, and disaster and emergency management.

In future, the inverse co-occurrence frequency edge-weights can be used for keyphrase extraction using decomposition of Microblog WCN (Abilhoa *et al.,* 2014). The proposed BArank techniques can be implemented and validated over well-formed documents. Efforts can be made to solve other problem domains using network science properties and network models for different types of WCN. The spatio-temporal meta-data of Tweets can be used for SMA using the structure and dynamics of Microblog WCN in streaming social media data. The Microblog WCN can be studied by using *skip-1* and *skip-2* nearest neighbour edging in the network. The proposed event detection for social media data can be explored for huge amount of social media data to validate its scalability and reliability over different types of Tweets.